\documentclass[10pt]{emulateapj}
\usepackage{apjfonts}
\usepackage{graphicx}
\usepackage{psfig}
\newcommand{\begit}{\begin{itemize}}
\newcommand{\enit}{\end{itemize}}
\newcommand{\begen}{\begin{enumerate}}
\newcommand{\enen}{\end{enumerate}}

\newcommand       \be           {\begin{equation}}
\newcommand       \ee           {\end{equation}}
\newcommand       \bea          {\begin{eqnarray}}
\newcommand       \eea          {\end{eqnarray}}

\newcommand{\beqa}{\begin{eqnarray}} 
\newcommand{\eeqa}{\end{eqnarray}}

\begin{document}

\title{On the Conditions for Neutron-Rich Gamma-Ray Burst Outflows}

\author{Brian D. Metzger\altaffilmark{1,3}}\author{Todd A. Thompson\altaffilmark{2,4}}\author{Eliot Quataert\altaffilmark{1}}

\altaffiltext{1}{Astronomy Department and Theoretical Astrophysics Center, 601 Campbell Hall, Berkeley, CA 94720; bmetzger@astro.berkeley.edu, eliot@astro.berkeley.edu }
\altaffiltext{2}{Department of Astrophysical Sciences, Peyton Hall-Ivy Lane, Princeton University, Princeton, NJ 08544; thomp@astro.princeton.edu}
\altaffiltext{3}{Department of Physics, 366 LeConte Hall, University of California, Berkeley, CA 94720}
\altaffiltext{4}{Lyman Spitzer Jr. Fellow}

\begin{abstract}

We calculate the structure and neutron content of neutrino-heated magnetohydrodynamic winds driven from the surface of newly-formed magnetars (``proto-magnetars'') and from the midplane of hyper-accreting disks, two of the possible central engines for gamma-ray bursts (GRBs) and hyper-energetic supernovae (SNe).  Both the surface of proto-magnetars and the midplane of neutrino-cooled accretion flows (NDAFs) are electron degenerate and neutron-rich (neutron-to-proton ratio $n/p \gg 1$).  If this substantial free neutron excess is preserved to large radii in ultra-relativistic outflows, several important observational consequences may result.  Weak interaction processes, however, can drive $n/p$ to $\sim 1$ in the nondegenerate regions that obtain just above the surfaces of NDAFs and proto-magnetars.  Our calculations show that mildly relativistic (Lorentz factor $\Gamma \lesssim 10$) neutron-rich outflows from NDAFs are possible in the presence of a strong poloidal magnetic field.  However, neutron-rich winds possess a minimum mass-loss rate that likely precludes simultaneously neutron-rich and ultra-relativistic ($\Gamma \gtrsim 100$) NDAF winds accompanying a substantial accretion power.  In contrast, proto-magnetars are capable of producing neutron-rich long-duration GRB outflows $\sim 10-30$ seconds following core bounce for sub-millisecond rotation periods; such outflows would, however, accompany only extremely energetic events, in which the GRB + SN energy budget exceeds $\sim 4\times 10^{52}$ ergs.  Neutron-rich highly relativistic outflows may also be produced during some $short$-duration GRBs by geometrically thick accretion disks formed from compact object mergers.  The implications for $r$-process nucleosynthesis, optical transients due to non-relativistic neutron-rich winds, and Nickel production in proto-magnetar and NDAF winds are also briefly discussed.

\end{abstract}


\section{Introduction}


\label{section:intro}

The rapid variability and large energies that characterize cosmological gamma-ray bursts (GRBs) strongly implicate stellar-mass compact objects as their central engines.  Indeed, the association of several long-duration GRBs (LGRBs) with Type Ibc supernovae (SNe) suggests that LGRBs result from relativistic outflow accompanying rapid accretion onto a newly-formed black hole (a ``collapsar''; Woosley 1993; MacFadyen $\&$ Woosley 1999) or the spin-down of a newly-formed magnetar (e.g., Usov 1992; Thompson 1994; Blackman $\&$ Yi 1998; Wheeler et al.~2000; Thompson, Chang, $\&$ Quataert 2004, hereafter TCQ04).  Short-duration GRBs may result from black hole accretion-powered outflows following the tidal disruption and coalescence of compact binaries (Paczy{\'n}ski 1986, 1991; Eichler et al.~1989; Narayan et al.~1992; Ruffert et al.~1997; Janka et al.~1999).  

A unique property of both the surface of a newly-formed neutron star (a ``proto-neutron star'' or PNS; Burrows $\&$ Lattimer 1986) and in some cases the midplane of hyper-accreting disks is a significant excess of free neutrons (electron fraction $Y_{e} \ll 0.5$), resulting from $\beta$-equilibrium under electron degeneracy (Pruet et al.~2003; Beloborodov 2003a, hereafter B03a).\footnote{In this paper we define a neutron excess as a neutron-to-proton ratio $n/p = (1-Y_{e})/Y_{e} > 1$, where we assume free nucleons and $Y_{e}$ is the proton-to-baryon ratio or ``electron fraction.''}  While degeneracy at a PNS's neutrinosphere is assured, hyper-accreting disks are viscously heated and only possess a neutron excess in places where they are sufficiently dense and efficiently neutrino-cooled.  If present at all, these neutrino-dominated accretion flows (or NDAFs) are thus geometrically-thin and confined to small radii in the disk (Popham et al.~1999).  Recent neutrino-cooled $\alpha$-disk calculations show that an NDAF only forms outside the last stable orbit for steady-state mass accretion above a critical ``ignition'' rate, given by
\be
\dot{M}_{\rm ign} \approx 0.07(0.02)M_{3}^{4/3}\alpha_{0.1}^{5/3} M_{\sun}{\rm\,s^{-1}} 
\label{mdotign}
\ee
for black hole mass $M= 3 M_{3} M_{\sun}$ and spin $a = 0(0.95)$, where $\alpha = 0.1\alpha_{0.1}$ is the disk viscosity parameter (Chen $\&$ Beloborodov 2007, hereafter CB07).  Disk mass accretion rates ($\dot{M}_{D}$) greater than $\dot{M}_{\rm ign}$, and thus neutron-rich NDAFs, are plausible in both collapsar and binary merger scenarios (e.g., Popham et al.~1999).

For lower accretion rates or larger radii than characterize NDAFs, cooling is inefficient and the accretion is geometrically-thick and quasi-virial, forming an advection dominated accretion flow (Narayan $\&$ Yi 1994; Narayan et al.~2001).  Although NDAFs must come into $\beta$-equilibrium before accreting (B03a), this is not necessarily the case for thick disks; the neutron content of a thick disk may therefore depend on the matter that feeds it.  We discuss the likely neutron content of thick disk winds more in $\S\ref{section:thickdiskwinds}$ and $\S\ref{section:thickdisks}$.

If the neutron excess that characterizes proto-magnetars and NDAFs can be preserved to large radii in the outflows that they power (where $Y_{e}$ obtains its asymptotic value $Y_{e}^{a}$; see Table \ref{table:defs} for definitions of commonly used variables), observable consequences may result.  The dynamics of the GRB jet may be altered by the presence of a dominant neutron component (Derishev, Kocharovsky, and Kocharovsky 1999), which at proton-neutron decoupling could result in the emission of multi-GeV photons and neutrinos (Bahcall \& M{\'e}sz{\'a}ros 2000; M{\'e}sz{\'a}ros $\&$ Rees 2000; Razzaque $\&$ M{\'e}sz{\'a}ros 2006a) and cause a significant reheating of the outflow (Rossi et al.~2006).  Furthermore, a large neutron-to-proton ratio $n/p$ may reduce the fireball's asymptotic baryon contamination (Fuller et al.~2000; Vlahakis et al.~2003), contribute an additional component to the GRB light curve (Pruet \& Dalal 2002; Bulik, Sikera, $\&$ Moderski 2002; Fan $\&$ Wei 2004; Fan et al.~2005; Dermer $\&$ Atoyan 2006) and a unique beta-decay signature (Razzaque $\&$ M{\'e}sz{\'a}ros 2006b), alter the subsequent afterglow emission (Beloborodov 2003b), and affect the fireball's nucleosynthetic yield (Lemoine 2002; Pruet et al.~2002; B03a).  Although not all of these consequences strictly require a neutron excess, most are far more conspicuous when $n/p \gg 1$, partly because only excess neutrons will ultimately remain free if $\alpha$-particle formation is efficient (e.g., Lemoine 2002).  Identification or strong upper limits on any of these effects would teach us much about the composition of GRB outflows.

In this paper we examine the processes that shape the neutron content of outflows from GRB progenitors, motivated by the promise that the distinctive neutron-rich GRB signature holds as a tool for constraining the elusive central engine, whose properties are masked by the outflow's otherwise rather generic dynamical evolution (e.g., the ``fireball'' model; Rees $\&$ M{\'e}sz{\'a}ros 1992).  In particular, we focus on NDAFs and proto-magnetars rather than thick accretion disks, not because these models are necessarily favored to produce GRBs, but rather because for neutron-rich central engines an asymptotically neutron-rich outflow is plausible $a$ $priori$.  The goal of our analysis is to determine the conditions under which and degree to which these neutron-rich central engines can produce equally neutron-rich outflows.

Determining the asymptotic neutron content of winds driven from the neutron-rich base of PNSs and NDAFs is nontrivial because the neutron fraction will evolve due to weak interactions under the comparatively nondegenerate conditions that characterize scales immediately larger than that of the central engine.  In $\S \ref{section:deneutronization}$ we discuss the relevant processes that may ``deneutronize'' the outflow, driving $n/p$ back to $\sim 1$.  Indeed, in $\S \ref{section:thermal}$ we show that thermally-driven outflows from PNSs and NDAFs are generally deneutronized by, if nothing else, electron neutrino absorption.  Although thermally-driven winds of this kind possess little or no neutron excess, they also cannot produce GRBs because they do not reach ultra-relativistic speeds (Lorentz factor $\Gamma \gg 1$).  Winds driven directly from the surface of PNSs or the midplane of NDAFs require an energetically-dominant Poynting flux to reach large $\Gamma$ because of the significant mass-loss driven by viscous and neutrino heating (e.g., Levinson $\&$ Eichler 1993); this requires rapid rotation and a strong magnetic field.  If, through such enhanced magnetic acceleration matter is advected from the PNS surface or the NDAF midplane sufficiently rapidly to avoid deneutronization, the outflow will retain the large neutron-to-proton ratio that characterizes its degenerate base.  The magnetocentrifugal acceleration required to maintain low $Y_{e}$ (high $n/p$), however, also enhances the wind's mass-loss rate over its purely thermally-driven value (TCQ04).  This raises the question of whether simultaneously neutron-rich and ultra-relativistic outflows are possible under physically realizable conditions.  

To address these issues quantitatively we have calculated the neutron content of magnetically-driven winds from proto-magnetars and hyper-accreting NDAF disks by solving the equations of one-dimensional neutrino-heated magnetohydrodynamics (MHD).  We have studied in detail the effects of magnetic fields and rotation on PNS winds in a previous work, assuming a constant $Y_{e}$ (Metzger, Thompson, $\&$ Quataert 2007, hereafter MTQ07).  In $\S \ref{section:magnetar}$ we include the evolution of $Y_{e}$ in these calculations in order to determine the asymptotic electron fraction $Y_{e}^{a}$ from proto-magnetar outflows.  In $\S \ref{section:disk}$ we adapt our calculations to the NDAF context by following outflow from the accretion disk midplane for several flux tube angles, employing the $\alpha$-disk NDAF models of CB07 as boundary conditions.  We present a summary of our results in $\S\ref{section:discussion}$, including a discussion of the prospect for neutron-rich outflows from central engines of both long and short-duration GRBs.  Our analysis and conclusions can be summarized in the broadest terms as follows: the conditions for $n/p \gg 1$ are simultaneously the conditions for short advection timescale, large mass-loss rate, and low asymptotic Lorentz factor.  Thus only under very restrictive conditions do both $n/p \gg 1$ and high $\Gamma$ obtain.  The most promising possibilities appear to be outflows from sub-millisecond proto-magnetars and from geometrically thick disks with $\dot{M}_{D} \lesssim \dot{M}_{\rm ign}$ (see Table \ref{table:scenarios}).

\subsection{Deneutronizing Processes}

\label{section:deneutronization}

Despite the presence of neutron-rich material at the base of NDAF and PNS outflows, this neutron excess may not be preserved.  Because conditions above the PNS surface and the accretion disk midplane are typically nondegenerate, equilibrium between the pair-capture reactions 
\be 
e^{-}+p\rightarrow n+\nu_{e},
\label{eprate}
\ee
\be 
e^{+}+n \rightarrow p+\bar{\nu}_{e}
\label{enrate}
\ee 
favor $Y_{e}>0.5$ in the potentially pair-rich atmosphere through which the wind must accelerate.  Furthermore, in the presence of an electron neutrino(antineutrino) energy density $u_{\nu_{e}}$($u_{\bar{\nu}_{e}}$) the inverse, neutrino absorption reactions 
\be
\nu_{e}+n \rightarrow e^{-}+p,
\label{nunrate}
\ee 
\be
\bar{\nu}_{e}+p \rightarrow e^{+}+n,
\label{nuprate}
\ee 
which dominate pair-captures abruptly above the outflow's launching surface, drive the system toward an asymptotic electron fraction given by
\be 
Y_{e}^{\nu} \equiv \left(1+\frac{u_{\bar{\nu}_{e}}}{u_{\nu_{e}}}\frac{\langle\epsilon_{\bar{\nu}_{e}}\rangle-2\Delta + 1.2\Delta^{2}/\langle\epsilon_{\bar{\nu}_{e}}\rangle}{\langle\epsilon_{\nu_{e}}\rangle+2\Delta+1.2\Delta^{2}/\langle\epsilon_{\nu_{e}}\rangle}\right)^{-1}, 
\label{yeanu}
\ee
where $\Delta = 1.293$ MeV is the neutron-proton mass difference, $\langle\epsilon_{\nu_{e}}\rangle$($\langle\epsilon_{\bar{\nu}_{e}}\rangle$) is the mean electron neutrino(antineutrino) energy (Qian et al.~1993; Qian $\&$ Woosley 1996, hereafter QW96), and the superscript $\nu$ denotes that the electron fraction given by equation (\ref{yeanu}) is set solely by the properties of the local neutrino radiation field.  

Equation (\ref{yeanu}) shows that when the electron neutrino and antineutrino fluxes are comparable and have a similar spectrum, as is generically the case for NDAFs and PNSs during the latter's Kelvin-Helmholtz cooling phase (Burrows $\&$ Lattimer 1986), equilibrium between neutrino absorptions also favors a comparatively neutron-poor state ($Y_{e} \sim 0.5$).\footnote{During deleptonization, efficiently neutrino-cooled central engines release slightly more $\nu_{e}$'s than $\bar{\nu}_{e}$'s; in addition, the $\nu_{e}$ and $\bar{\nu}_{e}$ spectra (and thus mean energies) also differ slightly due to the difference between the mean $e^{-}$ and $e^{+}$ energies, the $e^{-}$ and $e^{+}$ capture cross sections, and, in the neutrino optically thick case, the $\nu_{e}$ and $\bar{\nu}_{e}$ neutrinosphere temperatures and geometries.  Although $Y_{e}^{\nu}\lesssim 0.5$ is possible in some cases, these relatively modest effects are unlikely to yield $Y_{e}^{\nu} \ll 0.5$.}  Such a deneutronizing luminosity of neutrinos must be present, self-consistently, for matter to cool to the dense, degenerate conditions required for low $Y_{e}$ in the first place.  The total neutrino luminosity from a neutron-rich NDAF, for instance, must exceed
\be
L_{\nu,{\rm ign}} \equiv \eta\dot{M}_{\rm ign}c^{2} \approx 5\times 10^{51}M_{3}^{4/3}\alpha_{0.1}^{5/3}{\rm\,ergs\,\,s^{-1}},
\label{lign}
\ee
where $\eta \approx 0.04(0.15)$ for $a = 0(0.95)$ (CB07) and we have used equation (\ref{mdotign}) for $\dot{M}_{\rm ign}$.  Similarly, detailed numerical calculations show that a cooling PNS's electron neutrino/antineutrino luminosity is approximately given by
\be
L_{\nu} \approx 10^{52}(t/\rm {\,1\,s})^{-1}{\rm\,ergs\,\,s^{-1}} 
\label{lnupns}
\ee
from a time $t \approx 1$ s after core bounce until the end of the Kelvin-Helmholtz epoch at $t = \tau_{\rm KH} \sim 10-100$ s (see, e.g., Pons et al.~1999, Fig.~14).

To contrast the large neutron fraction in an NDAF's midplane with the much lower value favored in equilibrium with the NDAF's neutrino flux, Figure \ref{plot:ye_eq} shows the midplane electron fraction $Y_{e}^{D}$ ($dashed$ $line$) and the equilibrium electron fraction set by neutrino absorption $Y_{e}^{\nu}$ ($solid$ $line$) as a function of disk cylindrical radius $R_{0}$ (in units of gravitational radii $R_{g} \equiv GM/c^{2}$) for a steady-state NDAF solution taken from CB07 with $\alpha = 0.03$, $M = 3M_{\sun}$, $a=0$, and $\dot{M}_{D} = 0.2 M_{\sun}$ s$^{-1}$.  The local neutrino energy densities ($u_{\nu_{e}}$,$u_{\bar{\nu}_{e}}$) and mean energies ($\langle\epsilon_{\nu_{e}}\rangle$,$\langle\epsilon_{\bar{\nu}_{e}}\rangle$) used to calculate $Y_{e}^{\nu}$ from equation (\ref{yeanu}) were obtained by integrating the total flux incident on a given position just above the disk midplane at radius $R_{0}$, where $Y_{e}^{a}$ for an outflow launched near $R_{0}$ is set.  The disk neutrino emission is assumed to originate from axisymmetric annuli of negligible vertical height with a radial structure taken from CB07's one-dimensional, height-integrated calculations.  We assume that all relevant lines of site are optically thin in calculating $Y_{e}^{\nu}$; this is a good approximation because the vertical neutrino optical depth through the disk is $\lesssim 1$ at all radii and because the disc scale height increases more rapidly than $\propto R_{0}$ so that the outer disk's atmosphere has an unobstructed view of the interior flow.  Differential gravitational redshifts between emission and absorption radii are taken into account, although geodesic bending is ignored.   

Figure \ref{plot:ye_eq} shows that the disk midplane is very neutron-rich for $R_{0} \lesssim 30-100 R_{g}$ (the NDAF portion of the disk), reaching a neutron-to-proton ratio as large as $\sim 30$ at small radii.  However, Figure \ref{plot:ye_eq} also shows that $Y_{e}^{\nu} \sim 0.5$ at all radii, so that if the outflow comes into equilibrium with the disk's neutrino luminosity it will be driven back to a relatively neutron-poor state ($n/p \sim 1$).  Also note that $Y_{e}^{\nu} > 0.485$ at all radii, allowing possible $^{56}$Ni synthesis in a disk wind, again should the nucleons come into equilibrium with the neutrino flux (Pruet et al.~2004).\footnote{Although we find that $Y_{e}^{\nu} > 0.485$ at all disk radii using mean neutrino energies from CB07's height-integrated disk calculations, precisely whether $Y_{e}^{\nu} > 0.485$ or $Y_{e}^{\nu} < 0.485$ is difficult to determine with confidence because the neutrino spectra are sensitive to the disk's vertical temperature profile, which is theoretically uncertain, and to neutrino transport if the disk midplane is neutrino optically thick (Sawyer 2003).  If viscous heating is important in the wind, additional entropy deposition may drive $Y_{e}^{a} \gtrsim 0.485$, allowing $^{56}$Ni production even if $Y_{e}^{\nu} < 0.485$.}  Since the outer disk is radiatively inefficient and therefore particularly prone to large mass outflows (Blandford $\&$ Begelman 1999), disk wind-aided stellar explosions provide one way to produce optically luminous SNe in collapsar models for LGRBs (MacFadyen $\&$ Woosley 1999), where Ni masses up to $\sim 0.5 M_{\sun}$ have been inferred (e.g., GRB980425/SN1998bw; Iwamoto et al.~1998; Woosley, Eastman, $\&$ Schmidt 1999).

\section{Thermally-Driven Winds}
\label{section:thermal}

\begin{figure*}
\centerline{\hbox{\psfig{file=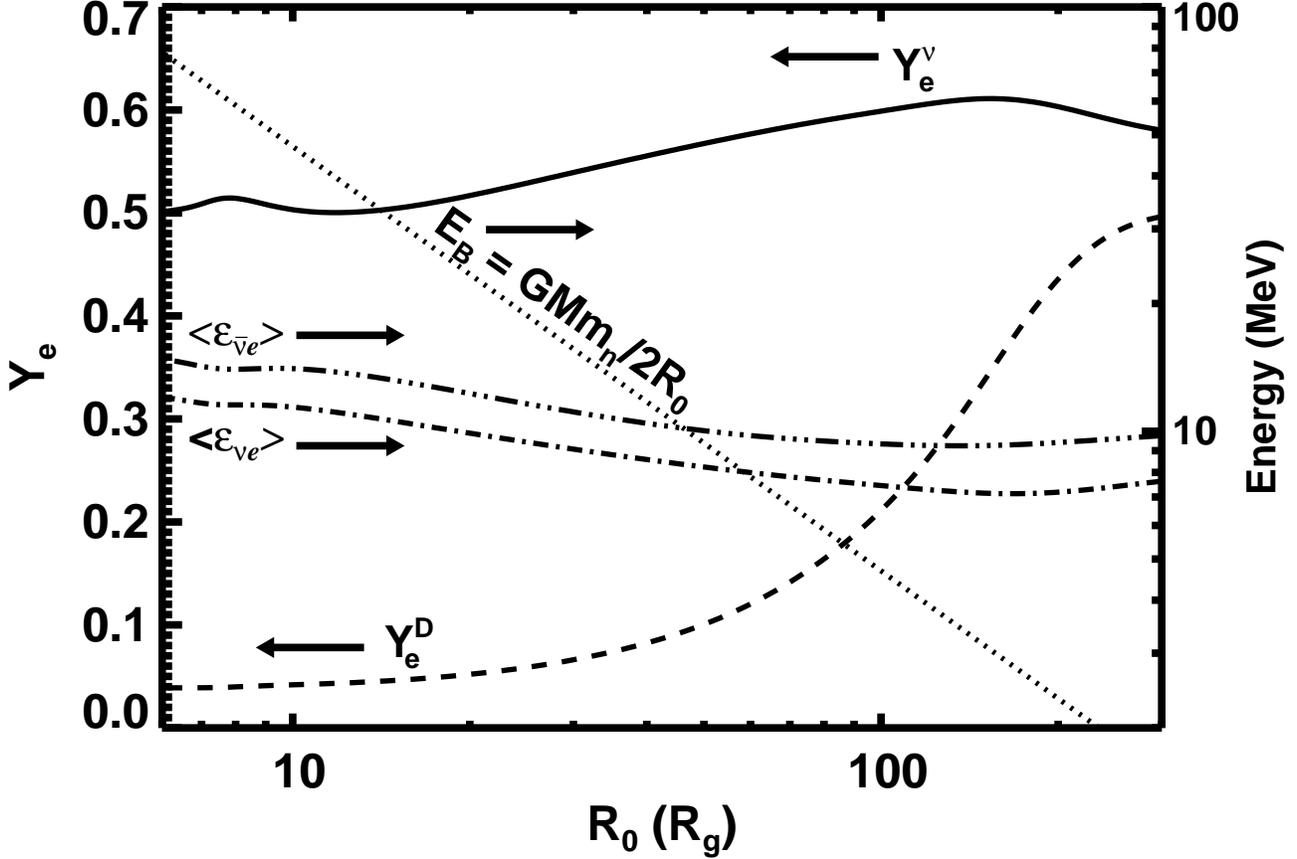,width=18cm}}}
\figcaption[x]{[LEFT AXIS] Electron fraction in neutrino absorption equilibrium $Y_{e}^{\nu}$ (eq.~[\ref{yeanu}]; $solid$ $line$) as a function of the wind launching radius $R_{0}$ (in units of gravitational radii $R_{g} = GM/c^{2}$) calculated from the thin, height-integrated $\alpha-$disk accretion model of Chen $\&$ Beloborodov (2007) with $\alpha = 0.03$, black hole mass $M = 3 M_{\sun}$, black hole spin $a = 0$, and accretion rate $\dot{M}_{D}$ = 0.2 $M_{\sun}$ s$^{-1}$.  Also shown is the midplane electron fraction $Y_{e}^{D}$ ($dashed$ $line$) taken directly from CB07's calculation.  Note that the disk midplane is very neutron-rich inside $R_{0} \sim 30 R_{g}$, with a neutron-to-proton ratio $n/p = (1-Y_{e})/Y_{e}$ exceeding 10.  However, if outflow driven from the disk comes into equilibrium with the disk's neutrino radiation then the asymptotic neutron-to-proton ratio is driven back to $\sim$ 1 because $Y_{e}^{\nu} \sim 0.5$ at all radii.  Also note that neutrino equilibrium favors $Y_{e}^{a} > 0.485$ at large radii, a requisite condition for producing $^{56}$Ni in neutrino-driven outflows. [RIGHT AXIS] The dot-dash(triple-dot-dash) line shows the mean electron neutrino(antineutrino) energy incident on the base of outflows driven from radius $R_{0}$.  Also shown ($dotted$ $line$) is the binding energy $E_{B}$ of a nucleon in a Keplerian thin disk, where the black hole's gravitational potential is assumed to be Newtonian for simplicity.  The disk is the most neutron-rich at radii where $E_{B} \gtrsim \langle\epsilon_{\nu_{e}}\rangle, \langle\epsilon_{\bar{\nu}_{e}}\rangle$ and thus where more than one neutrino absorption is required to unbind a nucleon.  This shows that purely neutrino-driven winds from small radii in the disk must come into equilibrium with neutrino absorptions and will thus obtain $Y_{e}^{a} \simeq Y_{e}^{\nu} \sim 0.5$ ($n/p \sim 1$).  At larger radii ($R_{0} \sim 50-200 R_{g}$), $E_{B} \lesssim \langle\epsilon_{\nu_{e}}\rangle, \langle\epsilon_{\bar{\nu}_{e}}\rangle$, and so deneutronization of a wind by neutrino absorption is less likely.
\label{plot:ye_eq}}
\end{figure*}


In spite of nondegenerate pair captures and neutrino absorptions (eqs.~[\ref{eprate}]$-$[\ref{nuprate}]), the high neutron fraction in NDAFs and proto-magnetar outflows will be preserved if deneutronization proceeds slower than a fluid element's advection from the surface.  In evaluating this possibility in the case of NDAFs, previous investigations have assumed that the disk is turbulent and that the relevant outflow rate is of the order of the turnover frequency of a typical turbulent eddy ($\sim \Omega_{\rm K}$, the Keplerian rotation rate), concluding that low $Y_{e}$ is preserved under most conditions appropriate to hyper-accreting disks (Pruet et al.~2003; B03a).  Specifically, Pruet et al.~(2003) argue that because NDAFs are generally dense, degenerate, and gas pressure-dominated, if a fluid element is carried out of the midplane adiabatically in an eddy turnover time, pair creation is somewhat suppressed and low $Y_{e}$ is preserved (i.e., $Y_{e}^{a} \approx Y_{e}^{D}$ obtains).  B03a also compares $\Omega_{\rm K}$ to the neutrino absorption rates (eqs.~[\ref{nunrate}]$-$[\ref{nuprate}]), reaching a similar conclusion to Pruet et al.~(2003) except for very high disk neutrino luminosities ($L_{\nu} \gtrsim 10^{53}$ erg s$^{-1}$), corresponding to $\dot{M}_{D} \gtrsim 1 M_{\sun}$ s$^{-1}$.  This leaves a wide range of astrophysically plausible accretion rates ($\dot{M}_{\rm ign} \lesssim \dot{M}_{D} \lesssim 1M_{\sun}$ s$^{-1}$) for which outflow, if it escapes the disk in an eddy turnover time, would be asymptotically neutron-rich.

Several arguments can, however, be raised against the conclusion that low $Y_{e}$ is preserved in NDAF outflows.  Although some degree of turbulence and turbulent mixing must accompany the accretion torque (e.g., via the MRI; Balbus $\&$ Hawley 1998), its scale and efficiency are unclear.\footnote{NDAFs cool efficiently and are not unstable to radial convection, except perhaps at small radii in the case of neutrino optically thick accretion (Lee et al.~2005); it is also not clear that NDAFs are vertically-convective (Hirose, Krolik, $\&$ Stone 2006), as this depends on the unknown vertical viscous dissipation profile (see the discussion in Blaes 2007).}  Even if present, large-scale turbulence is not likely to drive outflow from a thin disk; thus, an eddy turnover time is not the relevant timescale to compare to the weak interaction rates.  Outflow from the surface of a thin accretion disk must be heated, passing through a sonic point as it accelerates out of the black hole's potential.  Thus, the advection time of a self-consistent, viscous or neutrino-heated wind sets the residence time for a fluid element in regions of potential deneutronization.  This timescale is generally much longer than $\Omega_{{\rm K}}^{-1}$.  

Indeed, the neutrino-driven PNS winds that accompany the Kelvin-Helmholtz cooling of slowly rotating, non-magnetized PNSs generically come into equilibrium with the neutrino absorption rates, obtaining $Y_{e}^{a} = Y_{e}^{\nu}$ (QW96; Thompson et al.~2001, hereafter T01).  To see that the neutron fraction of a purely neutrino-driven NDAF wind must also come into equilibrium with the neutrino rates we adapt an argument first presented by QW96 in the PNS context.  First, note that a typical nucleon with mass $m_{\rm n}$ launched in a wind from radius $R_{0}$ of a thin Keplerian disk requires an energy $E_{B} \simeq GMm_{\rm n}/2R_{0} \approx 80 (R_{0}/6R_{g})^{-1}$ MeV to escape the black hole's gravitational potential, where we have assumed Newtonian gravity for simplicity.  In addition to $Y_{e}^{D}$ and $Y_{e}^{\nu}$, in Figure $\ref{plot:ye_eq}$ we show for comparison $E_{B}$ and the mean electron neutrino and antineutrino energies used to calculate $Y_{e}^{\nu}$.  The mean neutrino energies vary only weakly with radius because the neutrino flux released from an NDAF's inner disk is relatively hot (Popham et al.~1999; CB07) and dominates the neutrino heating at all outflow radii (relative to the neutrino emission from larger radii in the disk).  By balancing viscous heating and optically thin neutrino cooling at the radius $R_{\rm p} \sim 1-10 R_{g}$ where most of the disk's emission originates, we find that an NDAF's mean neutrino energy incident on any disk radius is approximately given by 
\be \langle\epsilon_{\nu}\rangle \approx 5.1T_{D} \approx 18\alpha_{0.1}^{1/5}M_{3}^{-1/5}(R_{\rm p}/6R_{g})^{-3/10}{\rm\,MeV},
\label{enuthin}
 \ee
where $T_{D}$ is the disk's midplane temperature at $R_{\rm p}$.  Equation (\ref{enuthin}), which agrees well with our calculation of $\langle\epsilon_{\nu}\rangle$ in Figure $\ref{plot:ye_eq}$, is valid so long as the inner disk is optically thin to neutrinos ($\tau_{\nu} \lesssim 1$), which remains true for accretion rates below $\approx 0.7(0.05)\alpha_{0.1}M_{\sun}{\rm\,s^{-1}}$ for electron neutrinos and $\approx 2.0(0.14)\alpha_{0.1}M_{\sun}{\rm\,s^{-1}}$ for electron antineutrinos for $a = 0(0.95)$ and $M = 3M_{\sun}$ (CB07).  For larger $\dot{M}_{D}$ the disk is opaque to neutrinos ($\tau_{\nu_{e}},\tau_{\bar{\nu}_{e}} \gtrsim 1$) and the mean neutrino energy is approximately given by its Fermi-Dirac blackbody value
\be \langle\epsilon_{\nu}\rangle \approx 3.2T_{\nu} \approx 16M_{3}^{-1/2}\dot{M}_{D,0.1}^{1/4}(R_{\rm p}/6R_{g})^{-1/2}{\rm\,MeV}, 
\label{enuthick}
\ee
where $\dot{M}_{D} = 0.1\dot{M}_{\rm D,0.1}M_{\sun}{\rm \,s^{-1}}$ and $T_{\nu}$ is the temperature at the disk's neutrinosphere. 

Despite the relatively high mean neutrino energies shown in Figure $\ref{plot:ye_eq}$ and implied by equations $(\ref{enuthin})-(\ref{enuthick})$, $\langle\epsilon_{\nu}\rangle$ is still much less than $E_{B}(\sim R_{\rm p})$ because NDAFs are efficiently neutrino-cooled.  Thus, each nucleon in a thermally-driven outflow originating from an inner NDAF must absorb several neutrinos in the process of being unbound from the black hole's potential.  Through these multiple absorptions the wind's electron fraction is unavoidably brought into equilibrium with the neutrino flux at $Y_{e}^{\nu}$ (QW96), a consequence of the fact that neutrino absorptions (eqs.~[\ref{nunrate}]$-$[\ref{nuprate}]) both dominate the wind heating and determine $Y_{e}^{a} \simeq Y_{e}^{\nu}$.

This conclusion may hold even for winds driven from near the NDAF's relatively loosely-bound outer edge.  NDAFs only exist interior to an ``ignition'' radius $R_{\rm ign}$, which we estimate as the location at which optically thin cooling by nondegenerate pair capture on free nuclei (eqs.~[\ref{eprate}]$-$[\ref{enrate}]) balances viscous heating for a thick disk:\footnote{Equation (\ref{rign}) overestimates $R_{\rm ign}$ for large $R_{\rm ign}$ (i.e., large $\dot{M}_{D}$ and low $\alpha$) because cooling via electron capture requires a threshold energy $\Delta - m_{e}c^{2} \sim 1$ MeV that exceeds the disk temperature at large radii.  Furthermore, NSE favors composite nuclei for $R_{0} > R_{\alpha} \sim 100 R_{g}$; nuclear disintegration is a significant heat source around $R_{0} \sim R_{\alpha}$ and we have overestimated cooling for $R_{0} > R_{\alpha}$ because pair capture rates are lower on composite nuclei.}
\be
R_{\rm ign} \approx 10\dot{M}_{D,0.1}^{6/5}M_{3}^{-8/5}\alpha_{0.1}^{-2}R_{\rm g}.
\label{rign}
\ee
For the disk parameters associated with the solution in Figure \ref{plot:ye_eq} ($\alpha = 0.03, M = 3M_{\sun}, \dot{M}_{D} = 0.2M_{\sun}$ s$^{-1}\gg \dot{M}_{\rm ign}$), equation (\ref{rign}) gives $R_{\rm ign} \approx 100 R_{\rm g}$, corresponding to the radius interior to which the disk possesses a significant neutron excess ($Y_{e}^{D} \ll 0.5$).  By requiring that $R_{\rm ign}$ exceed the radius of the innermost stable circular orbit, $R_{\rm isco} \simeq 6(1)R_{g}$ for black hole spin $a = 0(0.95)$, one recovers the numerically-determined value for $\dot{M}_{\rm ign}$ given in equation (\ref{mdotign}) to reasonable accuracy.   

Using equations (\ref{enuthin}) and (\ref{enuthick}), the ratio of the binding energy at $R_{\rm ign}$ to the mean neutrino energy from the inner disk is approximately given by 
\begin{eqnarray}
\frac{E_{B}(R_{\rm ign})}{\langle\epsilon_{\nu}\rangle} & & \sim 3\dot{M}_{\rm D,0.1}^{-6/5}M_{3}^{9/5}\alpha_{0.1}^{9/5}\left(\frac{R_{\rm p}}{6R_{g}}\right)^{3/10} :  \tau_{\nu} \lesssim 1 \nonumber \\
& &  \sim 3\dot{M}_{\rm D,0.1}^{-29/20}M_{3}^{21/10}\alpha_{0.1}^{2}\left(\frac{R_{\rm p}}{6R_{g}}\right)^{1/2}:  \tau_{\nu} \gtrsim 1 \nonumber \\ 
\label{eratio}
\end{eqnarray}
Equation (\ref{eratio}) shows that $E_{B}(R_{\rm ign})\gtrsim \langle\epsilon_{\nu}\rangle$ for $\alpha \gtrsim 0.1$ and for most physical values of $\dot{M}_{D}$; thus, winds thermally-driven from even $R_{0} \sim R_{\rm ign}$ would have $Y_{e}^{a} \approx Y_{e}^{\nu} \sim 0.5$.  Although equation (\ref{eratio}) implies that winds driven from the outer portions of high-$\dot{M}_{D}$, low-$\alpha$ NDAFs may remain neutron-rich, the astrophysical situations most likely to result in extended disks, collapsars and He core-black hole mergers (Fryer $\&$ Woosley 1998), are also likely to have lower values of $\dot{M}_{D}$.  Furthermore, NDAFs become unstable to self-gravity (Toomre Q < 1) at $R_{\rm ign}$ for $\dot{M}_{D} \gtrsim 2\alpha_{0.1}^{10/7}M_{3}^{3/11}M_{\sun}$ s$^{-1}$, comparable to the accretion rates for which $E_{B}(R_{\rm ign}) < \langle\epsilon_{\nu}\rangle$.  The dynamics at these radii is not well understood so it is difficult to draw definitive conclusions concerning the neutron content of outflows launched from large radii in high-$\dot{M}_{D}$, low-$\alpha$ disks.  

In addition to neutrino heating, viscous heating may thermally drive outflows from the surface of NDAFs.  Pruet et al.~(2004) have shown, in the context of a wind driven by viscous heating supplied through a simple $\alpha$ prescription, that entropy added to the outflow lifts electron degeneracy before the outflow falls out of $\beta$-equilibrium.  This drives $n/p$ to a value near unity via nondegenerate pair captures, even without the aid of neutrino absorption.  However, the ability of viscous heating to generically drive outflows from thin disks is unknown, as this depends in detail on where energy is deposited, a major uncertainty in current thin disk theory (Blaes 2007).  Recent radiation MHD thin disk simulations suggest that very little energy dissipation occurs in the disk corona (Turner 2004; Hirose, Krolick, $\&$ Stone 2006; Blaes et al.~2006b; Krolik et al. 2007), where a wind would most likely be launched.  Regardless of potential viscous entropy contributions, the existence of the neutrino flux discussed in the preceding paragraphs is assured.  We thus conclude that purely thermally-driven winds from the vicinity of compact objects, even those with a significant surface neutron excess, are unlikely to be neutron-rich asymptotically.

\section{Magnetically-Driven Winds}

\label{section:magnetically}
Winds driven directly from the surfaces of NDAFs and PNSs solely by viscous or neutrino heating are inherently mass-loaded and thus make poor candidates for producing GRBs.  The high energy-to-baryon ratio required of ultra-relativistic GRB outflows (asymptotic wind Lorentz factor $\Gamma \gtrsim 10-100$; e.g., Lithwick $\&$ Sari 2001) can, however, be achieved if rapid rotation and a strong poloidal magnetic field supply an energetically dominant Poynting flux.  The ratio of Poynting flux to kinetic energy flux at the light cylinder $R_{\rm L} \equiv c/\Omega$ is given by (for $\sigma > 1$)
\be 
\sigma \equiv \left.\frac{B^{2}}{4\pi \rho c^{2}}\right|_{R_{\rm L}},
\label{magnetization} 
\ee
where $B$ is the magnetic field strength, $\rho$ is the rest mass density, $\Omega$ is the rotation rate of the central star or disk, and we have assumed that the outflow is moving only mildly relativistically at $R_{\rm L}$.  If the magnetic energy is fully converted into the kinetic energy of bulk motion, either directly or through thermalization and subsequent thermal or magnetic pressure-driven expansion (e.g., Spruit $\&$ Drenkhahn 2002), then $\Gamma \sim \sigma$; Poynting-flux dominated GRB outflows therefore require $\sigma \gtrsim 10-100$.   

In the case of magnetically-driven, high-$\sigma$ NDAF or PNS winds, matter may be advected from the wind's base sufficiently quickly to remain effectively adiabatic, with its initial degeneracy never lifted due to insufficient heating.  The pair-capture reactions (eqs.~[\ref{eprate}]$-$[\ref{enrate}]) then continue to favor low $Y_{e}$ above the PNS surface or disk midplane.  In the PNS case this ineffective heating becomes manifest as an exponential drop in the asymptotic wind entropy with increasing $\Omega$, which occurs for rotation periods $P = 2\pi/\Omega \lesssim 2-3$ ms (see eq.~[37] of MTQ07) in the presence of a sufficiently strong surface dipole magnetic field ($B^{\rm dip}_{\nu} \sim 10^{14}-10^{15}$ G); note that the large fields and rapid rotation required are similar to those required for proto-magnetars to produce LGRBs in the first place.  It is therefore likely that GRB jets from proto-magnetars or NDAFs are not completely deneutronized by pair captures.  

Even with degeneracy intact and pair production suppressed, to remain neutron-rich a GRB-producing wind must advect material from its base sufficiently rapidly to overcome neutrino absorptions.  This requires significant magnetocentrifugal support in the outflow's inner, hydrostatic atmosphere, where $Y_{e}^{a}$ is set.  For the reasons discussed in $\S\ref{section:thermal}$, magnetocentrifugal forces acting on a wind launched from a radius $R$ must contribute a factor $\sim GMm_{\rm n}/R\langle\epsilon_{\nu}\rangle \gg 1$ more energy than neutrino heating in unbinding the outflow to avoid deneutronization.  Inevitably, such acceleration in the wind's subsonic region leads to significantly enhanced mass-loss (TCQ04; MTQ07).  $Thus,$ $the$ $very$ $conditions$ $required$ $to$ $preserve$ $low$ $Y_{e}$ $in$ $a$ $proto-magnetar$ $or$ $NDAF$ $wind$ $threaten$ $to$ $simultaneously$ $over-pollute$ $the$ $outflow$ $with$ $baryons,$ $reducing$ $\sigma$ $and$ $stiflying$ $the$ $outflow's$ $GRB$ $potential.$ 

In the following sections we address some of these issues by calculating the neutron content of magnetized proto-magnetar and NDAF winds.  In particular, we consider the conditions under which outflows from proto-magnetars (\S\ref{section:magnetar}) and NDAFs (\S\ref{section:disk}) can remain neutron-rich while simultaneously maintaining sufficiently low mass-loading to remain plausible GRB central engines.

\section{Proto-magnetar Winds}

\label{section:magnetar}

\subsection{Evolution Equations and Numerical Procedure}

\label{section:magnetarnumerical}

As in our previous work (MTQ07), we calculate the structure of rapidly rotating PNS winds by solving the equations of one-dimensional neutrino-heated, ideal MHD in the equatorial plane of the PNS.  Using the time-dependent ``inhomogeneous'' 2N-RK3 scheme described in Brandenburg (2003), we solve for the wind density $\rho$, temperature $T$, radial velocity $v_{r}$, azimuthal magnetic field $B_{\phi}$, and azimuthal velocity $v_{\phi}$ as a function of radius $r$ according to equations (2)-(5) and equation (9) in MTQ07.  In this work we extend our previous calculations by simultaneously solving for the electron fraction $Y_{e}$, thus determining the wind's asymptotic neutron abundance.

The electron fraction $Y_{e}$ evolves as the wind emerges off the PNS surface due to weak interactions according to:
\begin{eqnarray}
\frac{d}{dt}Y_{e} = (1-Y_{e})(\lambda_{\nu_{e}n \rightarrow pe^{-}}+\lambda_{e^{+}n\rightarrow p\bar{\nu}_{e}}) - \nonumber \\
Y_{e}(\lambda_{\bar{\nu}_{e}p\rightarrow ne^{+}} + \lambda_{e^{-}p\rightarrow n\nu_{e}}),  
\label{yeevo}
\end{eqnarray}
where $d/dt \equiv \partial/\partial t + v_{r}(\partial/\partial r)$ and the $\lambda$'s are the weak interaction rates; we take the pair capture rates (eqs.~[\ref{eprate}]$-$[\ref{enrate}]) and the neutrino capture rates (eqs.~[\ref{nunrate}]$-$[\ref{nuprate}]) from B03a (neglecting proton/neutron recoil to good approximation; Strumia $\&$ Vissani 2003).  We have included the full effects of electron degeneracy, including the outgoing electron and positron blocking factors in calculating the neutrino capture rates.  In calculating all of the weak interaction rates we assume that the electrons and positrons are relativistic.  This is a good approximation for neutrino capture because the average neutrino(antineutrino) energy (and hence the resultant electron(positron) kinetic energy) always far exceeds the electron rest mass.  We set the pair capture rates equal to zero for $T < 0.5$ MeV in order to artificially account for the disappearance of pairs; the evolution of $Y_{e}$ is not sensitive to this cutoff because for $T \sim 0.5$ MeV pair capture is always dominated by neutrino absorption.  We neglect the effects that $\alpha$-particle formation has on the evolution of $Y_{e}$; this is a reasonable approximation because for cases in which the formed $\alpha-$particle fraction is the most significant (i.e., $Y_{e}^{a} \sim 0.5$) and is thus likely to have its greatest effect on the evolution of $Y_{e}$, $Y_{e}^{a}$ almost obtains by the radius where most $\alpha$-particles form.  We also neglect the effects that magnetic fields have on the electron and positron distribution functions and, hence, on the interaction rates and equation of state, although these effects become important for the largest field strengths that we consider ($B = 10^{16}$ G) and should be included in more detailed work (e.g., Lai $\&$ Qian 1998; Duan $\&$ Qian 2004).  Lastly, we neglect the small effect that general relativity (GR) has on the evolution of $Y_{e}$; in non-rotating PNS winds GR slightly increases $Y_{e}^{a}$ (Fuller $\&$ Qian 1996), primarily due to neutrino gravitational redshifts and the deeper gravitational potential.

Because we are interested in the wind structure along open magnetic flux, we assume a monopole radial field structure, $B_{r} = B_{\nu}(R_{\nu}/r)^{2}$, where $B_{\nu}$ is the surface magnetic field, and $R_{\nu}$ is the neutrinosphere radius.  Since $Y_{e}^{a}$ is set near the PNS surface, $Y_{e}$ is more sensitive to the surface field strength than to the field's precise radial scaling, and so our results are likely relatively insensitive to our monopole assumption.

For asymptotically relativistic PNS outflows, using the conserved magnetic flux $\Phi_{\rm B} = B_{\nu}R_{\nu}^{2} = B(R_{\rm L})R_{\rm L}^{2}$ and a spherically symmetric mass flux $\dot{M} = 4\pi \rho v_{r} r^{2} = 4\pi c R_{\rm L}^{2}\rho(R_{\rm L})$ we have that the magnetization from equation (\ref{magnetization}) is given by
\be
\sigma = B_{\nu}^{2}R_{\nu}^{4}\Omega^{2}/\dot{M}c^{3} 
\label{sigmaPNS}
\ee
Although for asymptotically non-relativistic outflows ($\sigma < 1$), $\sigma$ as defined by equation (\ref{sigmaPNS}) is no longer the ratio of Poynting-to-kinetic energy flux at $R_{\rm L}$, $\sigma$ remains a useful dimensionless quantity for characterizing the importance of the magnetic field in accelerating the outflow.   

With the exception of the evolution of $Y_{e}$ and its effects on the other thermodynamic wind variables, our boundary conditions and microphysics are similar to those used in MTQ07; thus, we review only the most essential aspects of these here.  Our neutrino heating(cooling) rates include charged-current neutrino absorption(pair capture) (eqs.~[\ref{eprate}]$-$[\ref{nuprate}]), neutrino(pair) annihilation, and inelastic neutrino-lepton scattering (QW96; T01), corrected for solid angle and redshift effects (Salmonson \& Wilson 1999; T01).  We index stages of the PNS thermal evolution in terms of the electron antineutrino luminosity $L_{\bar{\nu}_{e}}$.  We scale all other neutrino luminosities ($L_{\nu_e}, L_{\nu_{\mu}}, L_{\bar{\nu}_{\mu}}, L_{\nu_{\tau}}$, and $L_{\bar{\nu}_{\tau}}$) as in TCQ04: $L_{\nu_{e}} = L_{\bar{\nu}_{e}}/1.3 = 1.08 L_{\nu_{\mu}}$, where $\mu$ denotes each of the other four neutrino/antineutrino species.  Note that the total neutrino luminosity is then $L_{\nu} \simeq 4.6 L_{\bar{\nu}_{e}}$.  Following T01, all first energy moments at the neutrinosphere ($\langle\epsilon_{\nu}\rangle \equiv \langle E_{\nu}^{2}\rangle/\langle E_{\nu} \rangle$, where $E_{\nu}$ is the neutrino energy) were scaled with luminosity as $\langle\epsilon_{\nu}\rangle \propto L_{\nu}^{1/4}$, anchoring $\{\langle\epsilon_{\nu_{e}}\rangle,\langle\epsilon_{\bar{\nu}_{e}}\rangle,\langle\epsilon_{\nu_{\mu}}\rangle\}$ at $\{11,14,23\}$MeV for $L_{\bar{\nu}_{e},51} = 8,$ where $L_{\bar{\nu}_{e},51}$ is the electron antineutrino luminosity in units of $10^{51}$ ergs s$^{-1}$.  Higher energy moments necessary for the heating calculations ($\langle\epsilon^{\,\,n}_{\nu_{e}}\rangle, \langle\epsilon^{\,\,n}_{\bar{\nu}_{e}}\rangle$, etc.) are related to the first through appropriate integrals over the assumed Fermi-Dirac surface distribution.  With these scalings we note that electron antineutrino luminosities $L_{\bar{\nu}_{e},51} = \{8, 3.5, 1\}$ result in a neutrino-driven asymptotic electron fraction $Y_{e}^{\nu} = \{0.48,0.50,0.53\}$ according to equation (\ref{yeanu}); neutrino absorptions therefore favor $n/p \sim 1$ throughout the PNS's Kelvin-Helmholtz cooling epoch (T01).  Although we have assumed neutrino spectra based on current calculations of non-rotating PNSs, for rotating PNSs the mean neutrino energies along the equator may be lower due to gravity-darkening, which would increase $Y_{e}^{\nu}$ from the non-rotating value (e.g., Fryer $\&$ Heger 2000; Thompson, Quataert, $\&$ Burrows 2005; Dessart et al.~2006).

We assume Newtonian gravity for a fixed central PNS mass $M=1.4 M_{\sun}$ and a neutrinosphere radius $R_{\nu} = 10$ km which is characteristic of the PNS's final, cooled state.  Although $R_{\nu}$ is probably larger than $10$ km for the first few seconds following the launch of the SN shock (e.g., Buras et al.~2003), neutron-rich GRB-producing outflows, which are the focus of this paper, are only possible at relatively late times, once the PNS has fully contracted to its most rapidly rotating state (see $\S$\ref{section:implications}).  Our code is non-relativistic but we still calculate flows with $\sigma > 1$, which accelerate to relativistic speeds outside the light cylinder, because $\dot{M}$ and $Y_{e}$ are set very close to the PNS surface, where the wind still moves non-relativistically.  We set the neutrinosphere density of the wind $\rho_{\nu}$ so that the neutrino optical depth to infinity is $\tau_{\nu} \simeq \frac{2}{3}$; $\rho_{\nu}$ ranges from $\sim 10^{12}$ g cm$^{-3}$ for high luminosity, rapidly rotating solutions to $\gtrsim 10^{13}$ g cm$^{-3}$ for our lowest luminosity solutions (see Table $\ref{table:yea_magnetar}$).  In general, we find that $Y_{e}^{a}$ is relatively insensitive to $\rho_{\nu}$.  The electron fraction at the neutrinosphere $Y_{e}^{0}$ is chosen to ensure that equilibrium between the weak interaction rates in equation (\ref{yeevo}) is established at $R_{\nu}$.

The value of the magnetic flux $\Phi_{\rm B} = B_{\nu}R_{\nu}^{2}$, stellar rotation rate $\Omega$, and neutrino luminosity $L_{\nu}$ uniquely identify a wind solution.  By letting the wind come into steady-state, we obtain eigenvalues for the mass-loss rate $\dot{M}$ (normalized to $4\pi$ sr), specific angular momentum loss rate $\mathcal{L}$, and Bernoulli integral $\mathcal{B}$ (see eqs.~[6], [7], and [11] in MTQ07).  Because our code is non-conservative, the radial conservation of these quantities is used to verify the code's accuracy.  The total asymptotic energy lost in the wind is given by $\dot{E}^{a} = \mathcal{B}^{a}\dot{M}$, where $\mathcal{B}^{a}$ is the Bernoulli integral evaluated at large radii.  The rate of angular momentum loss in the wind is given by $\dot{J}_{W} = \mathcal{L}\dot{M} = \Omega R_{A}^{2}\dot{M}$, where $R_{A}$ is the Alfv\'{e}n radius defined by $B_{r}(R_{A})/\sqrt{4\pi\rho(R_{A})} = v_{r}(R_{A})$.     

\subsection{Numerical Results}
\label{section:pnsresults}

We have calculated the MHD structure of PNS winds for several combinations of surface monopole magnetic field strength $B_{\nu}$, rotation rate $\Omega$, and electron antineutrino luminosity $L_{\bar{\nu}_{e}} = L_{\bar{\nu}_{e},51}\times10^{51}$ ergs s$^{-1}$ in order to study the neutron fraction in proto-magnetar outflows.  Some of these results are summarized in Table $\ref{table:yea_magnetar}$.  

\begin{figure}[t]
\centerline{\hbox{\psfig{file=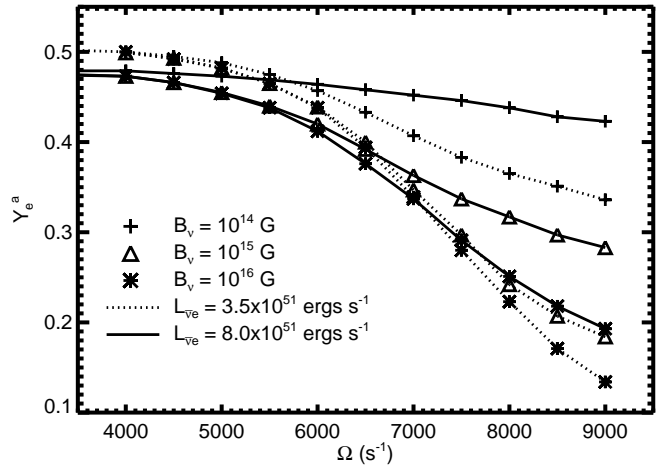,width=9.5cm}}}
\figcaption[x]{Asymptotic electron fraction $Y_{e}^{a}$ for magnetocentrifugally-driven PNS winds as a function of rotation rate $\Omega$ at $L_{\bar{\nu}_{e}} = 8\times 10^{51}$ ergs s$^{-1}$ and $L_{\bar{\nu}_{e}} = 3.5\times 10^{51}$ ergs s$^{-1}$ for $B_{\nu} = 10^{14}$ G, $B_{\nu} = 10^{15}$ G, and $B_{\nu} = 10^{16}$ G.  For slow rotation and weak magnetic fields, $Y_{e}^{a}$ approaches its neutrino absorption equilibrium value $Y_{e}^{\nu} \sim 0.5$ (eq.~[\ref{yeanu}]).  For more rapid rotation and stronger magnetic fields, $Y_{e}^{a}$ is reduced because matter is advected from the PNS surface sufficiently rapidly that it falls out of $\beta$-equilibrium before neutrino absorptions dominate degenerate pair captures (see Fig. \ref{plot:rates}).  For sufficiently large $B_{\nu}$, $Y_{e}^{a}$ no longer decreases with increasing $B_{\nu}$ because the wind corotates past a few scale heights above the PNS surface, where $Y_{e}^{a}$ is set.  An approximate fit to our numerical results in this limit is given by equation (\ref{ye_sat}).
\label{plot:magnetarye}}
\end{figure}

Figure \ref{plot:magnetarye} shows our calculation of the asymptotic electron fraction $Y_{e}^{a}$ of proto-magnetar winds as a function of $\Omega$ at neutrino luminosity $L_{\nu_{e},51} = 8$ and $3.5$ for $B_{\nu} = 10^{14},10^{15}$, and $10^{16}$ G.  Figure $\ref{plot:magnetarye}$ and Table $\ref{table:yea_magnetar}$ show that for slow rotation and low magnetic field strengths $Y_{e}^{a}$ obtains the neutrino absorption equilibrium value $Y_{e}^{\nu} \sim 0.5$ (eq.~[\ref{yeanu}]), as is expected from studies of slowly-rotating, non-magnetized PNSs (QW96; T01) and from the arguments given in $\S \ref{section:thermal}$.  As the rotation rate and magnetic field strength increase, $Y_{e}^{a}$ decreases because the material is advected sufficiently rapidly from the surface due to magnetocentrifugal slinging that $Y_{e}$ ``freezes out'' before coming into equilibrium with neutrino absorptions.  

Figure \ref{plot:magnetarye} also shows that for a sufficiently strong magnetic field, $B_{\nu} \gtrsim 10^{14}-10^{15}$ G, $Y_{e}^{a}(\Omega)$ no longer increases with increasing $B_{\nu}$, saturating to a profile $Y_{e}^{a,{\rm sat}}(\Omega)$ that we find is reasonably well-fit by a single empirical formula for all of the rotation rates and neutrino luminosities that we have considered ($\Omega \le 9000$ s$^{-1}$; 1 $\lesssim L_{\bar{\nu}_{e},51} \lesssim $ 10, corresponding to times $\sim 1-10$ s following bounce):
\be
Y_{e}^{a,{\rm sat}} = \frac{Y_{e}^{0}+Y_{e}^{\nu}}{2} + \frac{Y_{e}^{0}-Y_{e}^{\nu}}{2}{\rm tanh}\left[\frac{\Omega-\Omega_{\rm n}}{\Delta\Omega_{\rm n}}\right], 
\label{ye_sat}
\ee
where $\Omega_{\rm n} \approx 7800$ s$^{-1}$, $\Delta\Omega_{\rm n} \approx 2000$ s$^{-1}$, and $Y_{e}^{0}$ is the electron fraction at the neutrinosphere (typically, $Y_{e}^{0} \approx 0.01-0.05$; see Table $\ref{table:yea_magnetar}$).  Equation (\ref{ye_sat}) and Figure \ref{plot:magnetarye} show that proto-magnetars must have submillisecond ($P \lesssim P_{\rm n} \equiv 2\pi/\Omega_{\rm n} \approx 0.8$ ms), near break-up, rotation in order to produce asymptotically neutron-rich outflows; furthermore, because $\Delta\Omega_{\rm n}/\Omega_{\rm n} \ll 1$ the transition from $n/p \gg 1$ to $n/p \sim 1$ occurs over a very limited range in $\Omega$.  Therefore, because a PNS possesses a rotational energy 
\be
E_{\rm rot} = \frac{1}{2}I\Omega^{2} \approx 4\times 10^{52}\left(\frac{\Omega}{\Omega_{\rm n}}\right)^{2}\,{\rm ergs},
\label{erot}
\ee
where $I \approx (2/5)MR_{\nu}^{2}$ is the PNS moment of inertia, we conclude that neutron-rich outflows from magnetar birth would require a GRB plus SN energy totaling $\gtrsim 4\times 10^{52}$ ergs.


\begin{figure*}
\resizebox{\hsize}{!}{\includegraphics{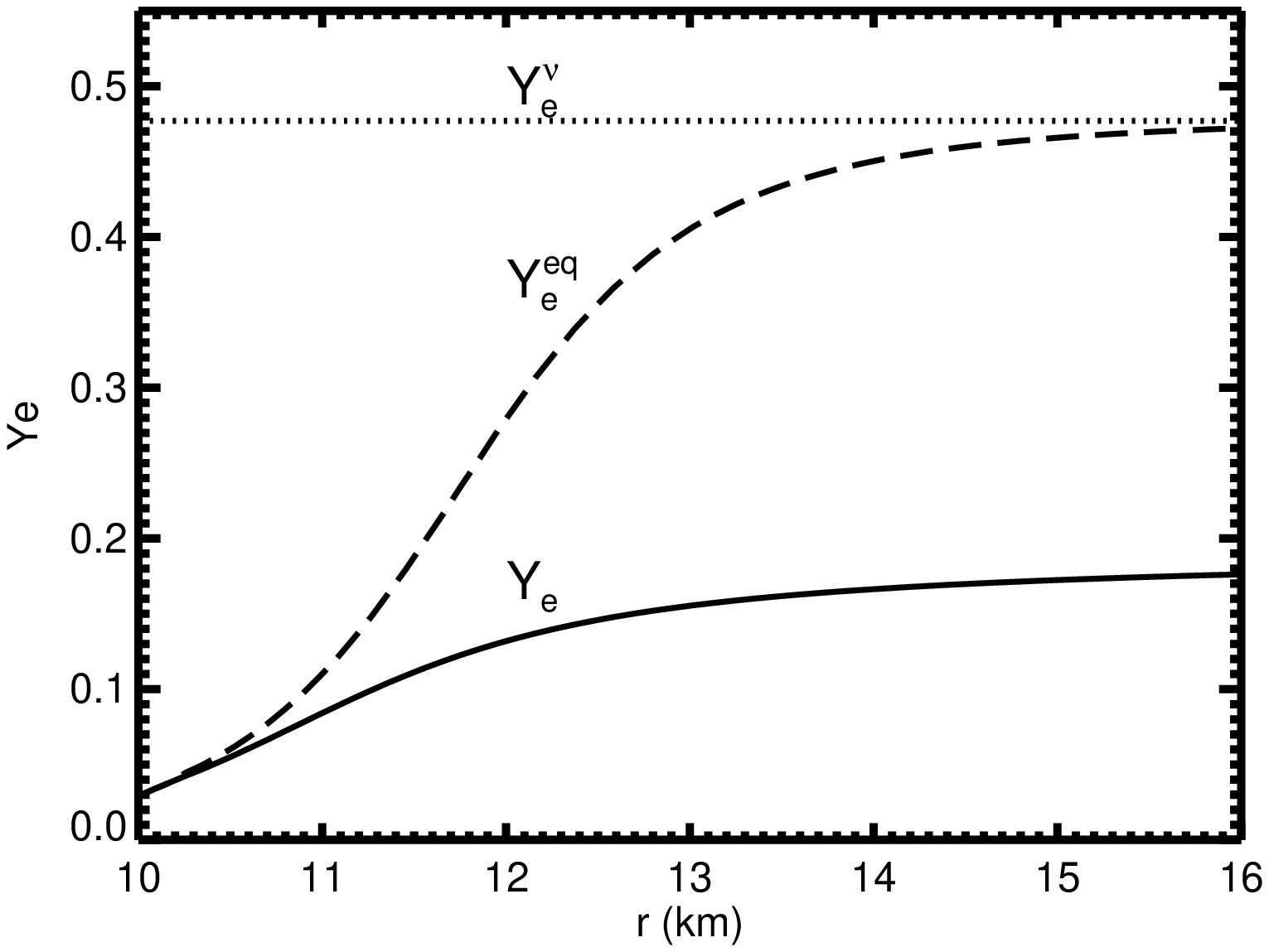}\includegraphics{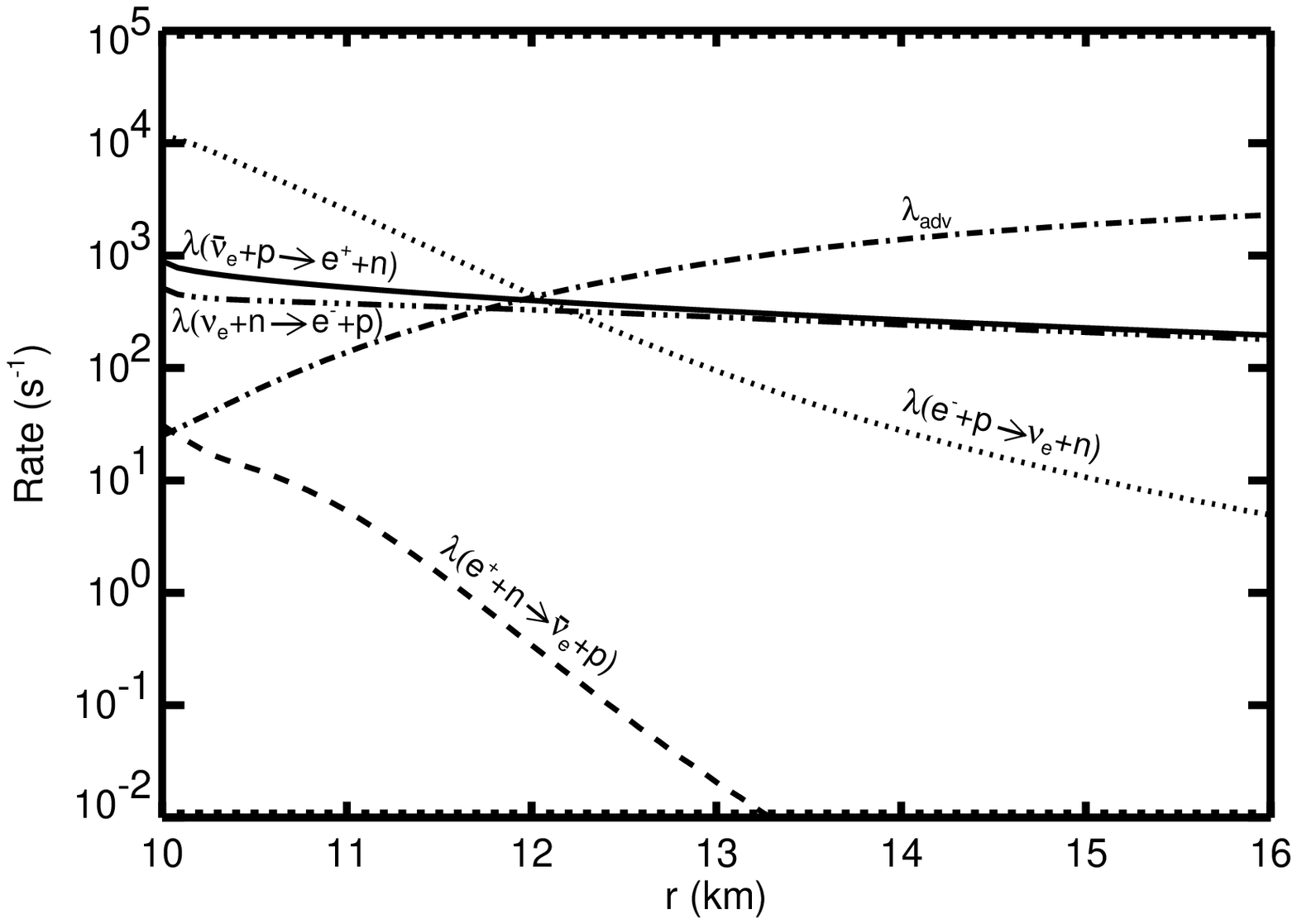}}
\caption{{$Left$}: Electron fraction $Y_{e}$ ($solid$ $line$) in the inner 6 km of a magnetically-driven PNS wind with $M = 1.4 M_{\sun}$, $R_{\nu} = 10$ km, $B_{\nu} = 10^{16}$ G, and $\Omega = 9000$ s$^{-1}$; also shown is the local equilibrium electron fraction $Y_{e}^{\rm eq}$ ($long$ $dashed$ $line$), obtained by setting the left hand side of equation (\ref{yeevo}) equal to zero.  The electron fraction is very small at the PNS surface ($Y_{e} \approx 0.05$).  As matter accelerates away from the surface, $Y_{e} \approx Y_{e}^{\rm eq}$ as long as matter moves sufficiently slowly to maintain local equilibrium with the weak interaction rates.  Because $Y_{e}$ is out of equilibrium by the time neutrino absorptions dominate pair captures, the asymptotic electron fraction $Y_{e}^{a}$ is  $\ll 0.5$, despite the fact that $Y_{e}^{\rm eq}$ asymptotes to $Y_{e}^{\nu} \approx 0.48$ at large radii.  {$Right$}: Rates in the wind that determine the radial evolution of $Y_{e}$.  The dotted(short dashed) lines show the electron(positron) absorption rates (eqs.~[\ref{eprate}]$-$[\ref{enrate}]).  The electron neutrino and antineutrino absorption rates (eqs.~[\ref{nunrate}]$-$[\ref{nuprate}]) are shown with a triple-dot-dash and solid line, respectively; the wind advection rate $\lambda_{\rm adv}\equiv v_{r}/r$ is shown with a dot-dashed line.  The asymptotic electron fraction obtained by the wind $Y_{e}^{a} \approx 0.19$ is appreciably lower than the neutrino absorption equilibrium value ($Y_{e}^{\nu} \approx 0.48$) that obtains for slower rotating or non-magnetized winds.  Lower $Y_{e}^{a}$ obtains because magnetocentrifugal acceleration is sufficiently strong that matter ``freezes out'' of $\beta$-equilibrium at small radii (corresponding to $\lambda_{\rm adv}$ rising above the weak interaction rates at $r \approx 12$ km), before the neutrino absorption rates begin to dominate the electron capture rate.  Equilibrium still favors neutron-rich matter at freeze-out because matter is advected from the PNS surface sufficiently quickly that neutrino heating is suppressed and degeneracy is never completely lifted.  
\\}
\label{plot:rates}
\end{figure*}


As an example of a neutron-rich wind solution, the left panel of Figure \ref{plot:rates} shows the electron fraction $Y_{e}$ as a function of radius just above the PNS surface for a solution with $L_{\bar{\nu}_{e},51}= 8, B_{\nu} = 10^{16}$ G, and $\Omega = 9000$ s$^{-1}$ (rotation period $P \approx 0.7$ ms).  This relatively high luminosity solution corresponds to a time $\approx$ 1 s after core bounce (see eq.~[\ref{lnupns}]).  Figure \ref{plot:rates} also shows $Y_{e}^{\nu}$ and the equilibrium electron fraction $Y_{e}^{\rm eq}$ obtained by setting the time derivative on the left hand side of equation (\ref{yeevo}) equal to zero.  The right panel of Figure \ref{plot:rates} shows the individual rates in the wind that determine the evolution of $Y_{e}$: electron absorption (eq.~[\ref{eprate}], $dotted$ $line$), positron absorption (eq.~[\ref{enrate}], $dashed$ $line$), electron neutrino absorption (eq.~[\ref{nunrate}], $triple-dot-dashed$ $line$), electron antineutrino absorption (eq.~[\ref{nuprate}]; $solid$ $line$), and the wind's advection rate $\lambda_{\rm adv} \equiv v_{r}/r$ ($dot-dashed$ $line$).   The slight difference between the neutrino and antineutrino absorption rates at large radii, although difficult to discern on the scale of this plot, implies that $Y_{e}^{\nu} \simeq 0.48$ for this solution.

Figure \ref{plot:rates} shows that the electron fraction at the base of the outflow is very neutron-rich ($Y_{e}^{0} \approx 0.05$).  This results from the equilibrium established between the electron and electron neutrino capture rates ($dotted$ and $triple-dot-dashed$ $lines$, respectively) under the degenerate surface conditions ($\rho_{\nu} \approx 10^{12}$ g cm$^{-3}$, T$_{\nu} \approx$ 4.2 MeV $\ll$ 14 MeV $\approx \mu_{e}$, where $\mu_{e}$ is the electron chemical potential at the neutrinosphere), which strongly suppresses the thermal positron population relative to electrons.  Thus our value of $Y_{e}^{0}$ is significantly larger than the pure pair-capture equilibrium value given in Figure 1 of B03a because we have included neutrino absorptions; for vanishing neutrino luminosity, however, the results of B03a Figure 1 would obtain.  Indeed, for solutions with lower $L_{\bar{\nu}_{e}} = 1.3L_{\nu_{e}}$  electron neutrino absorption becomes less important and $Y_{e}^{0}$ decreases (see Table $\ref{table:yea_magnetar}$).  

As matter accelerates off of the PNS surface, Figure \ref{plot:rates} shows that $Y_{e}$ rises above $Y_{e}^{0}$.  Because the wind initially moves slowly relative to the relevant weak interaction rates, at small radii $Y_{e}$ remains in approximate equilibrium at the value $Y_{e} \approx Y_{e}^{\rm eq} < Y_{e}^{\nu}$.  However, as magnetocentrifugal slinging accelerates matter away from the PNS surface, $\lambda_{\rm adv} = v_{r}/r$ also rises rapidly, eclipsing the weak interaction rates by $r \approx 12$ km.  At this point $Y_{e}$ ``freezes out'' at a value $Y_{e}^{a} \simeq 0.19$; this occurs before neutrino absorptions dominate the electron capture rates (which is why $Y_{e} \approx Y_{e}^{\rm eq}$ is still relatively low).

\subsection{Conditions for Neutron-Rich Outflows from Proto-Magnetars}
\label{section:conditionsPNS}

The mass-loss rate for purely thermal, neutrino-driven PNS winds $\dot{M}_{\rm th}$ can be derived analytically by requiring that the energy used to unbind a typical nucleon $E_{B} = GMm_{\rm n}/R_{\nu}$ be supplied entirely by the neutrino heating $Q_{\nu} \equiv \int_{R_{\nu}}^{\infty}\dot{q}_{\nu}dr/v_{r}$ that a nucleon experiences in accelerating from the PNS surface to large radii (QW96), where $\dot{q}_{\nu}$ is the net neutrino heating rate per baryon, which is dominated by neutrino absorption (eqs.~[\ref{nunrate}]$-$[\ref{nuprate}]).  By repeating this derivation but instead requiring that $Q_{\nu} \lesssim \langle\epsilon_{\nu}\rangle$ (see $\S\ref{section:thermal}$) we obtain the minimum mass-loss rate that must accompany a neutron excess over the neutrino-driven value
\be
\frac{\dot{M}}{\dot{M}_{\rm th}} \gtrsim \phi_{\rm n} \equiv \frac{GMm_{\rm n}}{R_{\nu}\langle\epsilon_{\nu}\rangle} \approx 20M_{1.4}\langle\epsilon_{\bar{\nu}_{e},10}\rangle^{-1}R_{10}^{-1},
\label{mdotneutron}
\ee
where (QW96)
\be
\dot{M}_{\rm th} \approx 10^{-6}L_{\bar{\nu}_{e},51}^{5/3}\langle\epsilon_{\bar{\nu}_{e},10}\rangle^{10/3}M_{1.4}^{-2}R_{10}^{5/3}\,{\rm M_{\sun}\,s^{-1}},
\label{mdotthermal}
\ee
$R_{\nu}=10R_{10}{\rm\,km}$, $M = 1.4M_{1.4}M_{\sun}$, and $\langle\epsilon_{\nu}\rangle \approx \langle\epsilon_{\bar{\nu}_{e}}\rangle = 10\langle\epsilon_{\bar{\nu}_{e},10}\rangle$ MeV.  In our numerical calculations we have assumed that $L_{\nu} \propto \langle\epsilon_{\nu}\rangle^{4}$ so that $\dot{M}_{\rm th} \propto L_{\nu}^{5/2}$.  

\begin{figure}[t]
\centerline{\hbox{\psfig{file=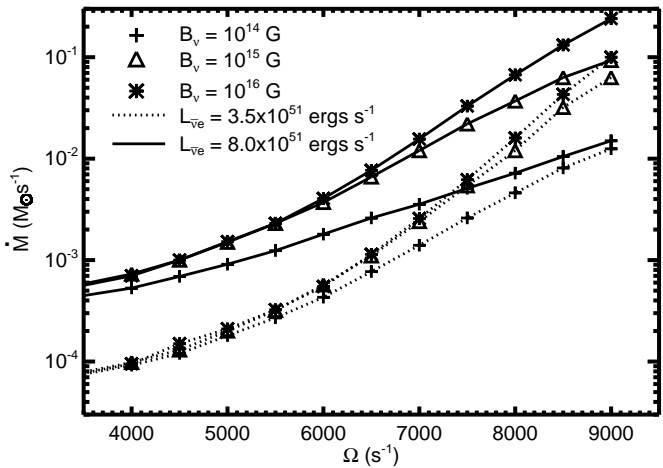,width=9.5cm}}}
\figcaption[x]{Mass loss rate $\dot{M}$ for the same solutions for which $Y_{e}^{a}$ is shown in Figure \ref{plot:magnetarye}.  Due to centrifugal slinging, $\dot{M}$ increases with $B_{\nu}$ and $\Omega$ and is substantially larger than its thermally-driven, non-rotating value, which is $\dot{M}_{\rm th} \approx 2\times 10^{-4}(2\times 10^{-5})M_{\sun}$ s$^{-1}$ for $L_{\bar{\nu}_{e},51} = 8(3.5)$.  By comparing $\dot{M}$ with $Y_{e}^{a}$ in Figure \ref{plot:magnetarye} note that winds with large $\dot{M}$ have low $Y_{e}^{a}$; in particular, $\dot{M} \gtrsim \phi_{\rm n}\dot{M}_{\rm th} \approx 15-20\dot{M}_{\rm th}$ (eq.~[\ref{mdotneutron}]) is required for $Y_{e}^{a}$ to noticeably decrease below its neutrino-driven value $Y_{e}^{\nu}$.      
\label{plot:magnetarmdot}}
\end{figure}

In Figure \ref{plot:magnetarmdot} we show $\dot{M}$ for the same wind solutions for which $Y_{e}^{a}$ is shown in Figure \ref{plot:magnetarye}.  Figure \ref{plot:magnetarmdot} shows that  $\dot{M}$ is substantially enhanced over its thermally-driven value $\dot{M}_{\rm th}$ in the presence of a strong magnetic field and rapid rotation.  This enhanced mass-loss occurs because a strong magnetic field forces the outflow to corotate above the PNS surface and rapid corotation brings the wind's sonic radius $R_{\rm s}$ much closer to the surface, into the corotating region; this increases the hydrostatic scale height and the wind density at $R_{\rm s}$, thereby increasing $\dot{M}$.   Because the magnetic field, as opposed to neutrinos, is primary responsible for unbinding matter from a rapidly rotating proto-magnetar, the outflow can have $\dot{M} > \dot{M}_{\rm th}\phi_{\rm n}$ and can thus remain neutron-rich.  Indeed, Figure \ref{plot:magnetarmdot} shows that $\dot{M}$ increases exponentially with $\Omega$ and a comparison with Figure \ref{plot:magnetarye} shows that $Y_{e}^{a}$ first noticeably decreases below $Y_{e}^{\nu}$ once equation (\ref{mdotneutron}) is satisfied.

Figure \ref{plot:magnetarmdot} also shows that for modest magnetic field strengths $\dot{M}$ increases with increasing $B_{\nu}$, but that for sufficiently large $B_{\nu}$, $\dot{M}$ saturates to a value $\dot{M}_{\rm cf}$ in the ``centrifugal limit'' (e.g., Lamers $\&$ Cassinelli 1999).  For $\Omega \approx \Omega_{\rm n}$ (or, equivalently, $Y_{e}^{a} \approx 0.25$) we find that
\be 
\dot{M}_{\rm cf}(\Omega_{\rm n}) \approx 6\times 10^{-2}(L_{\bar{\nu}_{e},51}/8)^{2.2}M_{\sun}{\,\rm s}^{-1},
\label{mdotcf}
\ee
which is over an order of magnitude greater than the mass-loss required, $\dot{M}_{\rm th}\phi_{\rm n} \approx 3\times 10^{-3}(L_{\bar{\nu}_{e},51}/8)^{2.25}M_{\sun}$ $s^{-1}$ (eq.~[\ref{mdotneutron}]), for just a mild neutron excess relative to the neutrino-driven value (i.e., $Y_{e}^{a} \lesssim Y_{e}^{\nu}$).  

Figures \ref{plot:magnetarye} and \ref{plot:magnetarmdot} show that at fixed $\Omega$, $\dot{M} \rightarrow \dot{M}_{\rm cf}$ and $Y_{e}^{a} \rightarrow Y_{e}^{a,{\rm sat}}$ (eq.~[\ref{ye_sat}]) for similar magnetic field strengths.  To understand why this occurs note that the requirement for fully centrifugally-enhanced mass-loss is that the magnetic field must be sufficiently strong to enforce corotation beyond the radius where $\dot{M}$ is set; because a wind approximately corotates to its Alfv\'{e}n radius $R_{A}$ and its mass-loss rate is set at its sonic radius $R_{\rm s}$ the condition for $\dot{M}\rightarrow\dot{M}_{\rm cf}$ is that $R_{A} \gtrsim R_{\rm s}$.  In analogy, because $Y_{e}^{a}$ obtains near the PNS surface (see Fig.~\ref{plot:rates}), the condition for $Y_{e}^{a} \rightarrow Y_{e}^{a,{\rm sat}}$ is that $R_{A} \gtrsim R_{\nu}$.  In the centrifugal limit the sonic radius for an equatorial wind is given by $R_{\rm s} = (GM/\Omega^{2})^{1/3}$ (e.g., Lamers $\&$ Cassinelli 1999), so that $R_{\rm s}(\Omega_{\rm n}) \approx 13$ km $ \sim R_{\nu}$.  Thus the conditions for fully enhanced mass-loss and fully enhanced neutron-richness are approximately the same because the sonic radius is close to the PNS surface for the near break-up rotation rates required to produce neutron-rich outflow. 

Finally, although we have focused on equatorial proto-magnetar winds, the properties of the outflow will vary with latitude.  In particular, $\dot{M}$ and $Y_{e}^{a}$ will decrease and increase with increasing angle $\theta$ from the equator, respectively, approaching $\dot{M}_{\rm th}$ and $Y_{e}^{\nu}$ for nearly polar outflow ($\theta \rightarrow 90^{\circ}$) because of the reduced centrifugal acceleration along nearly vertical field lines ($\dot{M}$ and $\dot{M}_{\rm th}$ here are per solid angle).  Thus, it may seem plausible that outflows driven from intermediate latitudes could maintain moderately low $Y_{e}$ yet become significantly more relativistic than matter driven from near the equator.  Even high-latitude outflows, however, require $\dot{M} \gg \dot{M}_{\rm th}$ (eq.~[\ref{mdotneutron}]) to attain $Y_{e}^{a} < Y_{e}^{\nu}$, making such a scenario unlikely, as we will show explicitly through NDAF disk wind calculations in $\S\ref{section:disk}$.  Furthermore, the limited solid angle occupied by such a hypothetical mid-latitude wind would preclude it from carrying a significant portion of the PNS's spin-down luminosity, which is primarily extracted by outflow originating near the equator (e.g., Bucciantini et al.~2006).
\\

\subsection{Implications for GRBs}
\label{section:implications}

We have shown that proto-magnetars born with $P < P_{\rm n} \approx 0.8$ ms produce energetic, neutron-rich winds; it therefore appears that the birth of very rapidly rotating proto-magnetars should produce neutron-rich GRBs.  This is, however, not necessarily the case.  Even strongly magnetized proto-magnetar winds are baryon-loaded at very early times following the launch of the SN shock because the PNS is very hot and its already substantial thermally-driven mass-loss is enhanced due to centrifugal slinging.  In fact, a proto-magnetar requires several seconds to contract and cool to the point that a wind launched from its surface achieves the high magnetization $\sigma \gtrsim$ 10-100 required to explain the Lorentz factors inferred from GRBs.  By the time a proto-magnetar cools to the point that its wind becomes ultra-relativistic, it may no longer rotate sufficiently rapidly to remain neutron-rich.

\begin{figure}[t]
\centerline{\hbox{\psfig{file=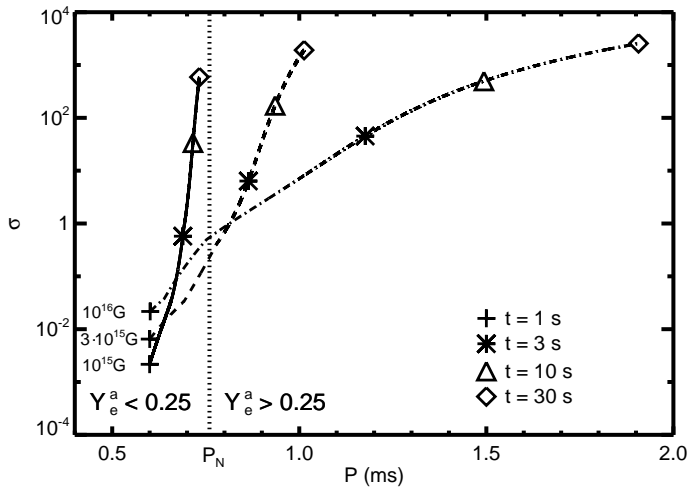,width=9.5cm}}}
\figcaption[x]{Magnetization $\sigma$ of a PNS's outflow (eq.~[\ref{sigmaPNS}]) as a function of the PNS rotation period $P$ from evolutionary spin-down calculations performed during the Kelvin-Helmholtz cooling of a proto-magnetar with initial rotation period $P_{0} = 0.6$ ms at a time $t= 1$ s after core bounce and for different fixed surface dipole magnetic field strengths: $B_{\nu}^{\rm dip} = 10^{15}$ G, $3\times 10^{15}$ G, and $10^{16}$ G; for reference, times $t = $ 1, 3, 10, and 30 seconds after core bounce are shown along the evolutionary tracks.  So long as the rotation period of the proto-magnetar remains less than $P \sim P_{\rm n} \approx 0.8$ ms its outflow remains neutron-rich (see eq.~[\ref{ye_sat}]).  The proto-magnetars with $B_{\nu}^{\rm dip} = 3\times 10^{15}$ G and $10^{16}$ G spin-down to a relatively neutron-poor state in just a few seconds, before cooling sufficiently to produce $\sigma \gg 1$ outflows.  In contrast, the proto-magnetar with $B_{\nu}^{\rm dip} = 10^{15}$ G spins down less rapidly, allowing the PNS to produce simultaneously neutron-rich and ultra-relativistic outflow $\sim 10-30$ s following core bounce. 
\\
\label{plot:sigma_period}}
\end{figure}

In Figure \ref{plot:sigma_period} we explore this possibility quantitatively by showing the magnetization $\sigma$ of a PNS's outflow (eq.~[\ref{sigmaPNS}]) as a function of the PNS rotation period $P$ from evolutionary calculations of a cooling, spinning down proto-magnetar with an initial period $P_{0} = 0.6$ ms $<P_{\rm n}$ and for three different fixed surface dipole magnetic field strengths:\footnote{The minimum stable neutron star rotation period $P_{\rm min}$ is uncertain theoretically because it depends on the uncertain supranuclear density equation of state (EOS); depending on the EOS, detailed studies find that $P_{\rm min} = 0.53-1.7$ ms for $M \simeq 1.4M_{\sun}$ (Cook, Shapiro, $\&$ Teukolsky 1994, Table 8).  Thus, although we take the specific value $P_{0} = 0.6$ ms in our calculations to illustrate the conditions under which proto-magnetars can produce neutron-rich GRBs, if $P_{\rm min} > P_{\rm n} \approx 0.8$ ms the conclusions of this section are much simpler: proto-magnetars cannot produce neutron-rich outflow.  Conversely, the detection of a neutron-rich GRB outflow from a confirmed magnetar birth would provide a constraint on the EOS.} $B_{\nu}^{\rm dip} = 10^{15}$ G, $3\times 10^{15}$ G, and $10^{16}$ G.  These evolutionary calculations, which are described in more detail in MTQ07 $\S4.1$, assume a PNS cooling evolution similar to that given in equation (\ref{lnupns}) and map the one-dimensional neutrino-heated monopole wind calculations of MTQ07 onto a more physical dipole geometry using the axisymmetric two-dimensional relativistic MHD calculations of Bucciantini et al.~2006 (hereafter B06).  Although Figure \ref{plot:sigma_period} only shows calculations for one initial rotation period $P_{0} = 0.6$ ms, the PNS's spin-down timescale is dominated by times at which it is slowly rotating; hence, the evolutionary tracks for PNSs born with $P_{0} \lesssim 0.6$ ms are similar to those shown in Figure \ref{plot:sigma_period} for $P > 0.6$ ms, following a brief initial spin-down phase. 

Figure \ref{plot:sigma_period} shows that PNSs born with $P<P_{\rm n}$ and $B_{\nu}^{\rm dip} \gtrsim 3\times 10^{15}$ G spin down to a relatively neutron-poor state ($P > P_{\rm n}$) in just a few seconds, before cooling sufficiently to produce an ultra-relativistic ($\sigma \gg 1$) outflow; these proto-magnetars could therefore not produce neutron-rich GRBs.  On the other hand, a PNS born with $B_{\nu}^{\rm dip} = 10^{15}$ G spins down more slowly and is thus capable of producing an ultra-relativistic, neutron-rich outflow $\sim 10-30$ seconds following core bounce.  Of the proto-magnetar's initial rotational energy of $E_{\rm rot} \sim 6\times 10^{52}$ ergs, a substantial portion ($\sim 2\times 10^{52}$ ergs) is extracted when $\sigma < 10-100$; although this early wind is potentially too mass-loaded to produce a GRB, it may enhance the energy and alter the morphology of the accompanying SN shock (TCQ04; MTQ07; Bucciantini et al.~2007a).  The remaining $\sim 4\times 10^{52}$ ergs, part of which could produce a neutron-rich GRB, emerges when $\sigma > 10-100$ on a somewhat longer spin-down timescale $\tau_{\rm J}$, which is $\sim 10^{2}-10^{3}$ s, depending on the fraction of the proto-magnetar's surface threaded by open magnetic flux (B06).

In $\S\ref{section:magnetically}$ we emphasized the need for magnetic acceleration to produce neutron-rich outflows from compact objects.  Despite the absence of magnetic fields, however, Dessart et al.~(2006) find neutron-rich ($Y_{e}^{a} \sim 0.25$), purely neutrino-driven outflows in 2D radiation-hydrodynamical simulations of the very early evolution ($t \lesssim $ 800 ms after core bounce) of a rapidly rotating neutron star newly-formed following the accretion-induced collapse (AIC) of a white dwarf.  Because these calculations show that very rapidly rotating PNSs can produce neutron-rich outflows without the aid of magnetic acceleration, it may appear that neutron-rich GRB outflows could be possible under less restrictive conditions than described in this section.  The purely neutrino-driven neutron-rich outflows found by Dessart et al.~(2006) are only possible, however, because at very early times following core bounce the PNS is inflated with respect to its final, cooled radius and because a PNS rotating near break-up is strongly oblately deformed.  Because the equatorial neutrinosphere radius $R_{\nu}$ can therefore exceed $\sim 100$ km at early times (see Dessart et al.~2006, Fig.~7), the gravitational binding energy of a nucleon on the PNS surface at a moderately low latitude is $E_{B} \sim 10(R_{\nu}/100{\rm\,km})^{-1}$ MeV, comparable to the mean energy of the neutrinos driving the outflow (typically $\approx 10-15$ MeV during the PNS's early cooling phase).  Thus, despite the fact that neutrino absorptions favor a $proton-rich$ composition ($Y_{e}^{\nu} > 0.5$) because the electron neutrino flux dominates the electron antineutrino flux during early deleptonization (especially in the presence of rapid rotation; Thompson, Quataert, $\&$ Burrows 2005), less than a single neutrino is required to unbind a typical nucleon (i.e., $\phi_{\rm n} \lesssim 1$; eq.~[\ref{mdotneutron}]); matter driven from low latitudes can therefore partially retain the neutron-rich composition of the PNS surface.  Although purely neutrino-driven neutron-rich outflows are thus possible from PNSs at $very$ $early$ $times$ ($\lesssim 1$ s) after core bounce, as the PNS cools and its radius shrinks $\phi_{\rm n}$ will increase; purely neutrino-driven neutron-rich outflows will therefore not be possible after a few seconds following core bounce, even if the PNS remains distorted by continuing to rotate near break-up.  In particular, once the PNS has cooled and contracted sufficiently to produce a GRB outflow, neutron-rich winds will only be possible if they are magnetically-driven and satisfy the constraints described in this section.

\section{Accretion Disk Winds}
\label{section:disk}

\begin{figure}[t]
\centerline{\hbox{\psfig{file=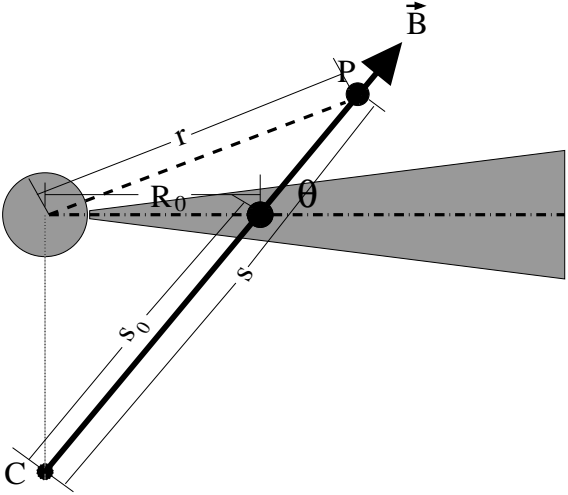,width=9.5cm}}}
\figcaption[x]{Geometry of NDAF disk winds.  The wind is calculated along a flux tube that is inclined at an angle $\theta$ with respect to the accretion disk midplane.  The calculation is started from a location just above the disk midplane a distance $R_{0}$ from the central black hole and at a distance $s_{0}$ from the monopole center (position ``C'').  A generic point ``P'' along the outflow is located at a distance $r$ from the central black hole and a distance $s$ from the monopole center.  
\\
\label{plot:diskfig}}
\end{figure}

In this section we describe calculations of the structure and neutron content of axisymmetric, one-dimensional MHD winds launched from NDAFs.  As shown schematically in Figure \ref{plot:diskfig}, we perform our calculations along a spherical, monopole flux tube centered about the position ``C'' a distance $R_{0}\tan\theta$ directly below the black hole along the disk rotation axis, where $R_{0}$ is the distance from the black hole to the wind's launching point just above the accretion disk midplane and $\theta \in [0^{\circ},90^{\circ})$ is the angle that the flux tube makes with respect to the midplane.  The distance from a given position ``P'' along the outflow to the black hole and monopole center are denoted $r$ and $s$, respectively.  The physical solid angle of the wind $\Delta\Omega$ is chosen such that the conserved mass outflow rate is given by $\dot{M} = \Delta\Omega \rho v_{\rm p}s^{2} =  4\pi \rho_{0}v_{\rm p,0}R_{0}^{2}$ so that $\Delta\Omega = 4\pi\cos^{2}\theta$, where $v_{\rm p}$ is the poloidal wind velocity, $\rho$ is the wind's density, and `0' denotes quantities evaluated at the base of the outflow;\footnote{We use `0' to denote quantities at the base of the outflow instead of `$\nu$' (as was used in $\S\ref{section:magnetar}$) because, unlike in the PNS case, the disk midplane can be neutrino transparent and therefore may not posses a neutrinosphere.} note that for equatorial outflow ($\theta = 0$) $\dot{M}$ reduces to the definition used in $\S\ref{section:magnetar}$ for proto-magnetar winds.  Although this choice for $\Delta\Omega$ is somewhat arbitrary, the quantities of most interest, $\sigma$ and $Y_{e}^{a}$, do not depend on our normalization for $\dot{M}$.

As in the proto-magnetar case, we assume an open poloidal magnetic field $B_{\rm p} = B_{0}(s_{0}/s)^{2}$, where $s_{0} = R_{0}/\cos\theta$ and $B_{0}$ is the strength of the poloidal field at the outflow's base.  As discussed by Levinson (2006), whose formulation and geometry are similar to ours, a more consistent approach would be to seek a self-similar solution to the trans-field equation (e.g., Li et al.~1992; Contopoulos 1994).  However, simultaneously including the wind's slow point topology and consequent neutrino-heated mass-loss in such a formalism is difficult without multi-dimensional neutrino-heated MHD calculations, a formidable numerical challenge.  In addition, the large-scale poloidal field threading NDAF disks is uncertain and so a more detailed calculation does not seem warranted.

As discussed in the proto-magnetar case, $Y_{e}^{a}$ is generally determined fairly close to where the wind is launched near the disk midplane and so the precise radial scaling of $B_{\rm p}$ ($\propto s^{-2}$) is likely less important than its magnitude near the base of the wind $\approx B_{0}$.  The properties of the wind, however, can depend strongly on $\theta$.  In particular for $\theta < 60^\circ$ matter that co-rotates at the Keplerian orbital velocity which is perturbed from the disk along the field line is unstable and can escape to infinity, even without additional energy deposition (Blandford $\&$ Payne 1982).  This results in significant mass-loading for field lines with $\theta < 60^\circ$; in addition, the field develops a substantial base toroidal component $B_{\phi,0}$.  For $\theta < 60^\circ$, NDAF winds are therefore slow, heavily mass-loaded, have $B_{\phi,0} \gg B_{0}$, and possess properties that are relatively independent of $\theta$.  On the other hand, more vertical flux tubes ($60^\circ < \theta < 90^\circ$), which resemble equatorial proto-magnetar winds more than equatorial disk winds, have lower $\dot{M}$, accelerate to faster asymptotic speeds, and possess properties that depend sensitively on $\theta$; in particular, the wind becomes increasing relativistic as $\theta \rightarrow 90^\circ$ (Daigne $\&$ Mochkovitch 2002), and ultra-relativistic outflows from nearly vertical field lines may be capable of producing GRBs.

In terms of $B_{0}$ and $\dot{M}$ the magnetization (eq.~[\ref{magnetization}]) of an NDAF wind launched from radius $R_{0}$ is given by
\be
\sigma = B_{0}^{2}R_{0}^{4}\Omega_{\rm K}^{2}/\dot{M}c^{3},
\label{magnetization_disk}
\ee
where we have assumed that the base of the outflow rotates at approximately the Keplerian rate $\Omega_{\rm K} = (GM/R_{0}^{3})^{1/2}$.  The rate of angular momentum lost through a non-relativistic ($\sigma < 1$) wind is given by $\dot{J}_{W} = \dot{M}R_{A}^{2}\Omega_{\rm K}$, while for relativistic ($\sigma > 1$) winds $R_{A} \simeq R_{\rm L}$ and $\dot{J}_{W} \simeq \dot{E}_{\rm mag}/\Omega_{\rm K} \simeq \sigma\dot{M}c^{2}/\Omega_{\rm K}$, where $\dot{E}_{\rm mag}$ is the wind's energy loss rate.  The ratio of $\dot{J}_{W}$ to the disk's angular momentum loss rate $\dot{J}_{D} = \dot{M}_{D}R_{0}^{2}\Omega_{\rm K}$ is therefore given by
\begin{eqnarray}
\frac{\dot{J}_{W}}{\dot{J}_{D}} & \simeq (\dot{M}R_{A}^{2}/\dot{M}_{D}R_{0}^{2}) =  (\dot{M}s_{A}^{2}/\dot{M}_{D}s_{0}^{2}):  \sigma < 1 \nonumber \\
& \simeq (B_{0}^{2}R_{0}^{2}/\dot{M}_{D}c):  \sigma > 1
\label{jdotratio}
\end{eqnarray} 
Although a wind with $\dot{J}_{W} < \dot{J}_{D}$ is physical because stresses internal to the disk can transfer angular momentum outwards, allowing accretion at $\dot{M}_{D}$ to proceed, a wind for which $\dot{J}_{W} > \dot{J}_{D}$ is unphysical because it violates conservation of angular momentum. 

\subsection{Numerical Procedure}

We calculate NDAF wind solutions starting from a location just above the accretion disk midplane at a fixed cylindrical radius $R_{0} = 14 R_{g} \approx 6\times 10^{6}$ cm, which is roughly the location of the disk's peak neutrino flux.  The wind's local neutrino luminosities and energies are determined from CB07's marginally neutrino optically thin $\dot{M}_{D} = 0.2M_{\sun}$ s$^{-1}$, $\alpha = 0.03$ NDAF solution.  We focus on winds driven from low $\dot{M}_{D}$, neutrino transparent disks because low $\dot{M}_{D}$ accretion, due to its lower neutrino-driven mass-loss, is probably more likely to produce neutron-rich, ultra-relativistic outflow.  When considering NDAFs with $\dot{M}_{D}< 0.2M_{\sun}$ s$^{-1}$ (down to $\dot{M}_{D} \approx \dot{M}_{\rm ign}$; eq.~[\ref{mdotign}]), we scale the boundary conditions and neutrino emission properties analytically from the $\dot{M}_{D} = 0.2M_{\sun}$ s$^{-1}$ solution, as discussed below.

We use Newtonian gravity, including just the point mass of a $M = 3M_{\sun}$ black hole.  The evolution equations for $v_{\rm p}$, $v_{\phi}$, $B_{\phi}$, $\rho$, and $T$ are the same as those used in the PNS case with $r$ replaced by $s$, except that the gravitational acceleration in the radial momentum equation (MTQ07, eq.~[7]) is now the component projected along the field line:
\be
g_{s} = \frac{GM(s-s_{0}\sin^{2}\theta)}{(s^{2}+\sin^{2}\theta[s_{0}^{2}-2ss_{0}])^{3/2}}.
\label{gs}
\ee   

Our microphysics is essentially identical to that in the PNS case.  In particular, the electron fraction is evolved according to equation (\ref{yeevo}), again with $r$ replaced by $s$; note that because of the disk's significantly lower density, degeneracy effects are less important near the base of the wind than in the PNS case.  The electron and anti-electron neutrino energy fluxes ($F_{\nu_{e}},F_{\bar{\nu}_{e}}$) used to calculate the neutrino absorption and heating rates near $R_{0}$ are computed from CB07's $\alpha-$disk solution in the same manner as the local neutrino energy densities that were used to calculate $Y_{e}^{\nu}$ for Figure \ref{plot:ye_eq}; to within a factor of $\sim 2$ the total spherically-equivalent neutrino luminosity obeys $L_{\nu} = L_{\nu_{e}} + L_{\bar{\nu}_{e}} \equiv 4\pi R_{0}^{2}(F_{\nu_{e}}+F_{\bar{\nu}_{e}}) \approx \eta\dot{M}_{D}c^{2}$, where $\eta \approx$ 0.04 is the total radiative efficiency for a non-rotating black hole from CB07.  In a similar manner as for PNSs, we take the neutrino flux intercepted by the outflow to be approximately constant with radius for $r \lesssim R_{0}$ and to decrease $\propto r^{-2}$ for $r\gtrsim R_{0}$ (T01 eq.~[24]; see Surman $\&$ McLaughlin 2004, 2005 for a more detailed treatment).  Although the geometry that we assume is simplistic, $Y_{e}^{a}$ is set relatively close to the base of the wind and thus depends primarily on the neutrino energy density in the vicinity of $R_{0}$.  Indeed, as Figure \ref{plot:rates} illustrates, the advection and pair capture rates rise and fall above the base of the outflow, respectively, much more rapidly than the neutrino absorption rates decrease; this would likely remain true for any realistic geometry.

Unlike PNSs, NDAFs do not produce significant tau or muon neutrino emission and so annihilations and neutral-current interactions from these neutrino species are ignored.  The mean electron and anti-electron neutrino energies near $R_{0}$, which are not expected to vary strongly with $\dot{M}_{D}$ for neutrino optically thin accretion (see eq.~[\ref{enuthin}]), are taken from Figure \ref{plot:ye_eq} as $\langle\epsilon_{\nu_{e}}\rangle = 10.5$ MeV and $\langle\epsilon_{\bar{\nu}_{e}}\rangle = 13.1$ MeV; from CB07's solution we also determine that $F_{\bar{\nu}_{e}} \simeq 1.2 F_{\nu_{e}}$ near $R_{0}$ so that $Y_{e}^{\nu} \approx 0.51$ (see eq.~[\ref{yeanu}] and Fig.~\ref{plot:ye_eq}).  We set the wind's inner density $\rho_{0}$ equal to the disk midplane density $\rho_{D}$ if the electron neutrino optical depth along $s$ to the midplane is $\tau_{\nu} \lesssim \frac{2}{3}$; otherwise, we instead choose $\rho_{\nu}$ to enforce $\tau_{\nu} \simeq \frac{2}{3}$ at $s_{0}$. 

Although viscous heating generally dominates neutrino heating in the midplane of NDAFs, recent radiation MHD simulations suggest that little energy is dissipated in the disk corona (Turner 2004; Hirose, Krolick, $\&$ Stone 2006; Blaes et al.~2006b; Krolik et al. 2006).  For this reason, we neglect viscous heating in the wind entirely and set the base temperature $T_{0}$ of the wind by balancing neutrino cooling with just neutrino heating, as in the PNS case; note, however, that significant viscous heating in the wind would likely result in both additional mass-loss and deneutronization.  The disk's midplane temperature $T_{D}$, which is set by the balance between viscous heating and neutrino cooling (see eqs.~[\ref{enuthin}]$-$[\ref{enuthick}]), is therefore generally higher than $T_{0}$.

Although NDAFs are efficiently cooled and geometrically-thin, radial pressure support is not completely negligible and so the disk's angular rotation frequency $\Omega$, which we use to set the wind's inner angular velocity, is slightly sub-Keplerian: $\Omega^{2} = \Omega_{\rm K}^{2}(1-H^{2}/R_{0}^{2})$, where $H \approx 0.2R_{0}$ is the disk scale height near $R_{0}$, which, like $T_{D}$, is approximately independent of $\dot{M}_{D}$ for neutrino optically thin accretion.   

Our disk wind calculations can be compared with similar one-dimensional flux tube calculations by Daigne $\&$ Mochkovitch 2002 (DM02), Pruet et al.~2004 (P04), and Levinson 2006 (L06).  DM02 calculated the requisite conditions for ultra-relativistic outflow from hyper-accreting disks, including neutrino heating and cooling and, in the neutrino optically thin case, a simplified viscous heating prescription.  Because DM02 was primarily concerned with obtaining $\dot{M}$ as a function of disk conditions, they only considered wind conditions near the sonic point; in addition, DM02 assumed co-rotation rather than accounting for the magnetic field explicitly.  L06 improved upon the calculations of DM02 by including the full equations of general relativistic MHD.  Although L06 explored the effects of finite $B_{0}$, L06 concentrated, like DM02, on the sub-slow magnetosonic regime and his calculations did not capture the Alfv\'{e}n or fast magnetosonic radii.  Although this approach allowed L06 to calculate $\dot{M}$ as a function of the open magnetic flux and $L_{\nu}$, the base toroidal field $B_{\phi,0}$ remained a free parameter in L06's formulation.  Because $B_{\phi,0}$ is associated with the conserved magnetic induction it is fixed in our calculations by the fact that our steady-state winds pass smoothly through all three MHD critical points.  L06 speculated on the potential deneutronization of magnetized NDAF winds by noting the similarity between the advection and relevant weak interaction rates; he did not, however, calculate the evolution of $Y_{e}$ explicitly.  

Finally, P04 investigated nucleosynthesis in collapsar disk winds by solving the equations of hydrodynamics and by evolving $Y_{e}$ from the disk midplane.  P04 included neutrino heating and viscous heating through an $\alpha-$prescription; although they neglected neutrino absorptions in evolving $Y_{e}$, they also argued for the generic deneutronization of thermally-driven winds.  Although P04 did not include the effects of magnetic fields on the wind explicitly, their outflows were calculated along well-defined, vertically-directed flux tubes and were artificially forced to co-rotate outside the base of the wind, presumably to mimic the effect of a strong poloidal magnetic field.  Our calculations are, to the best of our knowledge, the first to fully calculate the effects of MHD on the evolution of $Y_{e}$ in NDAF outflows and the first to capture all three MHD critical points.

\subsection{Numerical Results}

\begin{figure*}
\resizebox{\hsize}{!}{\includegraphics{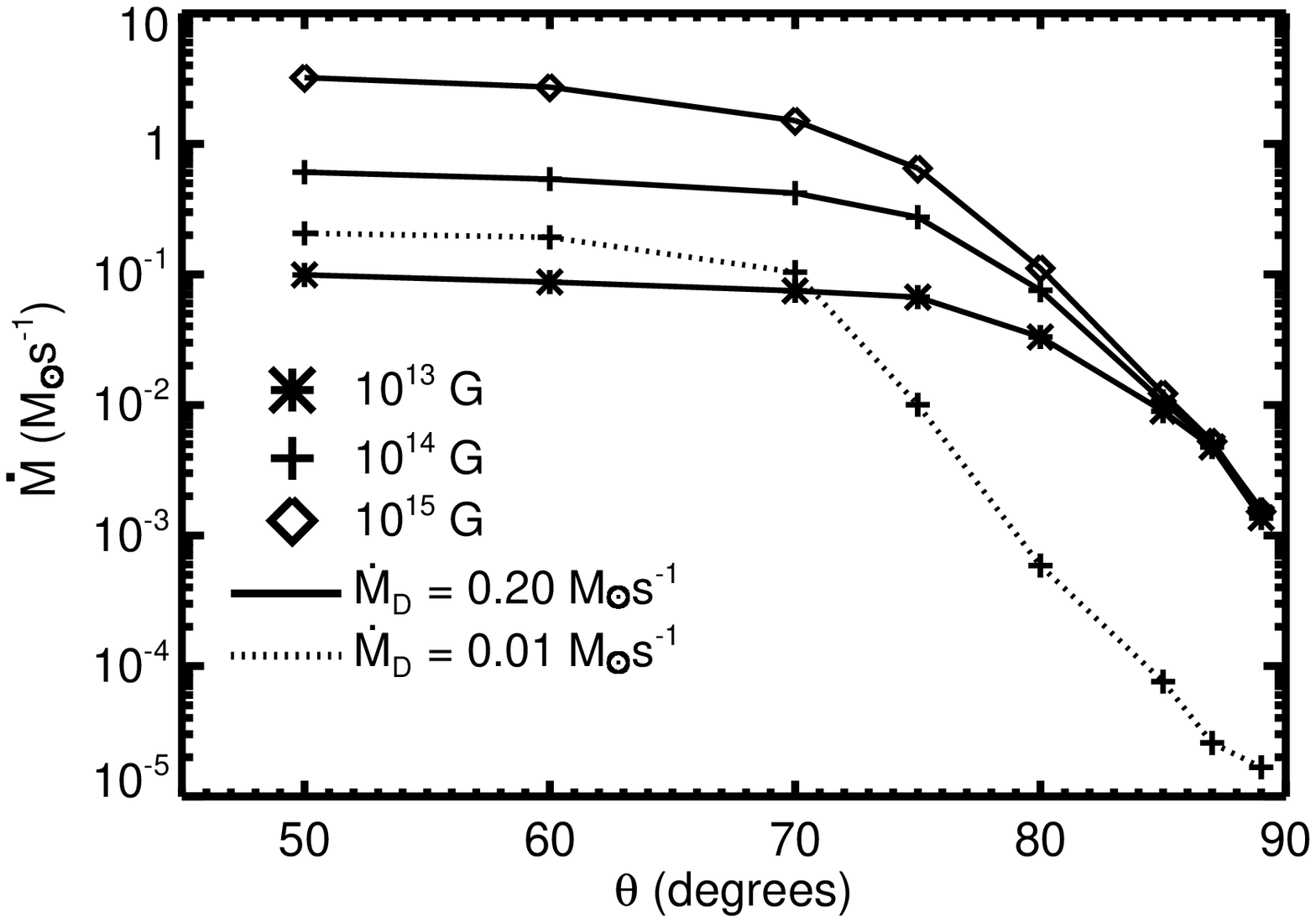}\includegraphics{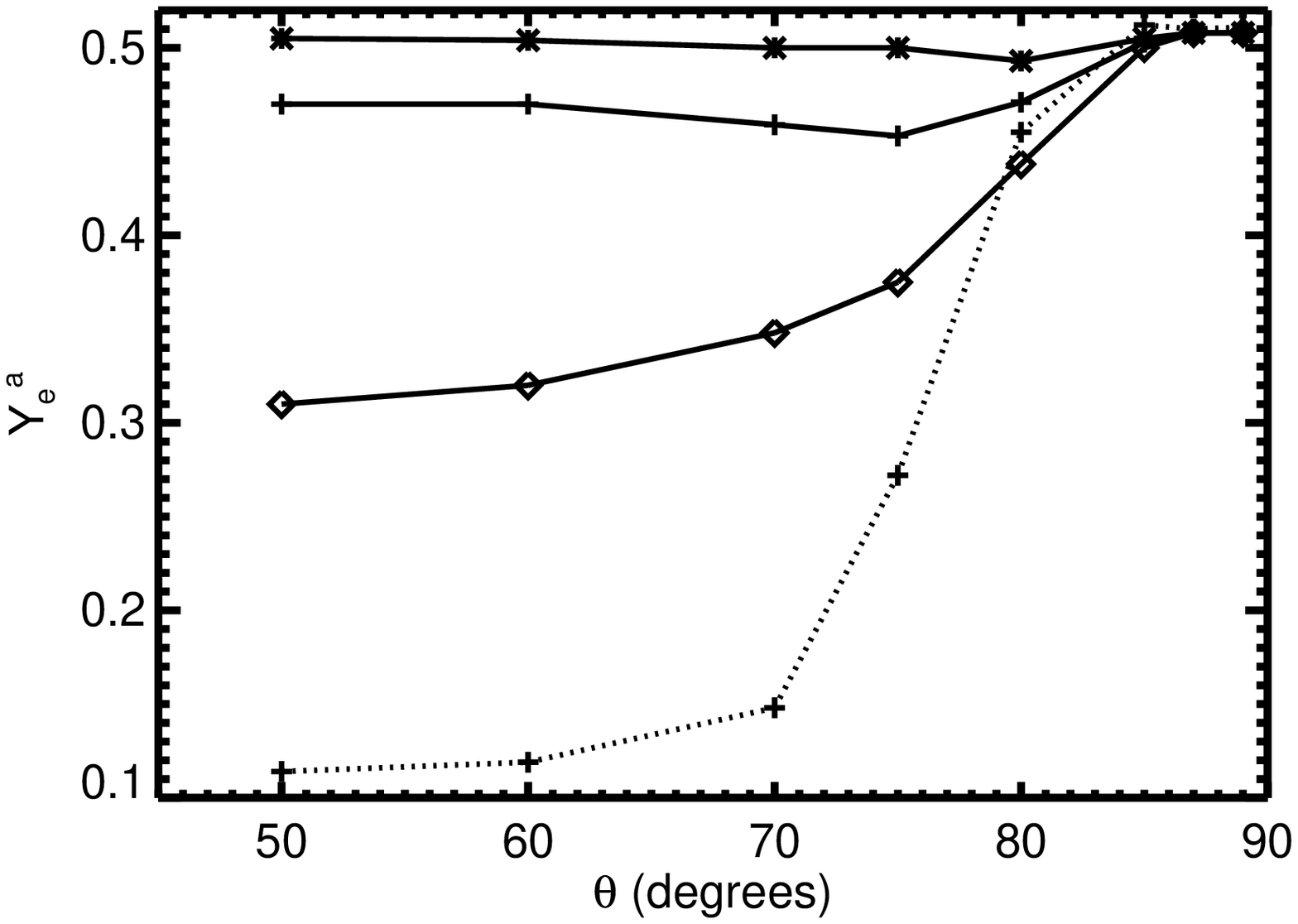}}
\resizebox{\hsize}{!}{\includegraphics{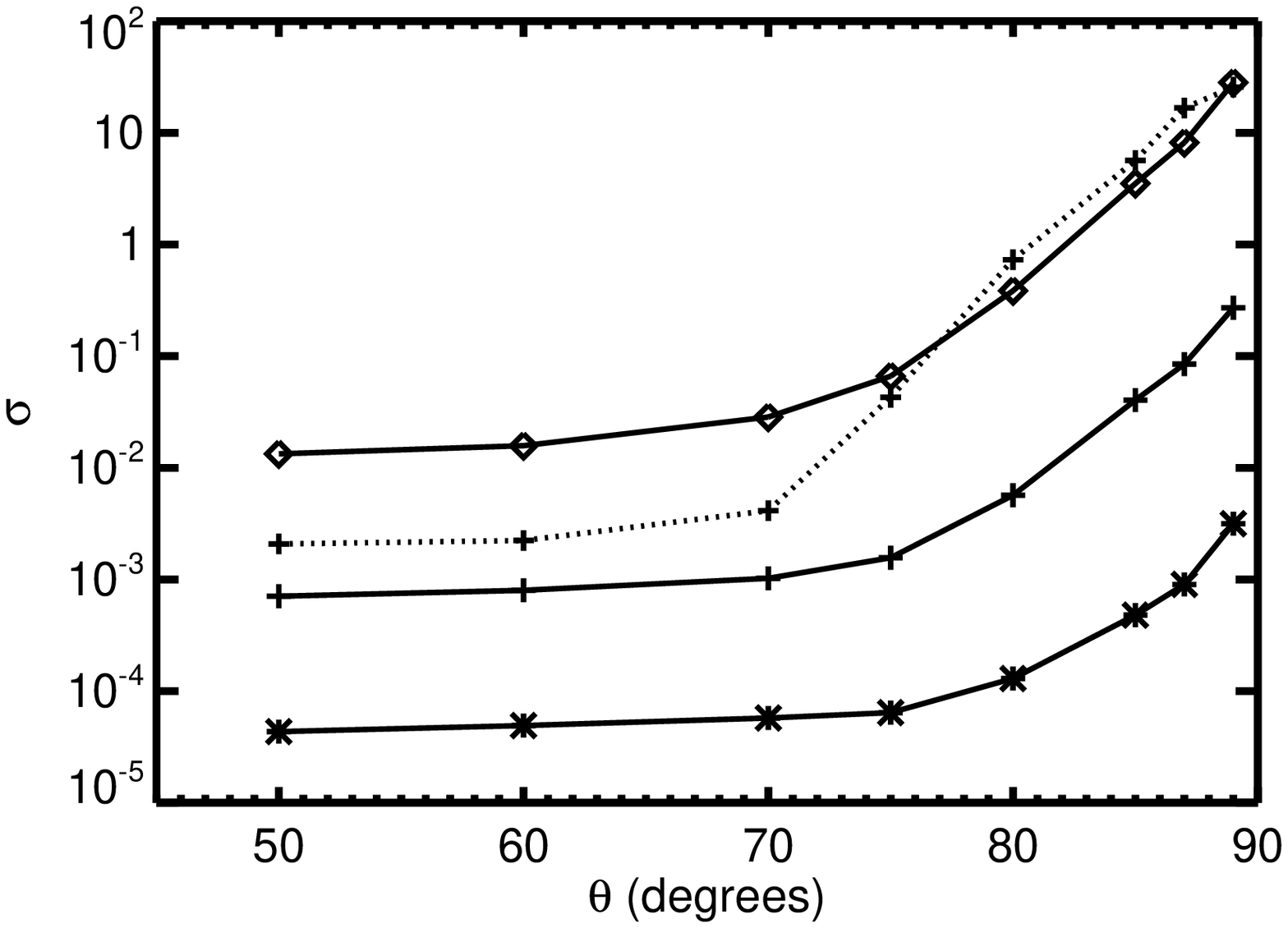}\includegraphics{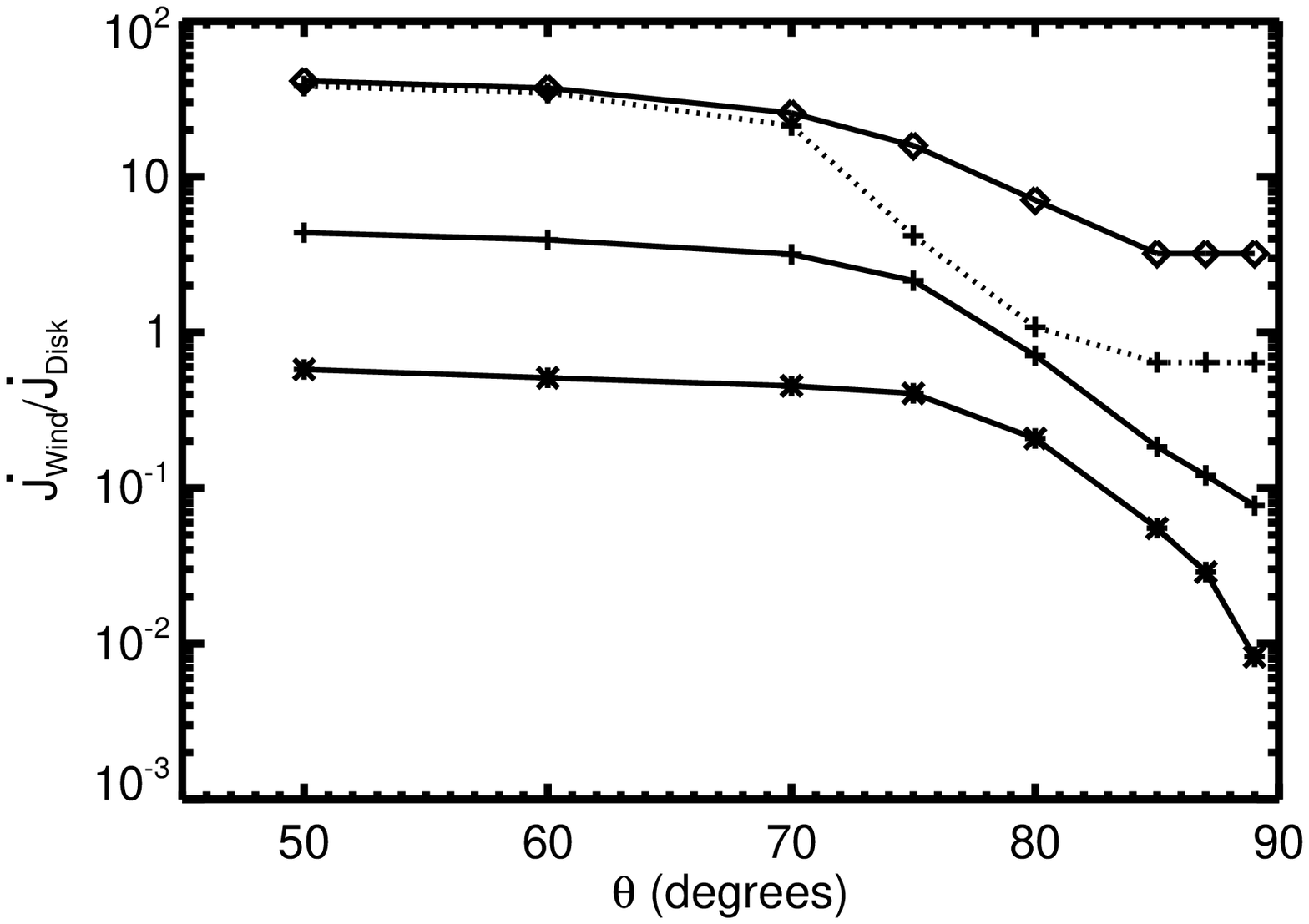}}
\caption{Mass loss rate $\dot{M}$, asymptotic electron fraction $Y_{e}^{a}$, magnetization $\sigma$ (eq.~[\ref{magnetization_disk}]), and the ratio of the wind to accretion angular momentum loss rates $\dot{J}_{W}/\dot{J}_{D}$ (eq.~[\ref{jdotratio}]) for NDAF winds as a function of the angle $\theta$ between the wind's flux tube and the disk midplane (see Fig.~\ref{plot:diskfig}); base poloidal magnetic field strengths $B_{0} = 10^{13}$ G ($asterisk$), $10^{14}$ G ($cross$), and $10^{15}$ G ($diamond$) are considered.  The solutions were calculated starting from a location just above the disk midplane at a radius $R_{0} = 14R_{g} \approx 6\times 10^{6}$ cm from the central $M=3M_{\sun}$, non-rotating ($a=0$) black hole and employed boundary conditions derived from the $\alpha-$disk NDAF solution of Chen $\&$ Beloborodov (2007) with $\alpha = 0.03$ and accretion rate $\dot{M}_{D} = 0.2M_{\sun}$ s$^{-1}$ ($solid$ $lines$).  Also shown are wind solutions for $\dot{M}_{D} \approx 10^{-2}M_{\sun}$ s$^{-1}$ and $B_{0} = 10^{14}$ G.  For $\theta \lesssim 60^{\circ}$ the wind properties are relatively independent of $\theta$ while for $\theta \gtrsim 60^{\circ}$, $\dot{M}$ decreases and $\sigma$ increases rapidly with $\theta$.  For the most vertical field lines ($\theta \rightarrow 90^{\circ}$), along which the outflow experiences minimal centrifugal support, $\dot{M}$ approaches its purely thermal, neutrino-driven value $\dot{M}_{\rm th}$ (eq.~[\ref{mdotthermalNDAF}]).  Of the solutions shown, only those with $\dot{J}_{W}/\dot{J}_{D} < 1$ are physical.  All of the physical solutions for $\dot{M}_{D} = 0.2M_{\sun}$ s$^{-1}$ are non-relativistic ($\sigma < 1$) and do not possess an asymptotic neutron excess ($Y_{e}^{a} \approx Y_{e}^{\nu} \simeq 0.51$).  Winds from the $\dot{M}_{D} = 10^{-2} M_{\sun}$ s$^{-1}$ NDAF have lower $\dot{M}$ and larger $\sigma$ than the winds driven from the higher $\dot{M}_{D}$ disk; however, the solutions for $\dot{M}_{D} = 10^{-2}M_{\sun}$ s$^{-1}$ with $\dot{J}_{W} < \dot{J}_{D}$ still have $Y_{e}^{a} \sim 0.5$.  
\\}
\label{plot:mdot1}
\end{figure*}
The solid lines in Figure \ref{plot:mdot1} show calculations of the mass-loss rate $\dot{M}$, asymptotic electron fraction $Y_{e}^{a}$, magnetization $\sigma$ (eq.~[\ref{magnetization_disk}]), and angular momentum loss rate $\dot{J}_{W}$ compared to that required for accretion $\dot{J}_{D}$ (eq.~[\ref{jdotratio}]) for NDAF winds with base poloidal field $B_{0} = 10^{13}$ G, $10^{14}$ G, and $10^{15}$ G as a function of flux tube angle $\theta$ for the $\dot{M}_{D} = 0.2$ $M_{\sun}$ s$^{-1}$, $\alpha = 0.03$, $M = 3 M_{\sun}$, $a=0$ NDAF solution of CB07 (the solution on which Figure \ref{plot:ye_eq} was based).  Because magnetic pressure at the base of the outflow exceeds the total thermal pressure in the midplane of the disk for $B_{0} \gtrsim 2\times 10^{15}$ G, fields much stronger than those we have considered in Figure \ref{plot:mdot1} are probably unphysical because they could not be self-consistently anchored to the disk.  Our disk wind calculations are summarized in Table \ref{table:disktable1}.

Figure \ref{plot:mdot1} shows that $\dot{M}$ is very large and relatively independent of $\theta$ for $\theta \lesssim 60^\circ$ but that for $\theta \gtrsim 60^\circ$ $\dot{M}$ decreases rapidly with increasing $\theta$.  The mass-loss rate also increases with increasing $B_{0}$, except for the largest angles, for which $\dot{M}$ saturates for sufficiently large $B_{0}$, no longer increasing with increasing $B_{0}$ (as in the PNS case; see Fig.~\ref{plot:magnetarmdot}).  There is no similar saturation for smaller $\theta$ because the large values of $\dot{M}$ preclude the outflow from co-rotating out to the sonic point for physical $B_{0}$.\footnote{Because the base of the wind rotates at a slightly sub-Keplerian rate due to radial pressure support in the disk, even for $\theta < 60^{\circ}$ mass-loss would saturate at $\dot{M} \approx 4\pi R_{0}^{2}\rho_{0}c_{s,0} \sim 10-100 M_{\sun}$ s$^{-1}$ for sufficiently large $B_{0}$ ($\sim 10^{16}$ G; the co-rotating limit), where $c_{s,0} \sim 0.1$ c is the sound speed near the base of the outflow.}  

Figure \ref{plot:mdot1} also shows that although $Y_{e}^{a}$ is relatively low ($\sim 0.3$) for the high-$B_{0}$, low-$\theta$ wind solutions, all of the solutions in Figure \ref{plot:mdot1} with $Y_{e}^{a} \ll Y_{e}^{\nu} \simeq 0.51$ are unphysical because they have $\dot{J}_{W} > \dot{J}_{D}$ and thus extract angular momentum at a rate exceeding that required for accretion through the disk from which the wind's boundary conditions were derived; in fact, invoking the criterion $\dot{J}_{W} < \dot{J}_{D}$ (see eq.~[\ref{jdotratio}]), the only physical solutions in Figure \ref{plot:mdot1} are those with $B_{0} = 10^{13}$ G for all $\theta$ and $B_{0} = 10^{14}$ G for $\theta \gtrsim 80^\circ$.  Also note from Figure \ref{plot:mdot1} that, in addition to having $Y_{e}^{a} \sim 0.5$, all of the physical solutions for $\dot{M}_{D} = 0.2 M_{\sun}$ s$^{-1}$ are non-relativistic ($\sigma < 1$).  Because our calculations have thoroughly spanned the physical parameter space of $B_{0}$ and $\theta$, this conclusion is robust, at least for $\dot{M}_{D} = 0.2 M_{\sun}$ s$^{-1}$. 

For lower $\dot{M}_{D}$ the deneutronizing neutrino luminosity and neutrino-driven mass-loss rate are lower and thus relativistic, neutron-rich outflow may be more likely.  To explore this possibility we have also calculated wind solutions for outflows driven from an NDAF with $\dot{M}_{D} = 10^{-2} M_{\sun}$ s$^{-1} \simeq \dot{M}_{\rm ign}(\alpha = 0.03)$, keeping the other parameters identical to those of the higher $\dot{M}_{D}$ case ($\alpha = 0.03$, $M = 3M_{\sun}$, $R_{0} = 14R_{g}$).  To compute these low $\dot{M}_{D}$ solutions we decreased the neutrino fluxes by a factor of $20$ from our $\dot{M}_{D} = 0.2M_{\sun}$ s$^{-1}$ calculation, left the mean neutrino energies unchanged (they are not expected to vary strongly with $\dot{M}_{D}$; see eq.~[\ref{enuthin}]), and decreased the disk midplane density $\rho_{0} \sim \rho_{D} \propto \dot{M}_{D}$.  We only calculated solutions with $B_{0} \lesssim 3\times 10^{14}$ G because for larger field strengths the magnetic pressure would exceed the thermal pressure in the disk.

In Figure \ref{plot:mdot1} we show the $\dot{M}_{D} = 10^{-2}M_{\sun}$ s$^{-1}$ calculations for $B_{0} = 10^{14}$ G with a dotted line for comparison with the higher $\dot{M}_{D}$ solutions.  Although the properties of these low $\dot{M}_{D}$ winds and their dependence on $\theta$ qualitatively resemble the higher $\dot{M}_{D}$ solutions, $Y_{e}^{a}$ and $\dot{M}$ are generally lower and $\sigma$ is higher than for the winds driven from the more neutrino-luminous disk.  Most notably, the high-$\theta$ solutions are now both physical ($\dot{J}_{W} < \dot{J}_{D}$) and relativistic ($\sigma > 1$), illustrating that nearly vertically-directed outflows from low-$\dot{M}_{D}$ NDAFs represent plausible GRB central engines.  However, these high-$\theta$, relativistic winds still possess no significant neutron excess ($Y_{e}^{a} \sim 0.5$); this means that, even for $\dot{M}_{D} \approx \dot{M}_{\rm ign}$, simultaneously neutron-rich and ultra-relativistic outflow appears unlikely.  One might think that $\dot{M}_{D}$ (and hence $L_{\nu}$) could be lowered further until neutron-rich, ultra-relativistic outflow was possible at some intermediate latitude; a key difference, however, between the NDAF problem considered here and the PNS problem considered in $\S\ref{section:magnetar}$ is that, unlike in the PNS case, the disk's neutrino luminosity cannot be decreased below $L_{\nu,{\rm ign}} = \eta\dot{M}_{\rm ign}(\alpha = 0.03)c^{2} \sim 10^{51}$ ergs s$^{-1}$ (eq.~[\ref{lign}]).  For $\dot{M}_{D} < \dot{M}_{\rm ign}$ the disk is geometrically thick, and the midplane is not necessarily degenerate and neutron-rich (see $\S\ref{section:thickdiskwinds}$).   

\subsection{Analytic Constraints}
\label{section:conditionsNDAF}

As in the PNS case, the purely thermal, neutrino-driven mass-loss rate from NDAFs can be estimated analytically by requiring that the energy used to unbind a typical nucleon $E_{B}$ be supplied entirely by neutrino heating; equation (\ref{mdotthermal}) can therefore also be used to estimate $\dot{M}_{\rm th}$ for NDAF disk winds, provided that $R_{\nu} \rightarrow R_{0}$, $\langle\epsilon_{\nu}\rangle$ is taken from equations (\ref{enuthin}) and (\ref{enuthick}), and that, as discussed below, we include the effects of the lower effective gravity $g_{\rm eff}$ near the base of the outflow than in the PNS case.  For neutrino optically thin accretion we find that $\dot{M}_{\rm th}$ for NDAFs is given by
\begin{eqnarray}
\lefteqn{\dot{M}_{\rm th} \approx 10^{-3}\left(\frac{\dot{M}_{D}}{\dot{M}_{\rm ign}}\right)^{5/3}\left(\frac{g_{\rm eff}/(GM/R_{0}^{2})}{H/R_{0}}\right)^{-1}} \nonumber \\
& & M_{3}^{119/90}\alpha_{0.1}^{301/90}\left(\frac{R_{\rm p}}{6R_{g}}\right)^{-1}\left(\frac{R_{0}}{6R_{g}}\right)^{79/60}\,M_{\sun}{\rm\,\,s^{-1}}
\label{mdotthermalNDAF}
\end{eqnarray}
where we have used equation (\ref{lign}) to scale the neutrino luminosity to the value $L_{\nu,{\rm ign}}$ associated with the ignition accretion rate $\dot{M}_{\rm ign}$ (eq.~[\ref{mdotign}]) because NDAFs only exist for $\dot{M}_{D} > \dot{M}_{\rm ign}$.  

The purely neutrino-driven mass-loss rate for PNSs (eq.~[$\ref{mdotthermal}$]) was derived for outflows emerging antiparallel to the PNS's gravitational field and by assuming substantially sub-break-up rotation.  NDAF winds, however, are driven from the disk midplane at an angle inclined with respect to the black hole's gravitational field (i.e., $\theta \ne 0$) and even non-magnetized centrifugal support can be important near the base of the outflow because the disk rotates at nearly the Keplerian rate.  This means that the effective surface gravity $g_{\rm eff}(\theta)$, which is the difference between the gravitational acceleration inward along $s$ ($g_{s}(\theta)$; eq.~[\ref{gs}]) and the centrifugal acceleration outward along $s$ ($v_{\phi}^{2}/s$), is always less than the gravitational acceleration for purely equatorial outflow ($g_{s}(\theta = 0) = GM/R_{0}^{2}$) that was used to derive $\dot{M}_{\rm th}$ for PNSs.  Assuming that the wind conserves angular momentum away from $s_{0}$ (i.e., $v_{\phi} = \Omega R_{0}(s/s_{0})^{-1}$) and concentrating on the wind's inner, quasi-hydrostatic atmosphere, we find that $g_{\rm eff}/(GM/R_{0}^{2})$ is roughly constant with $\theta$ and is approximately equal to $(H/R_{0}) \approx 0.3\alpha_{0.1}^{1/10}M_{3}^{-1/10}(R_{0}/6R_{g})^{7/20}$, where $H$ is the NDAF's vertical scale height near $R_{0}$; this calculation justifies our characteristic scaling for $g_{\rm eff}$ in equation (\ref{mdotthermalNDAF}).  

Although equation (\ref{mdotthermalNDAF}) represents the thermal, neutrino-driven mass-loss rate from an NDAF, only the nearly-vertical, $\theta = 89^{\circ}$ wind solutions shown in Figure \ref{plot:mdot1} (which experience minimal centrifugal support) in fact reach $\dot{M} \sim \dot{M}_{\rm th}$.  Thus, equation (\ref{mdotthermalNDAF}) should be taken as the absolute $minimum$ NDAF mass-loss rate; in the presence of even a modest magnetic field with a non-vertical inclination, $\dot{M}$ is significantly larger that $\dot{M}_{\rm th}$.  Because NDAFs possess an absolute minimum mass-loss rate, their outflows also posses an absolute maximum magnetization $\sigma_{\rm max}$, which is given by
\begin{eqnarray}
\lefteqn{\sigma_{\rm max} = \frac{B_{0}^{2}\Omega_{\rm K}^{2}R_{0}^{4}}{\dot{M}_{\rm th}c^{3}} \approx} \nonumber \\
& & 10(\alpha\beta_{\phi})^{-1}\left(\frac{\beta_{\phi}}{\beta_{\rm p}}\right)\left(\frac{\dot{M}_{D}}{\dot{M}_{\rm ign}}\right)^{-2/3}\left(\frac{g_{\rm eff}/(GM/R_{0}^{2})}{H/R_{0}}\right) \nonumber \\
& & \times M_{3}^{1/9}\alpha_{0.1}^{-16/9}\left(\frac{R_{\rm p}}{6R_{g}}\right)^{23/20}\left(\frac{R_{0}}{6R_{g}}\right)^{-199/60},
\label{sigma_maxNDAF}
\end{eqnarray}
where $\beta_{\rm p}$ is the ratio of the disk's midplane thermal pressure $P_{D}$ to the magnetic pressure associated with the wind's base poloidal field $B_{0}^{2}/8\pi$ and $\beta_{\phi}$ is similarly defined for the disk's midplane toroidal magnetic field $B_{\phi,D}$.  In equation (\ref{sigma_maxNDAF}) we have written $\sigma_{\rm max}$ in terms of $(\alpha\beta_{\phi})^{-1}$ because for angular momentum transport via the MRI, local shearing box simulations find that $\alpha\beta_{\phi} \sim 1$ (Hawley, Gammie, $\&$ Balbus 1995); we also write equation (\ref{sigma_maxNDAF}) in terms of $\beta_{\phi}/\beta_{\rm p} = B_{0}^{2}/B_{\phi,D}^{2}$ because if the open poloidal field is generated through a dynamo from the toroidal field this ratio is unlikely to exceed unity (and is probably much less).

Equation ($\ref{sigma_maxNDAF}$) shows that $\sigma_{\rm max} \propto \dot{M}_{D}^{-2/3}$, which explains why the low $\dot{M}_{D}$ solutions in Figure \ref{plot:mdot1} produced more relativistic outflows than the high $\dot{M}_{D}$ solutions under the physical constraint $\beta_{\rm p} > 1$ that we imposed on our calculations.  Equation (\ref{sigma_maxNDAF}) also shows that NDAFs can, in principle, produce ultra-relativistic ($\sigma \gg 10^{2}-10^{3}$) outflows from small radii ($R_{0} \sim R_{\rm isco}$), provided that a significant fraction of the magnetic energy present in the disk is associated with a large-scale, open poloidal field (i.e., $\beta_{\rm p} \sim \beta_{\rm t}$).  For $\alpha = 0.1$, $M = 3M_{\sun}$, $\beta_{\rm p} = \beta_{\phi}$, $\alpha\beta_{\phi} = 1$, and for accretion onto a rapidly rotating black hole (so that $R_{0} \approx R_{\rm p} \approx 2R_{g}$) equation (\ref{sigma_maxNDAF}) gives $\sigma_{\rm max} \approx 10^{2}(\dot{M}_{D}/\dot{M}_{\rm ign})^{-2/3}$.  In this case ultra-relativistic outflow could occur for $\dot{M}_{D} \sim \dot{M}_{\rm ign}$ and could thus accompany a substantial accretion power $\dot{E}_{\rm acc} \equiv \eta\dot{M}_{D}c^{2} \sim \eta\dot{M}_{\rm ign}(\alpha = 0.1)c^{2} \sim 10^{51}$ ergs s$^{-1}$.  Because $\dot{E}_{\rm acc}$ represents the maximum MHD luminosity of a GRB-producing jet, this shows that, under ideal conditions, disk winds from NDAFs form plausible GRB central engines.  

Equation (\ref{sigma_maxNDAF}) shows that under some circumstances ultra-relativistic outflow from NDAF disks is plausible.  However, the maximum magnetization for material with a $neutron$ $excess$ $\sigma_{\rm max}^{\rm n}$ is lower than $\sigma_{\rm max}$ because the minimum mass-loss rate that must accompany a neutron excess is substantially larger than $\dot{M}_{\rm th}$ according to the same arguments that were used in $\S\ref{section:conditionsPNS}$ for the PNS case; indeed, using equation (\ref{sigma_maxNDAF}) and the NDAF analog of equation (\ref{mdotneutron}) we find that
\begin{eqnarray}
\lefteqn{\sigma_{\rm max}^{\rm n} \equiv \frac{\sigma_{\rm max}\langle\epsilon_{\nu}\rangle 2R_{0}}{GMm_{\rm n}} \approx } \nonumber \\
& & 2(\alpha\beta_{\phi})^{-1}\left(\frac{\beta_{\phi}}{\beta_{\rm p}}\right)\left(\frac{\dot{M}_{D}}{\dot{M}_{\rm ign}}\right)^{-2/3}\left(\frac{g_{\rm eff}/(GM/R_{0}^{2})}{H/R_{0}}\right) \nonumber \\
& & \times M_{3}^{-4/45}\alpha_{0.1}^{-71/45}\left(\frac{R_{\rm p}}{6R_{g}}\right)^{17/20}\left(\frac{R_{0}}{6R_{g}}\right)^{-139/60}.
\label{sigma_maxNDAFneutron}
\end{eqnarray}
Equation (\ref{sigma_maxNDAFneutron}) shows that neutron-rich, relativistic outflow is very unlikely from NDAF disks.  For $\dot{M}_{D} = \dot{M}_{\rm ign}$, $\alpha = 0.1$, $M = 3M_{\sun}$, $\beta_{\rm p} = \beta_{\phi}$, $\alpha\beta_{\phi} = 1$, and for outflow launched from near $R_{\rm isco}$ of a rapidly spinning black hole ($R_{0} = R_{\rm p} = 2R_{g}$), equation (\ref{sigma_maxNDAFneutron}) gives $\sigma_{\rm max}^{\rm n} \sim 10$, insufficient to explain the ultra-relativistic outflows inferred from GRBs.  

If the fiducial scalings of equation (\ref{sigma_maxNDAFneutron}) are not adopted, ultra-relativistic outflow with a neutron excess may be possible in some circumstances.  For instance, if $\alpha = 0.01$ instead of $\alpha = 0.1$ (again for $\dot{M}_{D} = \dot{M}_{\rm ign}$, $M = 3M_{\sun}$, $\beta_{\rm p} = \beta_{\phi}$, $\alpha\beta_{\phi} = 1$, $R_{0} = R_{\rm p} = 2R_{g}$) then $\sigma_{\rm max}^{\rm n} \sim 10^{2}-10^{3}$.  In this case, however, $\dot{E}_{\rm acc}(\alpha = 0.01) \sim 10^{50}(\dot{M}_{D}/\dot{M}_{\rm ign})$ ergs s$^{-1}$ and so a very large fraction of the accretion energy would need to be deposited in the ultra-relativistic outflow to explain the observed luminosities of GRBs.  Likewise, although $\alpha\beta_{\rm p} \ll 1$ is possible if the poloidal field threading the disk is not the result of a local dynamo (instead resulting from, e.g., magnetic flux advected from large radii in the disk; e.g., Spruit $\&$ Uzdensky 2005), the constraint $\dot{J}_{W} < \dot{J}_{D}$ (eq.~[\ref{jdotratio}]) independently requires that $\alpha\beta_{\rm p} \gtrsim 2/3(R_{0}/R_{g})^{-1/2}(R_{0}/H) \approx 4\alpha_{0.1}^{-1/10}M_{3}^{1/10}(R_{0}/R_{g})^{-17/20}$; thus, outflow launched from near $R_{\rm isco}$ must have $\beta_{\rm p}\alpha \gtrsim 1$ because otherwise it would carry away angular momentum at a rate exceeding that required for matter to accrete at the rate $\dot{M}_{D}$.  Lastly, we reiterate that the true thermally-driven mass-loss rate can, in principle, far exceed the purely neutrino-driven value $\dot{M}_{\rm th}$ (eq.~[\ref{mdotthermalNDAF}]) if viscous heating is important near the base of the wind; if this were the case, $\sigma_{\rm max}^{\rm n}$ would be substantially reduced below the value given by equation (\ref{sigma_maxNDAFneutron}).  We thus conclude that neutron-rich GRB outflows are unlikely from NDAF disk winds.
\subsection{Cross-Field Neutron Diffusion}
\label{section:neutronpickup}

In the previous section we have argued that highly relativistic ($\sigma \gtrsim 100-1000$) winds driven from the innermost radii of NDAFs are unlikely to be intrinsically neutron-rich.  However, free neutrons are uncharged and may collisionally diffuse across magnetic field lines to the polar region (hereafter, the ``jet'') from an adjacent, more baryon-rich wind (Eichler $\&$ Levinson 1999; Levinson $\&$ Eichler 2003, hereafter LE03; McKinney 2005b).  If the total neutron mass diffusion rate $\dot{M}_{\rm n}^{\rm diff}$ dominates the mass-loading of the polar jet then the highly relativistic polar outflow will be significantly ``polluted'' by neutrons and may end up neutron-rich after all.  

Neutron diffusion into the polar jet from the adjacent mass-loaded wind is limited to a surface area $\sim 4\pi s_{\alpha}^{2}\varphi$, where $s_{\alpha}$ is the distance from the base of the wind to where free nucleons recombine into $\alpha-$particles (which are charged and therefore cannot efficiently diffuse across field lines) and we have assumed that $s_{\alpha} \gg R_{0}/\varphi$, where $\varphi = \pi/2 - \theta \ll 1$ is the opening angle of the jet.  Neutron diffusion is limited by elastic proton collisions, with a rate $\langle\sigma_{\rm n-p}v_{\rm rel}\rangle \simeq 10^{-15}$ cm$^{3}$ s$^{-1}$ and a corresponding mean free path $\lambda_{\rm n-p} \simeq v_{\rm th}/(n_{\rm p}\langle\sigma_{\rm n-p}v_{\rm rel}\rangle)$, where $n_{\rm p}$ and $v_{\rm th} \approx (kT/m_{\rm n})^{1/2}$ are the proton number density and the ion thermal speed, respectively.  Following LE03, we assume that the density gradient length scale separating the mass-loaded wind and the axial jet is given by $l \sim (v_{\rm th}\tau_{\rm dyn}\lambda_{\rm n-p})^{1/2}$, where $\tau_{\rm dyn} \equiv s/v_{\rm p}$ is the wind's dynamical timescale.\footnote{This choice for $l$ is appropriate for a very abrupt transition in the wind's density with cylindrical radius, such as between field lines threading the disk and those threading the black hole's event horizon.  A perhaps more natural (but less conservative) choice for $l$ is the cylindrical radius in the wind at $\alpha-$particle recombination $(\approx \varphi s_{\alpha}$), which would produce an even smaller $\dot{M}_{\rm n}^{\rm diff}$ than is given in equation (\ref{mdotdiff}).}  Using the Fick-diffusion number flux of neutrons into the jet of $F_{\rm n} \approx n_{\rm n}v_{\rm th}(\lambda_{\rm n-p}/l)$, we estimate that
\begin{eqnarray}
\lefteqn{\dot{M}_{\rm n}^{\rm diff} \sim 4\pi\varphi s_{\alpha}^{2}m_{\rm n}F_{\rm n}|_{s_{\alpha}} = \left(\frac{4\pi s_{\alpha}kT_{\alpha}\dot{M}}{\langle\sigma_{\rm n-p}v_{\rm rel}\rangle}\right)^{1/2}} \nonumber \\
& & \sim 10^{-8}\left(\frac{kT_{\alpha}}{\rm MeV}\right)^{1/2}\left(\frac{s_{\alpha}}{10^{8}{\rm\,\,cm}}\right)^{1/2}\left(\frac{\dot{M}}{0.1M_{\sun}\,{\rm s}^{-1}}\right)^{1/2}M_{\sun}{\rm s^{-1}} \nonumber \\
\label{mdotdiff}
\end{eqnarray}
where $\dot{M} \approx 4\pi\varphi^{2}m_{\rm n}v_{\rm p}(n_{\rm p}+n_{\rm n})s^{2}$, $n_{\rm n}$, and $T_{\alpha} \equiv T(s_{\alpha})$ are the wind's mass-loss rate, neutron number density, and temperature at $\alpha-$particle recombination, respectively, and we have assumed that $Y_{e} \simeq 0.5$ in the wind.  We evaluate equation (\ref{mdotdiff}) at $s_{\alpha}$ because $\dot{M}_{\rm n}^{\rm diff}$ is dominated by the largest radii at which the wind is still primarily free nucleons.

For the relatively moderate entropies ($S^{a} \lesssim 10^{2}$ k$_{B}$ baryon$^{-1}$) that characterize neutrino-heated, magnetocentrifugally-driven winds (see Tables \ref{table:yea_magnetar} and \ref{table:disktable1}), $\alpha-$particles form at a high temperature ($T_{\alpha} \sim 1$ MeV), which obtains relatively close to the base of the wind ($s_{\alpha}\lesssim 10^{7}$ cm).  In this case, even for $\dot{M} \sim M_{\sun}$ s$^{-1}$, equation (\ref{mdotdiff}) gives $\dot{M}_{\rm n}^{\rm diff} \lesssim 10^{-8}M_{\sun}$ s$^{-1}$.  If the axial jet is itself driven from the disk, $\dot{M}_{\rm n}^{\rm diff}$ is thus significantly lower than the minimum mass-loss already supplied by neutrino heating (eq.~[\ref{mdotthermalNDAF}]); hence, cross-field neutron diffusion is ineffective at segregating neutrons in low entropy NDAF winds.  

If, on the other hand, the axial jet is powered by $\nu-\bar{\nu}$ annihilation or the Blandford-Znajek process and has little or no intrinsic baryon-loading (such as if it threads the black hole's event horizon), then $\dot{M}_{\rm n}^{\rm diff}$, although small, may dominate the jet's mass loading.  For instance, for a polar jet power of $\dot{E} \sim 10^{50}-10^{51}$ erg s$^{-1}$ equation (\ref{mdotdiff}) shows that diffusive neutron mass-loading from an encasing wind with a mass-loss rate $\dot{M} \sim 10^{-2}M_{\sun}$ s$^{-1}$ would, by itself, limit the jet's asymptotic Lorentz factor to $\Gamma \sim 10^{4}-10^{5}$.  Asymptotically neutron-rich outflow may result in this case if the jet remains ``clean'' to large radii; elucidating the observable consequences of such very high-$\Gamma$ neutron-rich outflows will, however, require additional work.  Lastly, we note that although our calculations show that NDAF winds probably possess moderate entropy, previous works that have considered diffusion into the jet have focused on very high entropy outflows characteristic of hydrodynamic ``fireballs'' in the GRB literature (LE03; McKinney 2005b).  These calculations find larger $\dot{M}_{\rm n}^{\rm diff}$ than we have estimated in equation (\ref{mdotdiff}) in large part because $\alpha-$particles do not form until much larger radii in high entropy winds.  Furthermore, if $Y_{e}^{a} \lesssim 0.5$ in the encasing baryon-rich wind and some neutrons remain free to radii larger than $s_{\alpha}$, $\dot{M}^{\rm diff}_{\rm n}$ may be larger than predicted by equation (\ref{mdotdiff}).  Evaluating the free neutron fraction that remains in magnetically-driven disk winds following the freeze-out of any $\alpha$-process or $r$-process capture that may occur following $\alpha-$particle recombination is, however, beyond the scope of this paper. 

\subsection{Thick Accretion Disk Winds}
\label{section:thickdiskwinds}

For both proto-magnetars, where $L_{\nu} \propto t^{-1}$ starting $\sim 1$ s following core bounce (eq.~[\ref{lnupns}]), and NDAFs, where $L_{\nu} \propto \dot{M}_{D} \propto t^{-5/3}$ is expected at late times from ``fall-back'' accretion (Chevalier 1989; Woosley $\&$ Weaver 1995) or $\dot{M}_{D} \propto t^{-\alpha}$ with $\alpha \sim 1$ due to the viscous evolution of a disk of finite mass, the low neutrino luminosities attained at late times are the most favorable for ultra-relativistic, neutron-rich outflows.  However, while proto-magnetars must maintain sub-millisecond rotation for most of the Kelvin-Helmholtz cooling epoch to produce late-time neutron-rich GRB outflows, for NDAFs, which always rotate at nearly the Keplerian rate, the difficulty is more fundamental; for $L_{\nu} \lesssim L_{\nu,{\rm ign}}$ (eq.~[\ref{lign}]) the disk is no longer efficiently cooled and the disk midplane may no longer remain dense, degenerate, and neutron-rich.  For both collapsars and the accretion accompanying compact object mergers there will thus come a time $t_{\rm ign}$ after which $\dot{M}_{D} < \dot{M}_{\rm ign}$ and the disk will transition from an NDAF to an advection-dominated thick disk (Narayan $\&$ Yi 1994; Narayan et al.~2001).  

Although NDAFs enter $\beta$-equilibrium on an accretion timescale (B03a), thick disks generally do not.  Using equation (48) from B03a we find that $\beta-$equilibrium is only established in a thick disk at radii smaller than a critical $\beta$-equilibrium radius $R_{\beta}$, which is given by
\be R_{\beta} = 34(14)(\dot{M}_{D}/\dot{M}_{\rm ign})^{10/13}\alpha_{0.1}^{-4/39}M_{3}^{4/39}R_{g}
\label{rbeta}
 \ee
for accretion onto a black hole with spin $a = 0(0.95)$.  Because $R_{\beta} < R_{\rm isco}$ for $\dot{M}_{D} \lesssim 0.1(0.03)\dot{M}_{\rm ign}$, equation (\ref{rbeta}) shows that matter accreting through a thick disk at a rate $\dot{M}_{D} \ll \dot{M}_{\rm ign}$ is not in $\beta-$equilibrium at any radius.  The neutron content of the disk at small radii for $t \gg t_{\rm ign}$ (and thus of any potential late-time GRB-producing outflow) will therefore depend on the composition of the matter feeding it.  In particular, late-time GRB outflows from collapsar disks, which are continually fed from large radii by their progenitor's stellar He envelope (which has $Y_{e} \sim 0.5$), will not be neutron-rich. 

On the other hand, accretion disks formed from compact object mergers, which are usually fed from the tidal disruption of at least one neutron star\footnote{An exception are the mergers of black hole-white dwarf binaries, which may produce long-duration GRBs (Fryer et al.~1999b).} (e.g., Rosswog $\&$ Ramirez-Ruiz 2002), are initially neutron-rich.  However, the disks formed from compact object mergers are expected to be more compact than collapsar disks (with circularization radii $\lesssim 10-30R_{g}$) and are probably not continually supplied with substantial mass from large radii; the late-time neutron content of thick disks from compact object mergers therefore depends on the evolution of $Y_{e}^{D}$ immediately following the NDAF to thick disk transition at $t_{\rm ign}$.  Because equation (\ref{rbeta}) shows that a thick disk with $\dot{M}_{D} \lesssim \dot{M}_{\rm ign}$ does have sufficient time to enter $\beta-$equilibrium, the late-time electron fraction in a thick disk from a compact object merger depends on whether $\beta-$equilibrium near $R_{\rm isco}$ in a thick disk for $\dot{M} \sim \dot{M}_{\rm ign}$ favors a neutron-rich or a proton-rich composition.  Although NDAFs are always sufficiently dense and degenerate to favor a neutron-rich composition, $\beta-$equilibrium in a thick disk only favors $n/p > 1$ for radii $smaller$ than a critical neutron-rich radius $R_{\rm n}$; using B03a equation (50) we find that
\be R_{\rm n} = 24(2)(\dot{M}_{D}/\dot{M}_{\rm ign})^{2}\alpha_{0.1}^{4/3}M_{3}^{-4/3}R_{g}
\label{rn}
\ee  
for accretion onto a black hole with spin $a = 0(0.95)$.  Equation (\ref{rn}) shows that when a disk transitions from an NDAF to a thick disk, the matter near $R_{\rm isco}$ may be driven to either a neutron-rich or a proton-rich state, depending on $\alpha$, $M$, and the extent of the disk.  Thus, although disk winds from NDAFs are unlikely to produce neutron-rich GRB outflows, neutron-rich outflows may be possible from the thick disks associated with compact object mergers at late times ($t\gtrsim t_{\rm ign}$) or from collapsar disks at $t \sim t_{\rm ign}$ (i.e., after the thick disk transition but before the disk is fed by additional, neutron-poor material from large radii).

\section{Discussion}

\label{section:discussion}

By calculating the structure and neutron content of neutrino-heated MHD winds driven from the neutron-rich surfaces of proto-magnetars and NDAFs, we have delineated the conditions under which a large neutron excess can be preserved in these outflows.  We have focused on the conditions for simultaneously neutron-rich and ultra-relativistic outflows because magnetized winds from hyper-accreting disks and newly-formed magnetars are plausible GRB central engines; despite being difficult to distinguish on other grounds, each of these central engines may possess a distinctive nucleonic signature.  If the consequences of neutron-rich GRB outflows enumerated in $\S\ref{section:intro}$ can be identified or constrained, magnetar and black hole models for GRBs may thus be observationally distinguishable.

Although GRB central engines are often neutron-rich (Pruet et al.~2003; B03a), we find that ultra-relativistic neutron-rich outflows are possible only under surprisingly limited circumstances.  Central engines that are sufficiently dense and degenerate to be neutron-rich must be efficiently neutrino-cooled.  For the resulting sub-virial temperatures, several of the thermal neutrinos released by the central engine must be absorbed by a typical nucleon for it to escape the deep gravitational potential due to neutrino-driving alone.  Since neutrino absorptions from efficiently neutrino-cooled central engines usually favor an asymptotic electron fraction $Y_{e}^{a} \gtrsim 0.5$, purely neutrino-driven outflows are generally driven back to a relatively neutron-poor state, with the precise value of $Y_{e}^{a}$ determined by the neutrino spectrum of the central source (see eq.~[\ref{yeanu}] and surrounding discussion).

Additional forces (e.g., magnetocentrifugal) can prevent deneutronization by supplying most of the binding energy needed to escape the central engine's gravitational potential well.  However, $Y_{e}^{a}$ is set so close to the base of the outflow that the very inner, hydrostatic atmosphere of the wind must be altered for $Y_{e}^{a} \ll 0.5$ to obtain; this unavoidably increases the wind's mass-loss rate $\dot{M}$ (see eq.~[\ref{mdotneutron}]).  Indeed, a generic anti-correlation between $\dot{M}$ and $Y_{e}^{a}$ is evident in our numerical calculations shown in Figures \ref{plot:magnetarye}, \ref{plot:magnetarmdot}, and \ref{plot:mdot1}.  Neutron-rich GRB outflows are thus difficult to produce because the minimum neutron-rich mass-loss rate often precludes ultra-relativistic speeds.  Other heating (e.g., viscous) that may be present in the outflow in addition to neutrinos (which $must$ be present) will only further increase $\dot{M}$ and, through additional entropy deposition and pair creation, further deneutronize the outflow.  

Our conclusion that simultaneously ultra-relativistic and neutron-rich outflow is difficult to produce depends on the assumption that $Y_{e}^{\nu}$ is not significantly less than 0.5.  Therefore, to be more precise, the conclusion of this paper is that the nucleonic content of ultra-relativistic outflows driven from efficiently neutrino-cooled central engines is typically set by an equilibrium with neutrino absorptions.  Thus, if the neutrino spectra and luminosities of PNSs and NDAFs are significantly different from what current calculations find, and $Y_{e}^{\nu} \ll 0.5$, neutron-rich outflows from GRB central engines may be more common.  The next step in improving our understanding of magnetocentrifugal winds from GRB central engines is to include the effects of a strong magnetic field on the neutrino interactions and the equation of state for the leptons.  For the parameters considered in this work, the latter should be more important, modifying the pressure and entropy profiles.

With these general constraints in mind, we now discuss the prospects for neutron-rich outflows from individual central engines.  Our conclusions are summarized in Table \ref{table:scenarios}.

\subsection{Proto-Magnetars}
\label{section:magnetarconclusion}

Proto-magnetars with surface magnetic field strengths $B_{\nu} \gtrsim 10^{14}-10^{15}$ G can produce neutron-rich outflows, but only for rotation periods $P_{\rm} \lesssim P_{\rm n} \approx 0.8$ ms (see Fig.~\ref{plot:magnetarye}).  If the minimum stable neutron star rotation period $P_{\rm min}$ exceeds $P_{\rm n}$, then neutron star birth should not be accompanied by substantially neutron-rich outflow.\footnote{As discussed in $\S\ref{section:implications}$, an exception may arise for neutrino-driven winds produced at very early times after core bounce following the accretion-induced collapse (AIC) of a white dwarf (Dessart et al.~2006); similarly, in the absence of an overlying, accreting stellar mantle, an early-time ejection of $\sim 10^{-3}-10^{-1} M_{\sun}$ of low-$Y_{e}$ material may result from the core bounce shock's ``break-out'' or the subsequent neutrino-heated shock revival (e.g., Hillebrandt, Wolff, $\&$ Nomoto 1984; Fryer et al.~1999a)}  On the other hand, if $P_{\rm n} \lesssim P_{\rm min}$ and steady-state neutron-rich winds from magnetar birth are indeed possible, they should be restricted to events with a total GRB plus SN energy exceeding $E_{\rm rot}(P_{\rm n}) \approx 4\times 10^{52}$ ergs (absent significant gravitational wave losses).  Such extremely energetic events are likely rare, even among GRBs.  Furthermore, not all magnetar births that release $\gtrsim 4\times 10^{52}$ ergs will necessarily produce $simultaneously$ ultra-relativistic and neutron-rich outflows because proto-magnetar winds are heavily mass-loaded at early times following core bounce.  If the proto-magnetar's dipole field strength exceeds $B_{\nu}^{\rm dip} \sim 10^{15}$ G, spin-down is so rapid that neutron-rich outflow is unlikely by the time the PNS has cooled sufficiently that $\sigma \gtrsim 100-1000$ (see Fig.~\ref{plot:sigma_period}).  Since the PNS spin-down power for $P \approx P_{\rm n}$ and $B_{\nu}^{\rm dip} \approx 10^{15}$ G is $\sim 3\times 10^{49}$ ergs s$^{-1}$ (Spitkovsky 2006; B06), neutron-rich GRBs from proto-magnetars possess a maximum beaming-corrected luminosity $L_{\gamma} \sim 10^{49}(\epsilon/0.3)$ ergs s$^{-1}$, where $\epsilon$ is the efficiency for converting outflow to gamma-ray energy.

\subsection{NDAFs}

Although neutron-rich GRB outflows are possible from a subset of proto-magnetars, we find that they are unlikely to originate from NDAFs under any circumstances.  In agreement with previous works (e.g., Daigne $\&$ Mochkovitch 2002; Levinson 2006), we find that outflows with $\sigma \gtrsim 100$ are possible from the innermost radii of NDAFs around rapidly rotating black holes ($a \approx 1$), despite the minimum neutrino-driven mass-loss rate (eq.~[\ref{mdotthermalNDAF}]).  The significantly larger $\dot{M}$ required for a large neutron excess, however, precludes neutron-rich NDAF outflows from attaining $\sigma \gtrsim 1-10$ for significant accretion powers (eq.~[\ref{sigma_maxNDAFneutron}]).  Furthermore, because modest entropy, magnetocentrifugally-driven winds only possess free nuclei over a relatively limited range of radii, cross-field diffusion is ineffective at polluting otherwise neutron-poor axial jets with free neutrons from adjacent, more heavily baryon-loaded winds (see $\S$\ref{section:neutronpickup}).  

Although we find that neutron-rich winds from NDAFs are unlikely, several caveats should be discussed.  Pruet et al.~(2004) suggest that ``bubbles'' of neutron-rich material may escape the disk via chaotically-heated buoyant magnetic filaments, a picture similar to some models for GRBs (Narayan et al.~1992; Katz 1997; Klu{\'z}niak $\&$ Ruderman 1998).  Although this possibility cannot be ruled out, current thin disk simulations do not find significant energy deposition in low density, coronal regions (e.g., Hirose, Krolik, $\&$ Stone 2006), and whether such low-$Y_{e}$ bubbles can remain neutron-rich despite the pair-capture deneutronization that accompanies such chaotic heating is unclear.  If the Blandford-Znajek mechanism or $\nu-\bar{\nu}$ annihilation above the disk's rotation axis powers the GRB outflow instead of a disk wind, the base of the GRB-producing jet may be effectively baryon-free because the field lines would then thread the black hole's event horizon instead the disk midplane; what ultimately sets the wind's baryon-loading in this case is unclear.  Our calculations show that neutrino-heated disk winds will form a modest entropy ``sheath'' around such a baryon-free jet.  If, however, the wind encasing the jet possesses a much higher entropy (e.g., Pruet et al.~2001), cross-field neutron diffusion is more effective (Levinson $\&$ Eichler 2003; McKinney 2005b) and may result in asymptotically neutron-rich polar outflow.  Furthermore, even if the outflow's field lines don't thread the disk, it is in principle possible that some form of chaotic mass-loading may pollute the baryon-poor base of the jet with matter from the neutron-rich disk midplane.  We note, however, that current simulations find very little matter entraining the jet from the disk, thus requiring implementation of a numerical density floor along the polar axis (e.g., Proga $\&$ Begelman 2003; Proga et al.~2003).  

\subsection{Thick Disks}

\label{section:thickdisks}

Although NDAFs exist over a range of accretion rates that are relevant to both long and short-duration GRBs, GRB-producing outflows can also be powered by the accretion of matter that is not efficiently neutrino-cooled.  Such geometrically-thick, radiatively inefficient accretion flows (RIAFs) exist for both $\dot{M}_{D} < \dot{M}_{\rm ign}$ (``low-$\dot{M}_{D}$ RIAFs''), for which the density is too low for efficient neutrino cooling near $R_{\rm isco}$, and for $\dot{M}_{D} \gg \dot{M}_{\rm ign}$ (``high-$\dot{M}_{D}$ RIAFs''), for which matter advects into the black hole faster than it cools (e.g., Di Matteo, Perna, $\&$ Narayan 2002).  High $\dot{M}_{D}$ RIAFs, while possibly relevant to ``prompt'' collapsars (MacFadyen $\&$ Woosley 1999), are probably most relevant to compact object mergers given their smaller expected disk radii and shorter accretion timescales.  Low-$\dot{M}_{D}$ RIAFs are relevant to both the late stages of collapsars and compact binary mergers.

Like NDAFs, high-$\dot{M}_{D}$ RIAFs enter $\beta$-equilibrium before accreting and, although they are not as neutron-rich as NDAFs, they also typically have $Y_{e}^{D} \ll 0.5$ (e.g., Surman $\&$ McLaughlin 2004; Lee et al.~2005, hereafter L05).  High-$\dot{M}_{D}$ RIAFs are confined to radii in the disk smaller than the ``trapping'' radius $R_{\rm t}$, the point interior to which matter has insufficient time to cool before accreting; $R_{\rm t}$ exists outside $R_{\rm isco}$ for mass accretion rates greater than $\dot{M}_{\rm t} \approx 9(2)\alpha_{0.1}^{1/3}M_{\sun}$ s$^{-1}$ for $a= 0(0.95)$ and $M=3M_{\sun}$ (CB07).  Although radiatively inefficient on the whole, disks with $\dot{M}_{D} > \dot{M}_{\rm t}$ still release a substantial neutrino luminosity because $\dot{M}_{\rm t}$ is large and accretion is still efficiently cooled for radii larger than $R_{\rm t}$; indeed, for steady-state accretion, Di Matteo, Perna, $\&$ Narayan (2002) find that as $\dot{M}_{D}$ increases beyond $\dot{M}_{\rm t}$, $L_{\nu}$ saturates to a value $\sim 10^{53}$ ergs s$^{-1}$.  This substantial neutrino luminosity will drive significant neutrino-heated mass-loss, thereby severely limiting the asymptotic Lorentz factor $\Gamma$ of any outflow driven from the RIAF's surface.  Using the simulations of L05 (their Fig.~6), we estimate that the neutrinosphere temperature of a high-$\dot{M}_{D}$ RIAF near $R_{\rm t}$ is roughly $T_{\nu} \sim 3$ MeV (corresponding to a mean neutrino energy $\langle\epsilon_{\nu}\rangle \sim 10$ MeV); thus, from equation (\ref{mdotthermal}) we estimate that the minimum, neutrino-driven mass-loss rate from the innermost radii of high-$\dot{M}_{D}$ RIAFs is $\dot{M}_{\rm th} \approx 10^{-3}-10^{-2}M_{\sun}$ s$^{-1}$.  For a wind driven from the surface of a high-$\dot{M}_{D}$ RIAF to reach $\Gamma \gtrsim 100$ would thus require an MHD luminosity $\dot{E} \gtrsim 10^{53}-10^{54}$ ergs s$^{-1}$, which is comparable to the entire accretion power for $\dot{M}_{D} \sim \dot{M}_{\rm t} \sim 1-10M_{\sun}$ s$^{-1}$ and is substantially larger than the outflow powers typically inferred from short-duration GRBs (e.g., Bloom et al.~2005).  GRB-producing outflows, whether neutron-rich or not, are thus unlikely to originate from disk winds produced by high-$\dot{M}_{D}$ RIAFs. 

Unlike NDAFs and high-$\dot{M}_{D}$ RIAFs, low-$\dot{M}_{\rm D}$ RIAFs may not enter $\beta-$equilibrium on an accretion timescale (eq.~[\ref{rbeta}]); thus, the electron fraction in a low-$\dot{M}_{D}$ thick disk's midplane can remain approximately equal to that of the material used to form the disk initially.  Because of the low neutrino luminosity and positive Bernoulli parameters of low-$\dot{M}_{D}$ RIAFs, viscous heating likely dominates neutrino heating in outflows driven from low-$\dot{M}_{D}$ thick disks, thereby making deneutronization by neutrinos unlikely.  Although nondegenerate pair-captures may drive $Y_{e} \rightarrow 0.5$ depending on the precise viscous energy deposition profile in the wind, this possibility is less likely than for NDAF outflows because $\beta-$equilibrium is already slow in the disk midplane and because the accretion and outflow advection rates for thick disks are comparable.  Furthermore, if acceleration from the midplane is enhanced due to magnetocentrifugal slinging, pair-capture deneutronization would be further suppressed.  Thus, outflows driven from low-$\dot{M}_{D}$ RIAFs probably retain the electron fraction of the disk's original composition.  As a result, outflows driven at late times from collapsars, which are fed from the progenitor star's relatively neutron-poor envelope (e.g., $Y_{e} \simeq 0.5$ for a purely He composition), probably do not possess a significant neutron excess.  In this case, significant $^{56}$Ni could be produced in the disk's outflow, powering a bright SN (MacFadyen $\&$ Woosley 1999).

Although disks formed from compact object mergers are usually initially neutron-rich, they are compact and nearly all of the matter goes through an NDAF phase in which $\beta-$equilibrium modifies the initial composition of the disk; their composition at late times therefore depends on how they transition from an NDAF to a low-$\dot{M}_{D}$ thick disk.  Depending on $\alpha$, the size of the disk, and the mass of the black hole, the composition of the thick disk immediately following the NDAF to low-$\dot{M}_{D}$ RIAF transition at $t_{\rm ign}$ can be either neutron-rich or proton-rich (see eq.~[\ref{rn}]).  Although by no means assured, neutron-rich GRBs outflows may thus be possible from thick disks in both collapsars immediately following $t_{\rm ign}$ (especially if the variability imposed by the NDAF-thick disk transition actually causes the GRB; Giannios 2007) and in short-duration GRBs from compact object mergers.  In either case, neutron-rich GRB outflows should be restricted to events with accretion power (and thus maximum GRB luminosity) less than $L_{\nu,{\rm ign}}$ (eq.~[\ref{lign}]).

\subsection{Non-Relativistic Neutron-Rich Winds}
\subsubsection{Optical Transients}
\label{section:opticaltransients}

Relativistic, GRB-producing outflows from proto-magnetars and hyper-accreting disks are possible under specialized circumstances, but non-relativistic winds are also likely to be present, probably occur in a wider variety of progenitors, and probably carry more total energy.  The composition of such non-relativistic winds may also lead to an observable signature, most directly via radiation from ejecta that is reheated by the decay of radioactive elements in the wind.  One possibility in the case of magnetar birth is that the heavily mass-loaded wind that emerges at early times (carrying a total mass up to $\sim 0.01-0.1M_{\sun}$ for a millisecond rotator; Dessart et al.~2007; see MTQ07, Fig.~10) could produce a bright SN-like event.  However, by combining the mass-loss rate that accompanies a significant neutron excess (eq.~[\ref{mdotneutron}]) with the fiducial PNS cooling evolution $L_{\nu}(t)$ given in equation (\ref{lnupns}), we find that the total PNS mass-loss capable of being processed into $^{56}$Ni under NSE (which requires no significant neutron excess; Hartmann, Woosley, $\&$ El Eid 1985) cannot exceed $M_{\rm Ni}^{\rm max} \sim \int_{1\rm{\,s}}^{\tau_{\rm KH}}\dot{M}_{\rm th}\phi_{\rm n}dt \sim 10^{-3} M_{\sun}$, much too small to contribute appreciably to an optical light curve powered by $^{56}$Ni, and subsequent $^{56}$Co, decay (e.g., see Kulkarni 2005, Fig. 7).\footnote{Enhanced Ni production is still possible in the core collapse context via early energization of a successful SN shock due to rapid spin-down (TCQ04).  A proto-magnetar's spin-down luminosity is substantially enhanced over the vacuum-dipole rate at early times following the launch of the SN shock by the excess magnetic flux opened by neutrino-heated, centrifugally-driven mass-loss (B06).  In this case, the enhanced Ni yield is produced by additional shock heating of the stellar progenitor, not directly in the wind itself.}  

The modest optical luminosity associated with the decay of a mass $\sim M_{\rm Ni}^{\rm max}$ of Ni is consistent with the rather stringent upper limits on the SN component accompanying some short-duration GRBs (e.g., GRB050509B; Hjorth et al.~2005); this supports the viability of a model in which short-duration GRBs are powered by the rapid spin-down of a magnetized, rapidly rotating magnetar, formed following the accretion-induced collapse (AIC) of a white dwarf (Usov 1992) or resulting from the merger of a double neutron star binary (a ``super-pulsar''; Rosswog et al.~2003).  It is less clear how much of the substantial mass-loss driven from the accretion disk formed during the merger of two compact objects will be processed into $^{56}$Ni.  Quantifying this will require additional work.

Even if little mass is ejected with $Y_{e}^{a} \gtrsim 0.5$ (capable of producing $^{56}$Ni), both proto-magnetars and hyper-accreting disks could in principle produce detectable transients due to the presence of neutron-rich non-relativistic outflows.  For example, a proto-magnetar that is born rotating sufficiently rapidly to produce late-time, relativistic neutron-rich matter must also eject a total mass $M_{\rm n} \gtrsim 0.1M_{\sun}$ in slower ($v \gtrsim 0.3$ c) free neutrons at earlier times.   The detectability of such neutron-rich non-relativistic outflows is, however, uncertain.  Any neutrons that ultimately remain free in the wind will $\beta$-decay into protons on a timescale $\tau_{\rm \beta} \approx 900$ s at a radius $\gtrsim 10^{13}$ cm; thus, one observable manifestation of neutron-rich outflow may be ``Macronovae'' powered by the thermal energy released by this decay ($\approx 6\times 10^{49}(M_{\rm n}/0.1M_{\sun})$ ergs), which may be detectable on hour-day timescales following the birth of the central object (Li $\&$ Paczy{\'n}ski 1998; Kulkarni 2005).  

\subsubsection{$r$-process Nucleosynthesis}

Rather than leaving free neutrons, the decompression of slowly moving, modest entropy, and moderately neutron-rich ($0.1 \lesssim Y_{e}^{a} \lesssim 0.4$) matter is more likely to produce anomalous neutron-rich isotopes (e.g., $^{62}$Ni, $^{66}$Zn, $^{68}$Zn, $^{87}$Rb, $^{88}$Sr; Hartmann, Woosley, $\&$ El Eid 1985), and in some cases may be capable of producing $r$-process elements (e.g., Freiburghaus et al.~1999).  Thus, as a final application of our calculations, we briefly consider the possibility of $r$-process element synthesis in proto-magnetar and NDAF winds.  Given their intrinsic neutron-rich nature, outflows from neutron star formation have long been considered one of the most promising $r$-process sites (Woosley $\&$ Hoffman 1992; Meyer et al.~1992); however, the conditions necessary for a successful third-peak $r$-process have not been realized in detailed studies of non-rotating, non-magnetized PNS winds (QW96; Otsuki et al.~2000; Sumiyoshi et al.~2000; Wanajo et al.~2001; T01).  In MTQ07 we studied the effects of magnetic fields and rotation on the $r$-process in PNS winds for $P \gtrsim P_{\rm n}$ and constant $Y_{e}$; although we did not find solutions with a successful third-peak $r$-process (based on the criteria of Hoffman, Woosley, $\&$ Qian 1997, hereafter HWQ97), we did not consider the additional benefits of low $Y_{e}^{a}$ caused by very rapid rotation ($P < P_{\rm n}$).

Even accounting for the possibility of low $Y_{e}$ discussed in this paper, we find that all of the NDAF and most of the proto-magnetar wind solutions we have considered still fail to meet the criteria for third-peak $r$-process of HWQ97, mainly because the beneficial effects of low $Y_{e}$ are counteracted by the detrimental effects that the accompanying rapid advection has on the wind's asymptotic entropy $S^{a}$.\footnote{By the arguments given in $\S \ref{section:thermal}$, for a PNS wind to remain neutron-rich the thermal energy deposited in the wind per baryon cannot exceed the neutrinosphere's mean neutrino energy $\langle \epsilon_{\nu} \rangle$.  Thus, because most heating occurs near the PNS surface (at a temperature $\approx 0.7T_{\nu}$; QW96, eq.~[47]), the entropy added to a neutron-rich wind cannot exceed $\Delta S \sim \langle\epsilon_{\nu}\rangle/0.7T_{\nu} \approx 5$ $k_{B}$ baryon$^{-1}$ (compare $S^{a}$ and $Y_{e}^{a}$ in Tables \ref{table:yea_magnetar} and \ref{table:disktable1}), which is substantially less than that deposited in non-magnetized, non-rotating PNS winds $\Delta S \approx GMm_{\rm n}/0.7T_{\nu}R_{\nu} \approx 70(L_{\bar{\nu}_{e},51}/8)^{-1/4}$ $k_{B}$ baryon$^{-1}$ (QW96; T01).}  One possibility is that energy deposition by acoustic or MHD waves could raise the asymptotic entropy of low-$Y_{e}$ winds (Suzuki $\&$ Nagataki 2005; MTQ07).  In addition, despite their extremely low entropy, some of the most rapidly rotating and highly magnetized proto-magnetar solutions that we have calculated (e.g., the $B_{\nu} = 10^{16}$ G and $\Omega = 9000$ Hz solution shown in Figure \ref{plot:rates}) do eclipse the third-peak $r$-process threshold of HWQ97.  This occurs because $r$-process synthesis of nuclei with mean mass $A$ (typically $\sim 195$ for the third peak) by neutron captures on seed nuclei with mean proton number $\bar{Z}$ (typically $\sim 30-40$) is possible for $Y_{e}^{a} \lesssim \bar{Z}/A \approx 0.15-0.20$, even in outflows with vanishingly small $S^{a}$ (see the discussion in Wheeler et al.~1998).  The dynamical timescales for our very low $Y_{e}^{a}$ solutions are, however, much shorter than those considered by HWQ97 and several commonly made assumptions become suspect for such rapidly expanding outflows (Meyer 2002); thus, detailed nucleosynthesis calculations, which include all of the relevant rate equations, need to be completed in the regime of low $S^{a}$, low $Y_{e}^{a}$, and very rapid outflow before $r$-process success is assured.

The very rapid rotation and strong magnetic fields associated with the successful $r$-process winds that we find are extreme and will not accompany most core-collapse SNe; the substantial amount of neutron-rich material ($\gtrsim 0.1M_{\sun}$) ejected by such a proto-magnetar, however, means that just a few very rapidly rotating proto-magnetar births of this kind per Myr would noticeably affect the Galactic $r$-process abundance (Qian 2000).  Conversely, if the $r$-process yield from events like this do not resemble the observed abundances, the number of sub-millisecond magnetar births in our Galaxy can be strongly constrained; similar constraints can be placed on the incidence of the accretion-induced collapse (AIC) of a white dwarf, which produces a similar yield of low $Y_{e}$ material (e.g., Woosley $\&$ Baron 1992; Fryer et al.~1999a; Dessart et al.~2006, 2007).
\\

\acknowledgements We thank Niccolo Bucciantini and Jon Arons for helpful discussions.  We thank Wen-xin Chen and Andrei Beloborodov for making their accretion disk solutions available to us.  EQ and BDM were  supported in part by NASA grant NNG05GO22H, the David and Lucile Packard Foundation, and a NASA GSRP Fellowship to BDM.  Wind profiles are available upon request from BDM.

\newpage

\newpage
\newpage

\begin{table}
\begin{scriptsize}
\begin{center}
\vspace{0.05 in}\caption{Definitions for Commonly Used Variables}
\label{table:defs}

\begin{tabular}{ll}
\hline \hline
\\
\multicolumn{1}{c}{Variable} &
\multicolumn{1}{c}{Definition} \\
\\
\hline
\\
 $Y_{e}$ & Electron fraction (ratio of free protons to nucleons)  \\
 $Y_{e}^{\rm D}$ & Disk midplane electron fraction \\
 $Y_{e}^{\rm 0}$ & Electron fraction at the base of the wind \\
 $Y_{e}^{a}$ & Asymptotic electron fraction in the wind \\
 $Y_{e}^{\rm eq}(r)$ & Electron fraction in local weak equilibrium, defined via $dY_{e}/dt|_{Y_{e}^{\rm eq}} = 0$ (eq.~[\ref{yeevo}]) \\
 $Y_{e}^{\nu}$ & Asymptotic electron fraction in neutrino absorption equilibrium ($Y_{e}^{\rm eq}(r\rightarrow \infty) = Y_{e}^{\nu}$; eq.~[\ref{yeanu}]) \\
 $Y^{a,{\rm sat}}_{e}$ & Asymptotic electron fraction that obtains in the co-rotating, strong $B$ limit in our simulations (eq.~[\ref{ye_sat}])\\
 $\dot{M}$ & Wind mass-loss rate \\
 $\dot{M}_{\rm th}$ & Purely thermal, neutrino-driven mass-loss rate (eqs.~[\ref{mdotthermal}], [\ref{mdotthermalNDAF}])\\
 $\dot{M}_{D}$ & Disk mass accretion rate \\
 $\dot{M}_{\rm ign}$ & Minimum ``ignition'' NDAF accretion rate (eq.~[\ref{mdotign}]) \\
 $\dot{M}_{\rm n}^{\rm diff}$ & Total neutron mass diffusion rate into the accretion disk's polar region from an encasing baryon-rich wind (eq.~[\ref{mdotdiff}]). \\
 $r$ & Distance along the outflow to the center of the PNS or black hole \\
 $R_{\nu}$ & Radius of the PNS neutrinosphere and the base of the PNS wind \\
 $R_{0}$ & Distance from the black hole to the base of the NDAF wind (see Fig.~\ref{plot:diskfig})\\
 $s$ & Distance along the outflow to the monopole center in our NDAF wind calculations (see Fig.~\ref{plot:diskfig}) \\ 
 $s_{0}$ & Distance from the monopole center to the base of the NDAF wind (see Fig.~\ref{plot:diskfig}) \\
 $R_{g}$ & Black hole's gravitational radius ($GM/c^{2}$, where $M$ is the black hole mass) \\
 $R_{\rm isco}$ & Radius of the black hole's innermost stable circular orbit \\
 $R_{\rm ign}$ & ``Ignition'' radius interior to which accretion proceeds through a thin NDAF instead of a thick disk (eq.~[\ref{rign}]) \\
 $R_{\rm p}$ & Radius of the NDAF's peak integrated neutrino emission \\
 $R_{\beta}$ & Radius interior to which a thick disk enters $\beta-$equilibrium on an accretion timescale (eq.~[\ref{rbeta}]) \\
 $R_{\rm n}$ & Radius interior to which a thick disk favors a neutron-rich composition in $\beta-$equilibrium (eq.~[\ref{rn}])\\
 $\Omega$ & Rotation rate of the PNS and the base of the PNS wind \\
 $\Omega_{\rm n}$ & Rotation rate above which the PNS wind is significantly neutron-rich ($Y_{e}^{a} \lesssim 0.25$; see eq.~[\ref{ye_sat}]) \\
 $\Omega_{K}$ & Keplerian rotation rate of the accretion disk \\
 $\sigma$ & Magnetization (potential asymptotic Lorentz factor) of PNS/NDAF winds (eqs.~[\ref{magnetization}], [\ref{sigmaPNS}], [\ref{magnetization_disk}]) \\
 $\sigma_{\rm max}$ & Maximum magnetization of neutrino-heated NDAF winds (eq.~[\ref{sigma_maxNDAF}]) \\
 $\sigma_{\rm max}^{\rm n}$ & Maximum magnetization of neutrino-heated NDAF winds with a neutron excess (eq.~[\ref{sigma_maxNDAFneutron}]) \\ 
 \\
  \\
\hline
\hline
\\
\\
\\
\\
\\
\\
\\
\\
\\
\\
\\
\\
\\
\\
\\
\\
\\
\\
\end{tabular}
\end{center}
\end{scriptsize}
\end{table}

\begin{table}
\begin{scriptsize}
\begin{center}
\vspace{0.05 in}\caption{PNS Wind Properties}
\label{table:yea_magnetar}

\begin{tabular}{lccccccccccc}
\hline \hline

\\

\multicolumn{1}{c}{$L_{\bar{\nu}_{e}}$} &
\multicolumn{1}{c}{$B_{\nu}$} &
\multicolumn{1}{c}{$\Omega$} &
\multicolumn{1}{c}{$P$} &
\multicolumn{1}{c}{$\rho_{\nu}$ \tablenotemark{(a)}} &
\multicolumn{1}{c}{$Y_{e}^{0}$} &
\multicolumn{1}{c}{$Y_{e}^{a}$} &
\multicolumn{1}{c}{$\dot{M}$}&
\multicolumn{1}{c}{$\sigma$ }&
\multicolumn{1}{c}{$S^{a}$ \tablenotemark{(b)}}
\\

\multicolumn{1}{c}{$10^{51}$ ergs s$^{-1}$} &
\multicolumn{1}{c}{{\rm G}} &
\multicolumn{1}{c}{${\rm s}^{-1}$} &
\multicolumn{1}{c}{${\rm ms}$} &
\multicolumn{1}{c}{${\rm g\,cm^{-3}}$} &
\multicolumn{1}{c}{} &
\multicolumn{1}{c}{} &
\multicolumn{1}{c}{${\rm M_{\sun}\,s^{-1}}$} &
\multicolumn{1}{c}{} &
\multicolumn{1}{c}{$k_{B}$ baryon$^{-1}$} \\
\\

\\
\hline
\\
 8               & $10^{16}$ & 1000 & 6.3 & $4\times 10^{12}$ & 0.03 & 0.48 & $1.9\times 10^{-4}$ & 10 & 55 \\
 ..............  & $10^{16}$ & 4000 & 1.6 & $4\times 10^{12}$ & 0.03 & 0.47 & $7\times 10^{-4}$   & 40 & 28 \\
 ..............  & $10^{16}$ & 6000 & 1.0 & $3\times 10^{12}$ & 0.03 & 0.41 & $4\times 10^{-3}$   & 17 & 16 \\
 ..............  & $10^{16}$ & 8000 & 0.8 & $1.7\times 10^{12}$ & 0.04 & 0.26 & $6\times 10^{-2}$   & 1.9 & 10 \\
 ..............  & $10^{16}$ & 9000 & 0.7 & $1.3\times 10^{12}$ & 0.05 & 0.19 & $2.4\times 10^{-1}$   & 0.6 & 8 \\
\hline
 ..............  & $10^{15}$ & 4000 & 1.6 & $4\times 10^{12}$ & 0.03 & 0.47 & $7\times 10^{-4}$   & 0.4 & 28 \\
 ..............  & $10^{15}$ & 6000 & 1.0 & $3\times 10^{12}$ & 0.03 & 0.42 & $4\times 10^{-3}$   & 0.18 & 17 \\
 ..............  & $10^{15}$ & 8000 & 0.8 & $1.7\times 10^{12}$ & 0.04 & 0.32 & $4\times 10^{-2}$   & 0.03 & 11 \\
 ..............  & $10^{15}$ & 9000 & 0.7 & $1.3\times 10^{12}$ & 0.05 & 0.28 & $9\times 10^{-2}$   & 0.016 & 10 \\
\hline
 ..............  & $10^{14}$ & 4000 & 1.6 & $4\times 10^{12}$ & 0.03 & 0.48 & $7\times 10^{-4}$   & $4\times 10^{-3}$ & 35 \\
 ..............  & $10^{14}$ & 6000 & 1.0 & $3\times 10^{12}$ & 0.03 & 0.46 & $1.8\times 10^{-3}$ & $4\times 10^{-3}$ & 23 \\
 ..............  & $10^{14}$ & 8000 & 0.8 & $1.7\times 10^{12}$ & 0.04 & 0.44 & $7\times 10^{-3}$ & $1.6\times 10^{-3}$ & 18\\
 ..............  & $10^{14}$ & 9000 & 0.7 & $1.3\times 10^{12}$ & 0.05 & 0.42 & $1.5\times 10^{-2}$ & $1.0\times 10^{-3}$ & 16 \\
\hline
 3.5             & $10^{16}$ & 1000 & 6.3 & $7\times 10^{12}$ & 0.011 & 0.50 & $2.8\times 10^{-5}$ & 70 & 64 \\
 ..............  & $10^{16}$ & 4000 & 1.6 & $7\times 10^{12}$ & 0.011 & 0.50 & $1.0\times 10^{-4}$ & 300 & 31\\
 ..............  & $10^{16}$ & 6000 & 1.0 & $6\times 10^{12}$ & 0.011 & 0.44 & $6\times 10^{-4}$   & 120 & 17\\
 ..............  & $10^{16}$ & 8000 & 0.8 & $4\times 10^{12}$ & 0.014 & 0.22 & 0.016 & 7 & 9 \\
 ..............  & $10^{16}$ & 9000 & 0.7 & $3\times 10^{12}$ & 0.015 & 0.13 & 0.10 & 1.5 & 7 \\
\hline
 ..............  & $10^{15}$ & 4000 & 1.6 & $7\times 10^{12}$ & 0.011 & 0.50 & $1.0\times 10^{-4}$ & 3 & 31 \\
 ..............  & $10^{15}$ & 6000 & 1.0 & $6\times 10^{12}$ & 0.011 & 0.44 & $6\times 10^{-4}$   & 1.2 & 17 \\
 ..............  & $10^{15}$ & 8000 & 0.8 & $4\times 10^{12}$ & 0.014 & 0.24 & 0.014 & 0.08 & 10 \\
 ..............  & $10^{15}$ & 9000 & 0.7 & $3\times 10^{12}$ & 0.016 & 0.18 & 0.06   & 0.024 & 8 \\
\hline
 ..............  & $10^{14}$ & 4000 & 1.6 & $7\times 10^{12}$ & 0.011 & 0.50 & $9\times 10^{-5}$   & $3\times 10^{-2}$ & 32 \\
 ..............  & $10^{14}$ & 6000 & 1.0 & $6\times 10^{12}$ & 0.011 & 0.46 & $4\times 10^{-4}$ & $1.5\times 10^{-2}$ & 19 \\
 ..............  & $10^{14}$ & 8000 & 0.8 & $4\times 10^{12}$ & 0.014 & 0.37 & 5$\times 10^{-3}$ & $2.6\times 10^{-3}$ & 13 \\
 ..............  & $10^{14}$ & 9000 & 0.7 & $3\times 10^{12}$ & 0.016 & 0.34 & 0.013 & $1.2\times 10^{-3}$ & 12 \\
\hline 
 1               & $10^{15}$ & 7000 & 0.9 & $2.0\times 10^{13}$ & 0.003 & 0.42 & $9\times 10^{-5}$ & 10 & 14\\
 ..............  & $10^{15}$ & 8000 & 0.8 & $1.7\times 10^{13}$ & 0.003 & 0.23 & $1.0\times 10^{-3}$ & 1.1 & 9 \\
 ..............  & $10^{15}$ & 9000 & 0.7 & $1.2\times 10^{13}$ & 0.004 & 0.11 & 0.011 & 0.14 & 6 \\
\hline
\hline
\footnotetext[1]{Density at the base of the wind, set to enforce neutrino optical depth $\tau_{\nu} \approx 2/3$.}
\footnotetext[2]{The asymptotic entropy of the wind.}

\end{tabular}
\end{center}
\end{scriptsize}

\end{table}

\newpage

\begin{table}
\begin{scriptsize}
\begin{center}
\vspace{0.05 in}\caption{NDAF Wind Properties}
\label{table:disktable1}

\begin{tabular}{lccccccccccc}
\hline \hline

\\

\multicolumn{1}{c}{$\dot{M}_{D}$} &
\multicolumn{1}{c}{$B_{0}$} &
\multicolumn{1}{c}{$\theta$} &
\multicolumn{1}{c}{$B_{\phi,0}$ \tablenotemark{(a)}}&
\multicolumn{1}{c}{$\rho_{0}$ \tablenotemark{(b)}}&
\multicolumn{1}{c}{$Y_{e}^{0}$}  &
\multicolumn{1}{c}{$Y_{e}^{a}$}&
\multicolumn{1}{c}{$\dot{M}$}&
\multicolumn{1}{c}{$\sigma$}&
\multicolumn{1}{c}{$\dot{J}_{W}/\dot{J}_{D}$ \tablenotemark{(c)}}&
\multicolumn{1}{c}{$S^{a}$ \tablenotemark{(d)}}\\

\multicolumn{1}{c}{$M_{\sun}{\rm\,s^{-1}}$} &
\multicolumn{1}{c}{{\rm G}} &
\multicolumn{1}{c}{{\rm degrees}} &
\multicolumn{1}{c}{{\rm G}} &
\multicolumn{1}{c}{{\rm g cm$^{-3}$}} &
\multicolumn{1}{c}{} &
\multicolumn{1}{c}{} &
\multicolumn{1}{c}{${\rm M_{\sun}\,s^{-1}}$} &
\multicolumn{1}{c}{} &
\multicolumn{1}{c}{} &
\multicolumn{1}{c}{$k_{B}$ baryon$^{-1}$}\\

\\
\hline
\\
 0.2  & $10^{15}$ & 50 & 2.4$\times 10^{15}$ & 1.9$\times 10^{10}$ & 0.07 & 0.31 & 3.2 & 0.013 & 40 & 9  \\
 ..... & $10^{14}$ & 50 & 1.6$\times 10^{15}$ & 2.6$\times 10^{10}$ & 0.06 & 0.47 & 0.61 & 7$\times 10^{-4}$ & 4 & 13 \\
 .....  & $10^{13}$ & 50 & 1.1$\times 10^{15}$ & 4$\times 10^{10}$ & 0.05 & 0.51 & 0.10 & 4$\times 10^{-5}$ & 0.6 & 18 
\\
\hline
 .....  & $10^{15}$ & 60 & 2.2$\times 10^{15}$ & 1.6$\times 10^{10}$ & 0.09 & 0.32 & 2.7 & 0.016 & 40 & 10 \\
 ..... & $10^{14}$ & 60 & 1.4$\times 10^{15}$ & 2.2$\times 10^{10}$ & 0.06 & 0.47 & 0.54 & 8$\times 10^{-4}$ & 4 & 13 \\
 .....  & $10^{13}$ & 60 & 9$\times 10^{14}$   & 3$\times 10^{10}$ & 0.04 & 0.50 & 0.09 & 5$\times 10^{-5}$ & 0.5 & 18 \\
\hline
 .....  & $10^{15}$ & 70 & 1.6$\times 10^{15}$ & 1.4$\times 10^{10}$ & 0.11 & 0.35 & 1.5 & 0.028 & 40 & 11 \\
 ..... & $10^{14}$ & 70 & 1.1$\times 10^{15}$ & 2.0$\times 10^{10}$ & 0.08 & 0.46 & 0.42 & 1.0$\times 10^{-3}$ & 3 & 13 \\
 .....  & $10^{13}$ & 70 & 8$\times 10^{14}$   & 3$\times 10^{10}$ & 0.06 & 0.50 & 0.08 & 6$\times 10^{-5}$ & 0.5 & 18 \\
\hline
 .....  & $10^{15}$ & 80 & 5$\times 10^{14}$ & 3$\times 10^{10}$ & 0.05 & 0.44 & 0.11 & 0.39 & 7 & 10 \\
 ..... & $10^{14}$ & 80 & 2.8$\times 10^{14}$ & 3$\times 10^{10}$ & 0.05 & 0.47 & 0.08 & 6$\times 10^{-3}$ & 0.7 & 18 \\
 ..... & $10^{13}$ & 80 & 2.8$\times 10^{14}$   & 4$\times 10^{10}$ & 0.04 & 0.49 & 0.07 & 1.3$\times 10^{-4}$ & 0.21 & 24 \\
\hline
 .....  & $10^{15}$ & 85 & 1.9$\times 10^{14}$ & 4$\times 10^{10}$ & 0.04 & 0.50 & 0.012 & 4 & 3 & 10 \\
 ..... & $10^{14}$ & 85 & 1.1$\times 10^{14}$ & 4$\times 10^{10}$ & 0.03 & 0.50 & 0.011 & 0.04 & 0.18 & 18 \\
 ..... & $10^{13}$ & 85 & 6$\times 10^{13}$   & 4$\times 10^{10}$ & 0.03 & 0.50 & 0.009 & 5$\times 10^{-4}$ & 0.06 & 24 \\
\hline
 ..... & $10^{15}$ & 87 & 1.2$\times 10^{14}$ & 5$\times 10^{10}$ & 0.03 & 0.51 & 5$\times 10^{-3}$ & 8 & 3 & 16 \\
 ..... & $10^{14}$ & 87 & 8$\times 10^{13}$   & 5$\times 10^{10}$ & 0.03 & 0.51 & 5$\times 10^{-3}$ & 0.08 & 0.12 & 23 \\
 ..... & $10^{13}$ & 87 & 2.4$\times 10^{13}$   & 6$\times 10^{10}$ & 0.03 & 0.51 & 5$\times 10^{-3}$ & 9$\times 10^{-4}$ & 0.029 & 29 \\
\hline
..... & $10^{15}$ & 89 & 1.3$\times 10^{14}$ & 1.0$\times 10^{11}$ & 0.02 & 0.51 & 1.5$\times 10^{-3}$ & 28 & 3 & 40 \\
 ..... & $10^{14}$ & 89 & 5$\times 10^{13}$   & 1.0$\times 10^{11}$ & 0.02 & 0.51 & 1.6$\times 10^{-3}$ & 0.27 & 0.07 & 48 \\
 ..... & $10^{13}$ & 89 & 9$\times 10^{12}$   & 1.0$\times 10^{11}$ & 0.02 & 0.51 & 1.4$\times 10^{-3}$ & 3$\times 10^{-3}$ & 0.008 & 53 \\
\hline
\hline
 0.01 & $10^{14}$ & 50 & 8$\times 10^{14}$   & 5$\times 10^{9}$ & 0.06 & 0.11 & 0.21 & 2.1$\times 10^{-3}$ & 40 & 9 \\
 .... & $10^{14}$ & 60 & 8$\times 10^{14}$   & 5$\times 10^{9}$ & 0.06 & 0.12 & 0.19 & 2.2$\times 10^{-3}$ & 30 & 9 \\
 .... & $10^{14}$ & 70 & 5$\times 10^{14}$   & 5$\times 10^{9}$ & 0.06 & 0.15 & 0.10 & 4$\times 10^{-3}$ & 21 & 9 \\
 .... & $10^{14}$ & 75 & 1.3$\times 10^{14}$ & 5$\times 10^{9}$ & 0.06 & 0.27 & 0.010 & 0.04 & 4 & 12 \\
 .... & $10^{14}$ & 80 & 4$\times 10^{13}$   & 5$\times 10^{9}$ & 0.06 & 0.46 & 6$\times 10^{-4}$ & 0.7 & 1.1 & 17 \\
 .... & $10^{14}$ & 85 & 2.2$\times 10^{13}$   & 5$\times 10^{9}$ & 0.06 & 0.51 & 8$\times 10^{-5}$ & 6 & 0.6 & 36 \\
 .... & $10^{14}$ & 87 & 1.5$\times 10^{13}$   & 5$\times 10^{9}$ & 0.06 & 0.51 & 2.6$\times 10^{-5}$ & 17 & 0.6 & 55 \\
 .... & $10^{14}$ & 89 & 1.3$\times 10^{13}$ & 5$\times 10^{9}$ & 0.06 & 0.51 & 1.7$\times 10^{-5}$ & 26 & 0.6 & 60 \\

\hline
\hline
\footnotetext[1]{Azimuthal magnetic field at the base of the wind.}
\footnotetext[2]{Density at the base of the wind.}
\footnotetext[3]{The ratio of angular momentum lost in the wind to that lost through the disk (eq.~[\ref{jdotratio}]); solutions with $\dot{J}_{W} > \dot{J}_{D}$ are unphysical.}
\footnotetext[4]{The asymptotic entropy of the wind.}
\end{tabular}
\end{center}
\end{scriptsize}

\end{table}

\newpage

\begin{table}
\begin{scriptsize}
\begin{center}
\vspace{0.05 in}\caption{Neutron Content of Outflows from GRB Central Engines}
\label{table:scenarios}

\begin{tabular}{lll}
\hline \hline
\\
\multicolumn{1}{c}{Central Engine} &
\multicolumn{1}{c}{Neutron-Rich} &
\multicolumn{1}{c}{Conditions/Comments} \\
 & GRB Outflow? & 
\\
\hline
\\
 Magnetar, CC\tablenotemark{(a)}  & Sometimes & Subset with SN + long-duration GRB energy $\gtrsim 4\times 10^{52}$ ergs; \\
 & & restricted to GRB luminosities $\lesssim 3\times 10^{49}$ ergs s$^{-1}$; see $\S\ref{section:pnsresults}$ and $\S\ref{section:magnetarconclusion}$ \\
 & & $\sim 0.1 M_{\sun}$ of non-relativistic ($v \gtrsim 0.3$ c) free neutrons ejected prior to neutron-rich GRB outflow \\
 Magnetar, AIC\tablenotemark{(b)} & Sometimes & Same as CC; \\
& & additional $\sim 10^{-3}-10^{-1} M_{\sun}$ non-relativistic low-$Y_{e}$ matter ejected at early times;\tablenotemark{(c)}  \\
 & & SN-like component optically-dim due to $\lesssim 10^{-3}M_{\sun}$ total $^{56}$Ni production; see $\S\ref{section:opticaltransients}$ \\
 NDAF, CC & Unlikely & Disk midplane enters $\beta-$equilibrium;\tablenotemark{(d)} \\
 & &  outflow enters neutrino absorption equilibrium ($Y_{e}^{a} \simeq Y_{e}^{\nu}$); see $\S\ref{section:conditionsNDAF}$  \\
 NDAF, COM\tablenotemark{(e)} & Unlikely & Same as CC; \\
 & &  additional $\sim 10^{-3}-10^{-1} M_{\sun}$ non-relativistic low-$Y_{e}$ matter may be ejected at early times\tablenotemark{(f)} \\
 Low-$\dot{M}_{D}$ RIAF\tablenotemark{(g)}, CC & Unlikely & Disk midplane may not enter $\beta-$equilibrium, and stellar mantle feeding the disk has $Y_{e} \sim 0.5$; \\
 & & outflow likely viscously-driven with $Y_{e}^{a} \sim Y_{e}^{D}$; see $\S\ref{section:thickdiskwinds}$ and $\S\ref{section:thickdisks}$ \\
 Low-$\dot{M}_{D}$ RIAF, COM & Possible & Neutron star tidal debris feeding the disk has $Y_{e} \ll 0.5$, but composition altered by $\beta$-equilibrium at high $\dot{M}_{D}$; \\
 & & low $\dot{M}_{D}$-RIAF may remain neutron-rich during NDAF to RIAF transition; see $\S\ref{section:thickdiskwinds}$ \\
 & & due to low-$\dot{M}_{D}$, probably accompanies only relatively long short-duration GRBs \\
 High-$\dot{M}_{D}$ RIAF, CC & Unlikely & Disk midplane enters $\beta-$equilibrium; \\
 & & $\nu_{e}$-dominated radiation field accompanying deleptonization favors $Y_{e}^{\nu} > 0.5$;\tablenotemark{(h)}\\ 
 & & large neutrino-driven mass-loss likely precludes relativistic disk winds \\
 High-$\dot{M}_{D}$ RIAF, COM & Unlikely & Same as CC\\
 
 \\
  \\
\hline
\hline
\footnotetext[1]{Core Collapse (CC)}
\footnotetext[2]{Accretion-Induced Collapse (AIC).}
\footnotetext[3]{e.g., Hillebrandt, Wolff, $\&$ Nomoto (1984); Fryer et al.~(1999a); Dessart et al.~(2006, 2007).}
\footnotetext[4]{Pruet et al.~(2003); B03a.}
\footnotetext[5]{Compact Object Merger (COM).}
\footnotetext[6]{e.g., Rosswog et al.~(1999b).}
\footnotetext[7]{Radiatively Inefficient Accretion Flow (RIAF).}
\footnotetext[8]{e.g., Rosswog $\&$ Liebend{\"o}rfer (2003)}

\end{tabular}
\end{center}
\end{scriptsize}

\end{table}


\newpage

\begin{thebibliography}{}

\bibitem[Bahcall \& M{\'e}sz{\'a}ros(2000)]{2000PhRvL..85.1362B} Bahcall, 
J.~N., \& M{\'e}sz{\'a}ros, P.\ 2000, Physical Review Letters, 85, 1362 

\bibitem[Balbus \& Hawley(1998)]{1998RvMP...70....1B} Balbus, S.~A., \& 
Hawley, J.~F.\ 1998, Reviews of Modern Physics, 70, 1 

\bibitem[Beloborodov(2003)]{2003ApJ...588..931B} Beloborodov, A.~M.\ 2003a, 
\apj, 588, 931 

\bibitem[Beloborodov(2003)]{2003ApJ...585L..19B} Beloborodov, A.~M.\ 2003b, 
\apjl, 585, L19 

 \bibitem[Blackman \& Yi(1998)]{1998ApJ...498L..31B} Blackman, E.~G., \& Yi, 
I.\ 1998, \apjl, 498, L31 

\bibitem[Blaes et al.(2006)]{2006ApJ...645.1402B} Blaes, O.~M., Davis, 
S.~W., Hirose, S., Krolik, J.~H., \& Stone, J.~M.\ 2006b, \apj, 645, 1402 

\bibitem[Blaes(2007)]{2007astro.ph..3589B} Blaes, O.\ 2007, ArXiv 
Astrophysics e-prints, arXiv:astro-ph/0703589

\bibitem[Blandford \& Begelman(1999)]{1999MNRAS.303L...1B} Blandford, 
R.~D., \& Begelman, M.~C.\ 1999, \mnras, 303, L1 

\bibitem[Blandford \& Payne(1982)]{1982MNRAS.199..883B} Blandford, R.~D., 
\& Payne, D.~G.\ 1982, \mnras, 199, 883 

\bibitem[Bloom et al.(2005)]{2005AAS...206.4704B} Bloom, J.~B., et al.\ 
2005, Bulletin of the American Astronomical Society, 37, 793

\bibitem[Brandenburg, A.]{Brandenburg, A.} Brandenburg, A. 2003, in Advances in Nonlinear Dynamos: The Fluid Mechanics of Astrophysics \& Geophysics, Vol 9, ed. A. Ferriz-Mas \& M. Nunez (London, New York: Taylor \& Francis), 269

\bibitem[Bucciantini et al.(2007)]{2007arXiv0705.1742B} Bucciantini, N., 
Quataert, E., Arons, J., Metzger, B.~D., \& Thompson, T.~A.\ 2007, ArXiv 
e-prints, 705, arXiv:0705.1742 

\bibitem[Bucciantini et al.(2006)]{2006MNRAS.368.1717B} Bucciantini, N., 
Thompson, T.~A., Arons, J., Quataert, E., \& Del Zanna, L.\ 2006, \mnras, 
368, 1717 

\bibitem[Bulik et al.(2002)]{2002astro.ph..9339B} Bulik, T., Sikora, M., \& 
Moderski, R.\ 2002, ArXiv Astrophysics e-prints, arXiv:astro-ph/0209339 

\bibitem[Buras et al.(2003)]{2003PhRvL..90x1101B} Buras, R., Rampp, M., 
Janka, H.-T., \& Kifonidis, K.\ 2003, Physical Review Letters, 90, 241101 

\bibitem[Burrows \& Lattimer(1986)]{1986ApJ...307..178B} Burrows, A., \& 
Lattimer, J.~M.\ 1986, \apj, 307, 178 

\bibitem[Chen \& Beloborodov(2007)]{2007ApJ...657..383C} Chen, W.-X., \& 
Beloborodov, A.~M.\ 2007, \apj, 657, 383 

\bibitem[Chevalier(1989)]{1989ApJ...346..847C} Chevalier, R.~A.\ 1989, 
\apj, 346, 847 

\bibitem[Contopoulos(1994)]{1994ApJ...432..508C} Contopoulos, J.\ 1994, 
\apj, 432, 508 

\bibitem[Cook et al.(1994)]{1994ApJ...424..823C} Cook, G.~B., Shapiro, 
S.~L., \& Teukolsky, S.~A.\ 1994, \apj, 424, 823 

\bibitem[Daigne \& Mochkovitch(2002)]{2002A&A...388..189D} Daigne, F., \& 
Mochkovitch, R.\ 2002, \aap, 388, 189 

\bibitem[Derishev et al.(1999)]{1999ApJ...521..640D} Derishev, E.~V., 
Kocharovsky, V.~V., \& Kocharovsky, V.~V.\ 1999, \apj, 521, 640 

\bibitem[Dermer \& Atoyan(2006)]{2006NJPh....8..122D} Dermer, C.~D., \& 
Atoyan, A.\ 2006, New Journal of Physics, 8, 122 

\bibitem[Dessart et al.(2007)]{2007arXiv0705.3678D} Dessart, L., Burrows, 
A., Livne, E., \& Ott, C.\ 2007, ArXiv e-prints, 705, arXiv:0705.3678 

\bibitem[Dessart et al.(2006)]{2006ApJ...644.1063D} Dessart, L., Burrows, 
A., Ott, C.~D., Livne, E., Yoon, S.-C., \& Langer, N.\ 2006, \apj, 644, 
1063 

\bibitem[Di Matteo et al.(2002)]{2002ApJ...579..706D} Di Matteo, T., Perna, 
R., \& Narayan, R.\ 2002, \apj, 579, 706 

\bibitem[Drenkhahn \& Spruit(2002)]{2002A&A...391.1141D} Drenkhahn, G., \& 
Spruit, H.~C.\ 2002, \aap, 391, 1141 

\bibitem[Duan \& Qian(2004)]{2004PhRvD..69l3004D} Duan, H., \& Qian, Y.-Z.\ 
2004, \prd, 69, 123004

\bibitem[Eichler \& Levinson(1999)]{1999ApJ...521L.117E} Eichler, D., \& 
Levinson, A.\ 1999, \apjl, 521, L117 

\bibitem[Eichler et al.(1989)]{1989Natur.340..126E} Eichler, D., Livio, M., 
Piran, T., \& Schramm, D.~N.\ 1989, \nat, 340, 126 

\bibitem[Fan \& Wei(2004)]{2004ApJ...615L..69F} Fan, Y.~Z., \& Wei, D.~M.\ 
2004, \apjl, 615, L69

\bibitem[Fan et al.(2005)]{2005ApJ...628L..25F} Fan, Y.~Z., Zhang, B., \& 
Wei, D.~M.\ 2005, \apjl, 628, L25 

\bibitem[Freiburghaus et al.(1999)]{1999ApJ...525L.121F} Freiburghaus, C., 
Rosswog, S., \& Thielemann, F.-K.\ 1999, \apjl, 525, L121 

\bibitem[Fryer et al.(1999)]{1999ApJ...516..892F} Fryer, C., Benz, W., 
Herant, M., \& Colgate, S.~A.\ 1999a, \apj, 516, 892

\bibitem[Fryer \& Heger(2000)]{2000ApJ...541.1033F} Fryer, C.~L., \& Heger, 
A.\ 2000, \apj, 541, 1033 

\bibitem[Fryer \& Woosley(1998)]{1998ApJ...502L...9F} Fryer, C.~L., \& 
Woosley, S.~E.\ 1998, \apjl, 502, L9 

\bibitem[Fryer et al.(1999)]{1999ApJ...520..650F} Fryer, C.~L., Woosley, 
S.~E., Herant, M., \& Davies, M.~B.\ 1999b, \apj, 520, 650 

\bibitem[Fuller et al.(2000)]{2000PhRvL..85.2673F} Fuller, G.~M., Pruet, 
J., \& Abazajian, K.\ 2000, Physical Review Letters, 85, 2673 

\bibitem[Fuller \& Qian (1996)]{1996NucPhysA...606...167} Fuller, G.~M., \& Qian, Y.-Z. 1996, Nucl. Phys. A, 606, 167

\bibitem[Giannios(2007)]{2007arXiv0704.1659G} Giannios, D.\ 2007, ArXiv 
e-prints, 704, arXiv:0704.1659 

\bibitem[Hartmann et al.(1985)]{1985ApJ...297..837H} Hartmann, D., Woosley, 
S.~E., \& El Eid, M.~F.\ 1985, \apj, 297,

\bibitem[Hawley et al.(1995)]{1995ApJ...440..742H} Hawley, J.~F., Gammie, 
C.~F., \& Balbus, S.~A.\ 1995, \apj, 440, 742 

\bibitem[Hillebrandt et al.(1984)]{1984A&A...133..175H} Hillebrandt, W., 
Wolff, R.~G., \& Nomoto, K.\ 1984, \aap, 133, 175 

\bibitem[Hirose et al.(2006)]{2006ApJ...640..901H} Hirose, S., Krolik, 
J.~H., \& Stone, J.~M.\ 2006, \apj, 640, 901 

\bibitem[Hjorth et al.(2005)]{2005ApJ...630L.117H} Hjorth, J., et al.\ 
2005, \apjl, 630, L117 

\bibitem[Hoffman et al.(1997)]{1997ApJ...482..951H} Hoffman, R.~D., 
Woosley, S.~E., \& Qian, Y.-Z.\ 1997, \apj, 482, 951 

\bibitem[Iwamoto et al.(1998)]{1998Natur.395..672I} Iwamoto, K., et al.\ 
1998, \nat, 395, 672  

\bibitem[Janka et al.(1999)]{1999ApJ...527L..39J} Janka, H.-T., Eberl, T., 
Ruffert, M., \& Fryer, C.~L.\ 1999, \apjl, 527, L39 

\bibitem[Katz(1997)]{1997ApJ...490..633K} Katz, J.~I.\ 1997, \apj, 490, 633 

\bibitem[Klu{\'z}niak \& Ruderman(1998)]{1998ApJ...505L.113K} Klu{\'z}niak, 
W., \& Ruderman, M.\ 1998, \apjl, 505, L113 

\bibitem[Krolik et al.(2007)]{2007ApJ...664.1045K} Krolik, J.~H., Hirose, 
S., \& Blaes, O.\ 2007, \apj, 664, 1045 

\bibitem[Kulkarni(2005)]{2005astro.ph.10256K} Kulkarni, S.~R.\ 2005, ArXiv 
Astrophysics e-prints, arXiv:astro-ph/0510256 

\bibitem[Lai \& Qian(1998)]{1998ApJ...505..844L} Lai, D., \& Qian, Y.-Z.\ 
1998, \apj, 505, 844 

\bibitem[Lamers \& Cassinelli]{628} Lamers, H. \& Cassinelli, J.~P. 1999, Introduction to Stellar Winds (Cambridge: Cambridge University Press) 

\bibitem[Lee et al.(2005)]{2005ApJ...632..421L} Lee, W.~H., Ramirez-Ruiz, 
E., \& Page, D.\ 2005, \apj, 632, 421 

\bibitem[Lemoine(2002)]{2002A&A...390L..31L} Lemoine, M.\ 2002, \aap, 390, 
L31 

\bibitem[Levinson(2006)]{2006ApJ...648..510L} Levinson, A.\ 2006, \apj, 
648, 510 

\bibitem[Levinson \& Eichler(1993)]{1993ApJ...418..386L} Levinson, A., \& 
Eichler, D.\ 1993, \apj, 418, 386 

\bibitem[Levinson \& Eichler(2003)]{2003ApJ...594L..19L} Levinson, A., \& 
Eichler, D.\ 2003, \apjl, 594, L19 

\bibitem[Li \& Paczy{\'n}ski(1998)]{1998ApJ...507L..59L} Li, L.-X., \& 
Paczy{\'n}ski, B.\ 1998, \apjl, 507, L59

\bibitem[Li et al.(1992)]{1992ApJ...394..459L} Li, Z.-Y., Chiueh, T., \& 
Begelman, M.~C.\ 1992, \apj, 394, 459 

\bibitem[Lithwick \& Sari(2001)]{2001ApJ...555..540L} Lithwick, Y., \& 
Sari, R.\ 2001, \apj, 555, 540 

\bibitem[MacFadyen \& Woosley(1999)]{1999ApJ...524..262M} MacFadyen, A.~I., 
\& Woosley, S.~E.\ 1999, \apj, 524, 262 

\bibitem[McKinney(2005)]{2005astro.ph..6368M} McKinney, J.~C.\ 2005, ArXiv 
Astrophysics e-prints, arXiv:astro-ph/0506368 

\bibitem[M{\'e}sz{\'a}ros \& Rees(2000)]{2000ApJ...541L...5M} 
M{\'e}sz{\'a}ros, P., \& Rees, M.~J.\ 2000, \apjl, 541, L5

\bibitem[Metzger et al.(2007)]{2007ApJ...659..561M} Metzger, B.~D., 
Thompson, T.~A., \& Quataert, E.\ 2007, \apj, 659, 561 

\bibitem[Meyer et al.(1992)]{1992ApJ...399..656M} Meyer, B.~S., Mathews, 
G.~J., Howard, W.~M., Woosley, S.~E., \& Hoffman, R.~D.\ 1992, \apj, 399, 
656 

\bibitem[Meyer(2002)]{2002PhRvL..89w1101M} Meyer, B.~S.\ 2002, Physical 
Review Letters, 89, 231101 

\bibitem[Narayan \& Yi(1994)]{1994ApJ...428L..13N} Narayan, R., \& Yi, I.\ 
1994, \apjl, 428, L13 

\bibitem[Narayan et al.(1992)]{1992ApJ...395L..83N} Narayan, R., Paczy{\'n}ski, 
B., \& Piran, T.\ 1992, \apjl, 395, L83 

\bibitem[Narayan et al.(2001)]{2001ApJ...557..949N} Narayan, R., Piran, T., 
\& Kumar, P.\ 2001, \apj, 557, 949 

\bibitem[Otsuki et al.(2000)]{2000ApJ...533..424O} Otsuki, K., Tagoshi, H., 
Kajino, T., \& Wanajo, S.-y.\ 2000, \apj, 533, 424 

\bibitem[Paczy{\'n}ski(1986)]{1986ApJ...308L..43P} Paczy{\'n}ski, B.\ 1986, \apjl, 
308, L43 

\bibitem[Paczy{\'n}ski(1991)]{1991AcA....41..257P} Paczy{\'n}ski, B.\ 1991, Acta 
Astronomica, 41, 257 

\bibitem[Pons et al.(1999)]{1999ApJ...513..780P} Pons, J.~A., Reddy, S., 
Prakash, M., Lattimer, J.~M., \& Miralles, J.~A.\ 1999, \apj, 513, 780 

\bibitem[Popham et al.(1999)]{1999ApJ...518..356P} Popham, R., Woosley, 
S.~E., \& Fryer, C.\ 1999, \apj, 518, 356 

\bibitem[Proga \& Begelman(2003)]{2003ApJ...592..767P} Proga, D., \& 
Begelman, M.~C.\ 2003, \apj, 592, 767 

\bibitem[Proga et al.(2003)]{2003ApJ...599L...5P} Proga, D., MacFadyen, 
A.~I., Armitage, P.~J., \& Begelman, M.~C.\ 2003, \apjl, 599, L5 

\bibitem[Pruet \& Dalal(2002)]{2002ApJ...573..770P} Pruet, J., \& Dalal, 
N.\ 2002, \apj, 573, 770 

\bibitem[Pruet et al.(2001)]{2001ApJ...561..957P} Pruet, J., Fuller, G.~M., 
\& Cardall, C.~Y.\ 2001, \apj, 561, 957

\bibitem[Pruet et al.(2002)]{2002ApJ...580..368P} Pruet, J., Guiles, S., \& 
Fuller, G.~M.\ 2002, \apj, 580, 368 

\bibitem[Pruet et al.(2004)]{2004ApJ...606.1006P} Pruet, J., Thompson, 
T.~A., \& Hoffman, R.~D.\ 2004, \apj, 606, 1006 

\bibitem[Pruet et al.(2003)]{2003ApJ...586.1254P} Pruet, J., Woosley, 
S.~E., \& Hoffman, R.~D.\ 2003, \apj, 586, 1254

\bibitem[Qian et al.(1993)]{1993PhRvL..71.1965Q} Qian, Y.-Z., Fuller, 
G.~M., Mathews, G.~J., Mayle, R.~W., Wilson, J.~R., \& Woosley, S.~E.\ 
1993, Physical Review Letters, 71, 1965 

\bibitem[Qian \& Woosley(1996)]{1996ApJ...471..331Q} Qian, Y.-Z., \& 
Woosley, S.~E.\ 1996, \apj, 471, 331 

\bibitem[Razzaque \& M{\'e}sz{\'a}ros(2006)]{2006ApJ...650..998R} Razzaque, 
S., \& M{\'e}sz{\'a}ros, P.\ 2006a, \apj, 650, 998 

\bibitem[Razzaque \& M{\'e}sz{\'a}ros(2006)]{2006JCAP...06..006R} Razzaque, 
S., \& M{\'e}sz{\'a}ros, P.\ 2006b, Journal of Cosmology and Astro-Particle 
Physics, 6, 6 

\bibitem[Rees \& M{\'e}sz{\'a}ros(1992)]{1992MNRAS.258P..41R} Rees, M.~J., \& 
M{\'e}sz{\'a}ros, P.\ 1992, \mnras, 258, 41P 

\bibitem[Rossi et al.(2006)]{2006MNRAS.369.1797R} Rossi, E.~M., 
Beloborodov, A.~M., \& Rees, M.~J.\ 2006, \mnras, 369, 1797 

\bibitem[Rosswog \& Liebend{\"o}rfer(2003)]{2003MNRAS.342..673R} Rosswog, 
S., \& Liebend{\"o}rfer, M.\ 2003, \mnras, 342, 673

\bibitem[Rosswog et al.(1999)]{1999A&A...341..499R} Rosswog, S., 
Liebend{\"o}rfer, M., Thielemann, F.-K., Davies, M.~B., Benz, W., \& Piran, 
T.\ 1999, \aap, 341, 499 

\bibitem[Rosswog \& Ramirez-Ruiz(2002)]{2002MNRAS.336L...7R} Rosswog, S., 
\& Ramirez-Ruiz, E.\ 2002, \mnras, 336, L7 

\bibitem[Rosswog et al.(2003)]{2003MNRAS.345.1077R} Rosswog, S., 
Ramirez-Ruiz, E., \& Davies, M.~B.\ 2003, \mnras, 345, 1077 

\bibitem[Ruffert et al.(1997)]{1997A&A...319..122R} Ruffert, M., Janka, 
H.-T., Takahashi, K., \& Schaefer, G.\ 1997, \aap, 319, 122 

\bibitem[Salmonson \& Wilson(1999)]{1999ApJ...517..859S} Salmonson, J.~D., 
\& Wilson, J.~R.\ 1999, \apj, 517, 859 

\bibitem[Sawyer(2003)]{2003PhRvD..68f3001S} Sawyer, R.~F.\ 2003, \prd, 68, 
063001

\bibitem[Spitkovsky(2006)]{2006ApJ...648L..51S} Spitkovsky, A.\ 2006, 
\apjl, 648, L51 

\bibitem[Spruit \& Uzdensky(2005)]{2005ApJ...629..960S} Spruit, H.~C., \& 
Uzdensky, D.~A.\ 2005, \apj, 629, 960 

\bibitem[Strumia \& Vissani(2003)]{2003PhLB..564...42S} Strumia, A., \& 
Vissani, F.\ 2003, Physics Letters B, 564, 42 

\bibitem[Sumiyoshi et al.(2000)]{2000PASJ...52..601S} Sumiyoshi, K.,
> Suzuki, H., Otsuki, K., Terasawa, M., \& Yamada, S.\ 2000, \pasj, 52, 601

\bibitem[Surman \& McLaughlin(2004)]{2004ApJ...603..611S} Surman, R., \& 
McLaughlin, G.~C.\ 2004, \apj, 603, 611

\bibitem[Surman \& McLaughlin(2005)]{2005ApJ...618..397S} Surman, R., \& 
McLaughlin, G.~C.\ 2005, \apj, 618, 397  

\bibitem[Suzuki \& Nagataki(2005)]{2005ApJ...628..914S} Suzuki, T.~K., \& 
Nagataki, S.\ 2005, \apj, 628, 914 

\bibitem[Thompson(1994)]{1994MNRAS.270..480T} Thompson, C.\ 1994, \mnras, 
270, 480 

\bibitem[Thompson et al.(2001)]{2001ApJ...562..887T} Thompson, T.~A., 
Burrows, A., \& Meyer, B.~S.\ 2001, \apj, 562, 887 

\bibitem[Thompson et al.(2004)]{2004ApJ...611..380T} Thompson, T.~A., 
Chang, P., \& Quataert, E.\ 2004, \apj, 611, 380 

\bibitem[Thompson et al.(2005)]{2005ApJ...620..861T} Thompson, T.~A., 
Quataert, E., \& Burrows, A.\ 2005, \apj, 620, 861 

\bibitem[Turner(2004)]{2004ApJ...605L..45T} Turner, N.~J.\ 2004, \apjl, 
605, L45 

\bibitem[Usov(1992)]{1992Natur.357..472U} Usov, V.~V.\ 1992, \nat, 357, 472 

\bibitem[Vlahakis et al.(2003)]{2003ApJ...594L..23V} Vlahakis, N., Peng, 
F., Konigl, A.\ 2003, \apjl, 594, L23 

\bibitem[Wanajo et al.(2001)]{2001ApJ...554..578W} Wanajo, S., Kajino, T.,
> Mathews, G.~J., \& Otsuki, K.\ 2001, \apj, 554, 578

\bibitem[Wheeler et al.(1998)]{1998ApJ...493L.101W} Wheeler, J.~C., Cowan, 
J.~J., \& Hillebrandt, W.\ 1998, \apjl, 493, L101 

\bibitem[Wheeler et al.(2000)]{2000ApJ...537..810W} Wheeler, J.~C., Yi, I., 
H{\"o}flich, P., \& Wang, L.\ 2000, \apj, 537, 810 

\bibitem[Woosley(1993)]{1993ApJ...405..273W} Woosley, S.~E.\ 1993, \apj, 
405, 273 

\bibitem[Woosley \& Baron(1992)]{1992ApJ...391..228W} Woosley, S.~E., \& 
Baron, E.\ 1992, \apj, 391, 228 

\bibitem[Woosley et al.(1999)]{1999ApJ...516..788W} Woosley, S.~E., 
Eastman, R.~G., \& Schmidt, B.~P.\ 1999, \apj, 516, 788 

\bibitem[Woosley \& Hoffman(1992)]{1992ApJ...395..202W} Woosley, S.~E., \& 
Hoffman, R.~D.\ 1992, \apj, 395, 202 

\bibitem[Woosley \& Weaver(1995)]{1995ApJS..101..181W} Woosley, S.~E., \& 
Weaver, T.~A.\ 1995, \apjs, 101, 181 

\end{thebibliography}
\end{document}